\documentclass[a4paper,11pt]{article}
\pdfoutput=1 

\usepackage{jheppub} 
\usepackage{bm}
\usepackage{array}
\usepackage{multirow}

\usepackage[T1]{fontenc} 
\usepackage[pdftex]{color}

\title{\boldmath 
Resummation and renormalization of kinematical effects in 
inclusive $P$-wave quarkonium production 
}

\author[a]{Hee~Sok~Chung}

\affiliation[a]{Department of Physics, Korea University, Seoul 02841, Korea}

\emailAdd{neville@korea.ac.kr}

\abstract{
We investigate the renormalization properties of the shape function formalism
for inclusive production of $P$-wave heavy quarkonia, 
which arises from resumming
a class of corrections coming from kinematical effects associated with the
motion of the heavy quark and antiquark pair relative to the quarkonium. 
Such kinematical effects are encoded in the nonperturbative shape functions, 
which are normalized to the corresponding nonrelativistic QCD
long-distance matrix elements. 
By using the known ultraviolet divergences in the matrix elements, 
we derive the large-momentum asymptotic behavior of the shape functions. 
This strongly constrains the form of the shape functions
and significantly reduces the dependence on the nonperturbative model. 
Based on these results 
we show that the shape function formalism at loop level can be useful in
taming the threshold logarithms at large transverse momentum, 
and at small transverse momentum the kinematical corrections 
reduce the sizes of $\chi_c$ and $\chi_b$ cross sections 
which may improve agreement with measurements. 
}

\begin{document}

\maketitle

\flushbottom

\section{Introduction} 
\label{sec:intro}

First-principles based studies of inclusive heavy quarkonium production
phenomenology have usually been done in the nonrelativistic QCD (NRQCD) 
factorization formalism~\cite{Bodwin:1994jh}. 
In this formalism, the production rate of a heavy quarkonium is factorized into 
products of perturbatively calculable short-distance coefficients and
nonperturbative long-distance matrix elements. 
The predictive power of the factorization formalism comes from the 
universality and the nonrelativistic power counting of the matrix elements, 
which allow quantitative descriptions of production rates 
once values of a few nonperturbative parameters are determined from 
experiments. 
This approach had limited success in
deciphering the heavy quarkonium production mechanism: 
as have been discussed in the literature, results from the global fit analyses of 
$J/\psi$ production based mostly on total inclusive production and low
transverse momentum data~\cite{Butenschoen:2010rq, Butenschoen:2011yh,
Butenschoen:2011ks, Butenschoen:2012qh, Butenschoen:2012qr, Butenschoen:2012px} 
are in conflict with analyses mostly based on large transverse momentum 
data~\cite{Ma:2010yw, Gong:2012ug, Han:2014kxa, Bodwin:2015iua, Feng:2018ukp,
Chung:2018lyq, Brambilla:2022rjd, Chung:2022uih, Brambilla:2022ayc}
and with LHC measurements of polarized and polarization-summed hadroproduction 
rates at large transverse momentum. Similarly, in the case of 
$\psi(2S)$, NRQCD analyses generally have difficulty explaining production 
rates with transverse momentum smaller than 2 -- 3 times the heavy quarkonium 
mass~\cite{Han:2014kxa, Bodwin:2015iua, Butenschoen:2022qka, Brambilla:2022rjd, 
Brambilla:2022ayc}. 
Moreover, some matrix element determinations can lead to 
large uncertainties in the cross sections at very large transverse momentum
due to mixings of contributions
from different channels, which comes from renormalization of the long-distance
matrix elements~\cite{Butenschoen:2022qka}. 
These call for investigation of possible refinements to the NRQCD factorization
approach in 
quarkonium production phenomenology that can modify the transverse momentum
dependences of the cross sections. 

In an inclusive production process, the quark $Q$ and antiquark $\bar Q$
produced in a hard process can have nonzero total momentum relative to the
heavy quarkonium. 
This happens because the $Q \bar Q$ can emit an arbitrary number of soft 
particles before evolving into a heavy quarkonium, which carry small but
nonzero momenta. 
This kinematical effect gives rise to operators that are given in the form of
total derivatives of the $Q \bar Q$ bilinear~\cite{Mannel:1994xh,
Rothstein:1997ac, Mannel:1997uk}. 
While these operators are suppressed by powers of the heavy-quark velocity $v$ 
in the nonrelativistic power counting of NRQCD, they can be enhanced due to
threshold effects associated with the boundary of phase
space~\cite{Rothstein:1997ac, Mannel:1997uk}. 
The corrections from these operators can be partially resummed 
to all orders in $v$, which leads to the shape function formalism developed in
ref.~\cite{Beneke:1997qw}, 
where products of short-distance coefficients and NRQCD matrix elements 
are replaced by convolutions of short-distance coefficients and nonperturbative
shape functions. Here, shape functions are functions of the momentum of the
$Q \bar Q$ pair in the rest frame of the heavy quarkonium in the final state. 
A similar formalism based on the soft-collinear effective theory have been
developed in ref.~\cite{Fleming:2003gt}. 
It has been anticipated that the kinematical effects computed in the 
shape function formalism will have a significant impact on the
transverse-momentum differential cross sections of heavy quarkonia. 
However, phenomenological applications of this formalism have been limited, 
because shape functions are nonperturbative functions that are in general
unknown, other than the fact that they are normalized to the corresponding 
NRQCD matrix elements. 
That is, unless one can determine the shape functions from first principles, 
the shape function formalism has less predictive power than the NRQCD
factorization formalism. 

Another shortcoming of the shape function formalism is that, in its
original form, it does not correctly incorporate the renormalization of
the NRQCD matrix elements. Because the shape functions are normalized to the
NRQCD matrix elements, their normalizations must reproduce the same ultraviolet
(UV)
divergences in the matrix elements for the formalism to be consistent with 
NRQCD. 
This point has not been considered in the development of the 
formalism in refs.~\cite{Beneke:1997qw, Fleming:2003gt}, 
because at that time, phenomenological applications of
the NRQCD factorization formalism have usually been done at leading order in
the strong coupling. 

The UV divergences in the normalizations of the shape functions can
have important phenomenological implications. First, the known UV 
divergences fix the large-momentum asymptotic behaviors of the shape functions, 
which strongly constrain their forms. This
happens because, as we will see later, the UV divergence comes 
from the normalization integral of the shape function 
over the momentum of the $Q \bar Q$ in the
quarkonium rest frame. This significantly enhances the predictive power of the
shape function formalism, compared to existing studies where models for the 
shape functions had to be chosen arbitrarily. 
Second, the UV divergences induce momentum-dependent mixings between
channels, which modify the matching conditions that are needed to determine the
short-distance coefficients. This makes the matching conditions 
originally obtained in ref.~\cite{Beneke:1997qw} invalid beyond tree level, 
because there the mixing
effects were not taken into account. Since current-day phenomenological studies 
of quarkonium production are usually carried out at next-to-leading order (NLO) 
in $\alpha_s$, a correct derivation of the matching conditions incorporating
mixing effects is necessary. 

In this work, we consider the inclusive production of $P$-wave heavy
quarkonia in the shape function formalism at NLO accuracy. 
We focus on the simpler case of production of $P$-wave quarkonia, 
which involves only one unknown nonperturbative matrix element, 
and the renormalization of the matrix elements is less complicated 
compared to the $S$-wave case. The $P$-wave production rate is also important 
in understanding the feeddown contributions in $S$-wave quarkonium production. 
In order to correctly incorporate the renormalization of NRQCD matrix elements, 
we compute the shape functions in perturbative QCD at NLO accuracy, 
which lets us obtain one-loop matching conditions for the shape function
formalism that are necessary for computing the short-distance coefficients. 
This also gives us the large-momentum asymptotic forms of the shape functions, 
which, together with the normalization condition, help strongly constrain the 
nonperturbative form of the shape function
and significantly reduce the model dependence. 
Furthermore, we can compute the shape functions in terms of quarkonium
wavefunctions and universal gluonic operator vacuum expectation values using
the potential NRQCD (pNRQCD) formalism developed in refs.~\cite{Pineda:1997bj,
Brambilla:1999xf, Brambilla:2001xy, Brambilla:2002nu, Brambilla:2004jw,
Brambilla:2020ojz, Brambilla:2021abf, Brambilla:2022rjd, Brambilla:2022ayc}.
Although such a calculation does not yet lead to a 
first-principles determination of the shape functions due to our
lack of knowledge of the nonperturbative dynamics of gluons, the universality of
the gluon operator vacuum expectation values can enhance the predictive power
of the formalism, similarly to the pNRQCD calculations of NRQCD matrix
elements. 
Based on the results for the shape functions and the corresponding
short-distance coefficients we obtain in this work, we resum the
kinematical corrections to inclusive $\chi_c$ and $\chi_b$
hadroproduction rates at small and large transverse momentum. 

This paper is organized as follows. In sec.~\ref{sec:shape} we introduce the 
NRQCD and shape function formalisms for $P$-wave quarkonium production. 
In sec.~\ref{sec:diagram} we compute the shape functions in perturbative QCD,
which are necessary in deriving loop-level matching conditions for the shape
function formalism. In sec.~\ref{sec:pNRQCD} we compute the shape functions
nonperturbatively in pNRQCD and obtain expressions in terms of 
quarkonium wavefunctions and universal gluonic operator vacuum expectation 
values. In sec.~\ref{sec:matching} we establish the matching conditions for
the shape function formalism that allow us to obtain short-distance
coefficients from the known results in NRQCD. 
By using these results we discuss phenomenological applications of the shape
function formalism for hadroproduction of $\chi_c$ and $\chi_b$ at large
and small transverse momentum in sec.~\ref{sec:pheno}. We conclude in
sec.~\ref{sec:summary}.

\section{\boldmath Shape function formalism for $P$-wave production} 
\label{sec:shape}

We first review the NRQCD factorization formalism for production of a
$\chi_{QJ}$ for $J=0$, 1, and 2, where $Q = c$ or $b$. 
At leading order in $v$, the inclusive production cross section 
of $\chi_{QJ}$ is given by~\cite{Bodwin:1994jh}
\begin{equation}
\label{eq:nrqcdfac}
\sigma[\chi_{QJ}(P)] = (2 J+1) 
\left(
c_{^3P_J^{[1]}}(P) 
\langle {\cal O}^{\chi_{Q0}} (^3P_0^{[1]}) \rangle
+ c_{^3S_1^{[8]}} (P) 
\langle {\cal O}^{\chi_{Q0}} (^3S_1^{[8]}) \rangle \right), 
\end{equation}
where $P$ is the momentum of the $\chi_{QJ}$, $c_{^3P_J^{[1]}}(P)$ 
and $c_{^3S_1^{[8]}} (P)$ are short-distance coefficients, 
$\langle {\cal O} \rangle$ denote the vacuum expectation value of an operator
${\cal O}$, 
and the $\langle {\cal O}^{\chi_{Q0}} (^3P_0^{[1]}) \rangle$
and $\langle {\cal O}^{\chi_{Q0}} (^3S_1^{[8]}) \rangle$ 
are NRQCD long-distance matrix elements that describe the 
evolution of a $Q \bar Q$ in a specific color and angular momentum state 
into $\chi_{Q0}$+anything. 
We have used the heavy-quark spin symmetry relations, which are valid up to
corrections of order $v^2$, to write the $\chi_{QJ}$ matrix elements in terms
of the ones for $\chi_{Q0}$. 
The definitions of the NRQCD matrix elements read 
\begin{subequations}
\label{eq:MEdefs}
\begin{eqnarray}
\langle 
{\cal O}^{\chi_{Q0}} (^3P_0^{[1]}) 
\rangle 
&=&
\frac{1}{d-1}
\langle 
\chi^\dag ( -\tfrac{i}{2} \overleftrightarrow{\bm{D}}
\cdot \bm{\sigma} ) \psi {\cal P}_{\chi_{Q0}} \psi^\dag ( -\tfrac{i}{2}
\overleftrightarrow{\bm{D}} \cdot \bm{\sigma} ) \chi
 \rangle ,
\\
\langle {\cal O}^{\chi_{Q0}} (^3S_1^{[8]})  \rangle 
&=&
\langle 
\chi^\dag \sigma^i T^a \psi \Phi_\ell^{\dag ab} (0) {\cal P}_{\chi_{Q0}}
\Phi_\ell^{bc} (0) \psi^\dag \sigma^i T^c \chi
\rangle ,
\end{eqnarray}
\end{subequations}
where $d$ is the number of spacetime dimensions, 
$\psi$ and $\chi$ are Pauli spinor fields that destroy and create a heavy
quark and antiquark, respectively, $\bm{D} = \bm{\nabla}-i g \bm{A}$ 
is the gauge-covariant derivative, $\bm{A}$ is the gluon field, 
$\overleftrightarrow{\bm{D}}$ is defined through the relation
$\psi^\dag \overleftrightarrow{\bm{D}} \chi =
\psi^\dag \bm{D} \chi - (\bm{D} \psi)^\dag \chi$,
$\bm{\sigma}$ is a Pauli matrix, ${\cal P}_{\chi_{Q0}}$ is a
projection onto states that include the quarkonium $\chi_{Q0}$ at rest,
and $T^a$ is a $SU(3)$ color matrix.
In the color-octet operator ${\cal O}^{\chi_{Q0}} (^3S_1^{[8]})$,
the adjoint Wilson line $\Phi_\ell (x)= P \exp [ -i g \int_0^\infty d z
\ell \cdot A(x+\ell z)]$, where $P$ is the path ordering for color matrices and
$\ell$ is an arbitrary lightlike direction, is inserted to ensure the gauge
invariance of the matrix element~\cite{Nayak:2005rw, Nayak:2005rt, 
Nayak:2006fm}.
The $Q \bar Q$ bilinears on the right and left of the projection operator
${\cal P}_{\chi_{Q0}}$ create and destroy a $Q \bar Q$ in a specific color and
angular momentum state.

Equation~(\ref{eq:nrqcdfac}) follows from a factorization conjecture, where the
nonperturbative long-distance physics of scales below the heavy-quark mass $m$
is encoded in the matrix elements, while the short-distance coefficients depend
only on the short-distance process of the perturbative production of 
$Q \bar Q$. In this case, a $Q \bar Q$ in either a color-singlet or color-octet
state can evolve into a color-singlet quarkonium by emitting an arbitrary
number of light particles with soft momentum of scales $mv$ and lower. 
Hence, the $Q \bar Q$ produced in a short-distance process can have a relative
momentum of order $mv$ with respect to the produced quarkonium. 
The nonvanishing of this relative momentum can have a significant effect on 
quarkonium production observables, if the $Q \bar Q$ is produced in the
short-distance process predominantly near the boundary of phase space. 
In this case, logarithmic corrections to the short-distance coefficients from
radiation of soft gluons can become enhanced, which must be matched with
the effect of soft gluon emission in the NRQCD matrix elements. 
Also, in experimental observables kinematical cuts on the energy or momentum of
the quarkonium are often taken, which can be sensitive to the order-$mv$ change 
in the quarkonium momentum. 
In the NRQCD factorization formula in its usual form, the effect of the
relative momentum between the $Q \bar Q$ and the quarkonium
is not taken into account, because the scales $mv$ and lower do not affect the
short-distance coefficients. Hence, in NRQCD, the momentum of the $Q \bar Q$ produced in
the hard process is identified with the momentum of the quarkonium. 
Therefore, predictions based on the factorization formula in 
eq.~(\ref{eq:nrqcdfac}) can suffer from large corrections from kinematical
effects associated with the motion of the $Q \bar Q$ in the quarkonium rest
frame. 

The kinematical effects can be taken into account by including 
contributions from certain higher dimensional operators
that involve total derivatives of the $Q \bar Q$ bilinears 
in the factorization formula.
The motion of the $Q \bar Q$ along a momentum $l$ gives rise to the 
following forms of higher dimensional matrix elements~\cite{Mannel:1994xh,
Rothstein:1997ac, Mannel:1997uk}:
\begin{subequations}
\label{eq:highdim}
\begin{eqnarray}
&& 
\frac{1}{d-1}
\langle 
\chi^\dag ( -\tfrac{i}{2} \overleftrightarrow{\bm{D}} \cdot \bm{\sigma} ) \psi
{\cal P}_{\chi_{Q0}}
(l \cdot D)^n 
\psi^\dag ( -\tfrac{i}{2} \overleftrightarrow{\bm{D}} \cdot \bm{\sigma} ) 
\chi
 \rangle
,  \\
&&
\label{eq:highdimoctet}
\langle 
\chi^\dag \sigma^i T^a  \psi \Phi_\ell^{\dag ab} (0)
{\cal P}_{\chi_{Q0}} (l \cdot D)^n \Phi_\ell^{bc} (0)
\psi^\dag \sigma^i T^c \chi 
\rangle,
\end{eqnarray}
\end{subequations}
where the derivatives $D$ act to the right and 
$n$ is a positive integer. 
The operators involving derivatives on the left of the projector 
${\cal P}_{\chi_{Q0}}$ can be obtained by using Hermitian conjugation and the
invariance of the vacuum. 
Note that matrix elements of this kind do not appear in exclusive production
due to conservation of energy and momentum. 
As have been shown in refs.~\cite{Beneke:1997qw, Fleming:2003gt}, 
near the boundary of phase space, 
contributions from the above matrix elements associated with a lightlike
momentum $l$ collinear to the quarkonium momentum $P$ become enhanced. 
These matrix elements can be obtained from
moments of the ``shape functions'' defined by 
\begin{subequations}
\label{eq:shape}
\begin{eqnarray}
{\cal S}^{\chi_{Q0}}_{^3P_0^{[1]}} (l_+) &=& 
\frac{1}{d-1}
\langle 
\chi^\dag ( -\tfrac{i}{2} \overleftrightarrow{\bm{D}} \cdot \bm{\sigma} ) \psi
{\cal P}_{\chi_{Q0}}
\delta( l_+ - i D_+) 
\psi^\dag ( -\tfrac{i}{2} \overleftrightarrow{\bm{D}} \cdot \bm{\sigma} )
\chi
\rangle
,  \\
{\cal S}^{\chi_{Q0}}_{^3S_1^{[8]}} (l_+) &=& 
\langle 
\chi^\dag \sigma^i T^a  \psi \Phi_\ell^{\dag ab} (0)
{\cal P}_{\chi_{Q0}} \delta( l_+ - i D_+)
\Phi_\ell^{bc} (0) \psi^\dag \sigma^i T^c \chi 
\rangle,
\end{eqnarray}
\end{subequations}
where $\delta(x)$ is the Dirac delta function, and 
the $+$ direction is defined along the quarkonium momentum $P$. 
We take the convention $l_\pm = l_0 \pm l_z$, where $l_z = \bm{P} \cdot
\bm{l}/|\bm{P}|$ in the frame where the quarkonium three-momentum 
$\bm{P}$ is nonzero. 
From the above definitions, it is evident that they are formally normalized by 
$\int_0^\infty dl_+ {\cal S}^{\chi_{Q0}}_{^3P_0^{[1]}} (l_+)
= \langle {\cal O}^{\chi_{Q0}} (^3P_0^{[1]}) \rangle$
and 
$\int_0^\infty dl_+ {\cal S}^{\chi_{Q0}}_{^3S_1^{[8]}} (l_+)
= \langle {\cal O}^{\chi_{Q0}} (^3S_1^{[8]}) \rangle$. 
The shape functions defined in eqs.~(\ref{eq:shape}) can be interpreted as the
probabilities for a $Q \bar Q$ to evolve into a quarkonium after emitting soft
particles with total momentum $l$. 
The requirement that NRQCD factorization holds after inclusion of the higher
dimensional matrix elements in eqs.~(\ref{eq:highdim}) constrains the shape
functions to be defined only for $l_+ > 0$. This is because a negative $l_+$, 
corresponding to the case where the $Q \bar Q$ absorbs energy before evolving
into a quarkonium, implies that the soft interactions between the $Q \bar Q$ 
and the environment are not disentangled, 
and factorization is explicitly broken~\cite{Beneke:1999gq}. 

While the lightlike direction $\ell$ of the gauge-completion Wilson line is
arbitrarily chosen in the definition of the color-octet matrix element, its
origin is the direction of the heavy quarkonium momentum in a boosted
frame such as the hadron CM frame~\cite{Nayak:2005rw,Nayak:2005rt}, 
just like the direction of the lightlike momentum $l$. Although the
gauge-completion Wilson line will not play a role in the phenomenological
analysis of $\chi_{QJ}$ production at NLO accuracy, because
diagrams that involve the gauge-completion Wilson line begin to appear from two
loops~\cite{Nayak:2005rw,Nayak:2005rt, Nayak:2006fm, Bodwin:2019bpf}, 
for definiteness we will take the direction $\ell$ to be the same as $l$. 

The factorization assumption leads to the following form of the factorization
formula for the shape function formalism 
\begin{eqnarray}
\label{eq:shapefac}
\sigma[{\chi_{QJ} (P)}] &=& (2 J+1) \int_0^\infty dl_+
\left[ s_{^3P_J^{[1]}}(P+l) {\cal S}_{^3P_0^{[1]}}^{\chi_{Q0}}(l_+)
+ s_{^3S_1^{[8]}} (P+l) {\cal S}_{^3S_1^{[8]}}^{\chi_{Q0}}(l_+) \right],
\end{eqnarray}
where the $s_N (P+l)$ are the short-distance coefficients in the shape
function formalism, which must be functions of the $Q \bar Q$ momentum $P+l$. 
This follows from the fact that the shape function formalism is obtained 
by resumming contributions from higher dimensional NRQCD matrix elements 
of the form given in eqs.~(\ref{eq:highdim}), 
which implies that the coefficients $s_N(P+l)$ can be obtained from the 
standard NRQCD matching procedure by using the perturbative cross sections 
$\sigma[Q \bar Q (P+l)]$, formally by expanding in powers of $l$. 
In ref.~\cite{Beneke:1997qw}, it has thus been suggested that the
short-distance coefficients in the shape function formalism are simply given by
$s_N (P+l) = c_N (P+l)$. 
However, this can become invalid beyond tree level, because UV divergences in
the NRQCD matrix elements induce mixing between the two channels, 
which occur in different forms in NRQCD and shape function formalisms.

Let us compare the matching conditions for the two formalisms. 
The matching conditions that determine the NRQCD short-distance coefficients
read 
\begin{subequations}
\label{eq:NRQCDmatchform}
\begin{eqnarray}
\label{eq:NRQCDmatchform_singlet}
\sigma[Q \bar{Q}(^3P_J^{[1]}) (P)]
&=& 
(2 J\!+\!1)
\Big(
c_{^3P_J^{[1]}}(P)
\langle {\cal O}^{Q \bar{Q}(^3P_0^{[1]})} (^3P_0^{[1]}) \rangle
\nonumber \\&& \hspace{10ex} 
+ c_{^3S_1^{[8]}} (P)
\langle {\cal O}^{Q \bar{Q}(^3P_0^{[1]})} (^3S_1^{[8]}) \rangle \Big),
\\
\label{eq:NRQCDmatchform_octet}
\sigma[Q \bar{Q}(^3S_1^{[8]}) (P)]
&=&
c_{^3S_1^{[8]}} (P)
\langle {\cal O}^{Q \bar{Q}(^3S_1^{[8]})} (^3S_1^{[8]}) \rangle,
\end{eqnarray}
\end{subequations}
which we obtain from eq.~(\ref{eq:nrqcdfac}) by replacing the quarkonium state 
by $Q \bar Q$ states with definite color and angular momentum. 
We used the fact that, as we will see later, the matrix element 
$\langle {\cal O}^{Q \bar{Q}(^3S_1^{[8]})} (^3P_0^{[1]}) \rangle$ 
vanishes at current accuracy in $v$.
At tree level, only the matrix elements 
$\langle {\cal O}^{Q \bar{Q}(^3P_0^{[1]})} (^3P_0^{[1]}) \rangle$ and 
$\langle {\cal O}^{Q \bar{Q}(^3S_1^{[8]})} (^3S_1^{[8]}) \rangle$ are nonzero,
so that at this level the short-distance coefficients $c_N(P)$ are proportional
to the $Q \bar Q$ cross sections $\sigma[Q \bar{Q}(N)(P)]$ for 
$N = {}^3P_J^{[1]}$ and $^3S_1^{[8]}$. 
However, at one-loop level the color-octet matrix element 
$\langle {\cal O}^{Q \bar{Q}(^3S_1^{[8]})} (^3S_1^{[8]}) \rangle$
acquires a logarithmic UV divergence that is proportional to the 
color-singlet matrix element 
$\langle {\cal O}^{Q \bar{Q}(^3P_0^{[1]})} (^3P_0^{[1]}) \rangle$.
After the UV divergence is removed through renormalization, 
the short-distance coefficient $c_{^3P_J^{[1]}}(P)$ acquires a logarithmic 
scale dependence that is proportional to $c_{^3S_1^{[8]}} (P)$, 
which cancels the scale dependence of the color-octet
matrix element in the factorization formula. 
This kind of mixing occurs also in the shape function formalism, but in a
different, $l$-dependent manner. 
The matching conditions that determine the short-distance
coefficients $s_N(P+l)$ in the shape function formalism read
\begin{subequations}
\label{eq:shapematchform}
\begin{eqnarray}
\sigma[Q \bar{Q}(^3P_J^{[1]})( P)]
&=&
(2 J+1) \int_0^\infty dl_+
\bigg[ s_{^3P_J^{[1]}}(P+l) {\cal S}_{^3P_0^{[1]}}^{Q\bar{Q} (^3P_0^{[1]})}(l_+)
\label{eq:shapematchform_singlet}
\nonumber \\ && \hspace{20ex}
+ s_{^3S_1^{[8]}} (P+l) {\cal S}_{^3S_1^{[8]}}^{Q\bar{Q} (^3P_0^{[1]})}(l_+)
\bigg],
\\
\label{eq:shapematchform_octet}
\sigma[Q \bar{Q}(^3S_1^{[8]})( P)]
&=&
\int_0^\infty dl_+ \,
s_{^3S_1^{[8]}} (P+l) {\cal S}_{^3S_1^{[8]}}^{Q\bar{Q} (^3S_1^{[8]})}(l_+),
\end{eqnarray}
\end{subequations}
which we obtain from eq.~(\ref{eq:shapefac}) by replacing the quarkonium state 
by $Q \bar Q$ states with definite color and angular momentum. 
We have used the fact that, as we will show later, the shape function 
${\cal S}_{^3P_0^{[1]}}^{Q\bar{Q} (^3S_1^{[8]})}(l_+)$ vanishes at current
accuracy in $v$. 
Similarly to the matching calculation in NRQCD, determination of 
$s_N(P+l)$ at one-loop level requires computation of the shape functions 
${\cal S}_{^3P_0^{[1]}}^{Q\bar{Q} (^3P_0^{[1]})}(l_+)$,
${\cal S}_{^3S_1^{[8]}}^{Q\bar{Q} (^3S_1^{[8]})}(l_+)$, and
${\cal S}_{^3S_1^{[8]}}^{Q\bar{Q} (^3P_0^{[1]})}(l_+)$ at loop level. 
We will show later that to one-loop accuracy, we have 
${\cal S}_{^3P_0^{[1]}}^{Q\bar{Q} (^3P_0^{[1]})}(l_+)
= \delta(l_+) \langle {\cal O}^{Q \bar{Q} (^3P_0^{[1]})} (^3P_0^{[1]}) \rangle$
and 
${\cal S}_{^3S_1^{[8]}}^{Q\bar{Q} (^3S_1^{[8]})}(l_+)
=\delta(l_+) \langle {\cal O}^{Q \bar{Q} (^3S_1^{[8]})} (^3S_1^{[8]}) \rangle$,
because a $Q \bar Q$ can evolve into a $Q \bar Q$ in the same color and angular
momentum state without emitting any gluons, but the same process cannot occur
by exchange of a single gluon due to conservation of color. 
These relations give $s_{^3S_1^{[8]}} (P) = c_{^3S_1^{[8]}} (P)$ for the
color-octet channel. 
The $s_{^3P_J^{[1]}}(P)$ for the color-singlet channel on the other hand,
involves the color-octet shape function ${\cal S}_{^3S_1^{[8]}}^{Q\bar{Q}
(^3P_0^{[1]})}(l_+)$. 
Similarly to the NRQCD case, 
${\cal S}_{^3S_1^{[8]}}^{Q\bar{Q} (^3P_0^{[1]})}(l_+)$ acquires at one loop a
contribution that is proportional to 
$\langle {\cal O}^{Q \bar{Q}(^3P_0^{[1]})} (^3P_0^{[1]}) \rangle$. 
As we will see in sec.~\ref{sec:diagram}, an explicit calculation gives 
\begin{eqnarray}
\label{eq:octetshape_d4}
{\cal S}^{Q \bar Q(^3P_0^{[1]})}_{^3S_1^{[8]}}(l_+) \Big|_{d=4}
=
\langle {\cal O}^{Q \bar Q(^3P_0^{[1]})} (^3P_0^{[1]}) \rangle \times
\frac{4 \alpha_s C_F}{3 N_c \pi m^2} \frac{1}{l_+},
\end{eqnarray}
in $d=4$ spacetime dimensions. 
The $d=4$ result can actually be inferred from the UV divergence in the
normalization of the color-octet shape function, as we will explain shortly. 
The result in $d$ dimensions 
enters eq.~(\ref{eq:shapematchform_singlet}) and induces mixing between the
color-singlet and color-octet channels. That is, $s_{^3P_J^{[1]}}(P)$ will be
given by a linear combination of $c_{^3P_J^{[1]}}(P)$ and the convolution of 
$c_{^3S_1^{[8]}} (P+l)$ with eq.~(\ref{eq:octetshape_d4}). 
In order to properly regularize the divergences associated with the integral
over $l_+$ in the factorization formula, the $d$-dimensional expression must be
used in the matching condition, which requires an explicit calculation.
Explicit calculations of the color-octet shape function
${\cal S}_{^3S_1^{[8]}}^{Q\bar{Q} (^3P_0^{[1]})}(l_+)$,
as well as 
${\cal S}_{^3P_0^{[1]}}^{Q\bar{Q} (^3P_0^{[1]})}(l_+)
= \delta(l_+) \langle {\cal O}^{Q \bar{Q} (^3P_0^{[1]})} (^3P_0^{[1]}) \rangle$
and
${\cal S}_{^3S_1^{[8]}}^{Q\bar{Q} (^3S_1^{[8]})}(l_+)
=\delta(l_+) \langle {\cal O}^{Q \bar{Q} (^3S_1^{[8]})} (^3S_1^{[8]}) \rangle$, 
will be presented in section~\ref{sec:diagram}, 
so that we can obtain the short-distance coefficients $s_N$ from the 
NRQCD short-distance coefficients $c_N$ by comparing 
eqs.~(\ref{eq:NRQCDmatchform}) and (\ref{eq:shapematchform}). 

The result for the color-octet shape function in eq.~(\ref{eq:octetshape_d4})
in $d=4$ dimensions can be understood without explicit calculations 
in the following way. 
The normalization condition 
$\int_0^\infty dl_+ {\cal S}^{\chi_{Q0}}_{^3S_1^{[8]}} (l_+)
= \langle {\cal O}^{\chi_{Q0}} (^3S_1^{[8]}) \rangle$ implies that 
the one-loop expression for 
${\cal S}_{^3S_1^{[8]}}^{Q\bar{Q} (^3P_0^{[1]})}(l_+)$ must reproduce the UV
divergence in the color-octet matrix element when integrated over $l_+$. 
The UV divergence in the color-octet matrix element gives the following 
evolution equation~\cite{Bodwin:1994jh, Brambilla:2021abf}
\begin{equation}
\frac{d}{d \log \Lambda} \langle {\cal O}^{\chi_{Q0}} (^3S_1^{[8]}) \rangle
= \frac{4 \alpha_s C_F}{3 N_c \pi m^2} 
\langle {\cal O}^{\chi_{Q0}} (^3P_0^{[1]}) \rangle
+ O(\alpha_s^2),
\end{equation}
where $\Lambda$ is the scale for the renormalized matrix element
$\langle {\cal O}^{\chi_{Q0}} (^3S_1^{[8]}) \rangle$. 
In the perturbative calculation of this anomalous dimension, 
the divergence occurs from the integral over the momentum of the gluon 
emitted by the color-octet $Q \bar Q$ as it evolves into a color-singlet
$P$-wave state. Hence, the UV divergence in the normalization integral 
$\int_0^\infty dl_+ {\cal S}^{\chi_{Q0}}_{^3S_1^{[8]}} (l_+)$ 
must come from the $l_+$ dependence of the shape function\footnote{In the 
case of the color-singlet matrix element, its two-loop UV divergence 
comes from renormalization of the derivative of the $P$-wave wavefunction 
at the origin~\cite{Hoang:2006ty, Chung:2021efj}, and so is unrelated to the
$l_+$ dependence of the color-singlet shape function. 
}. 
This implies that at large $l_+ \gg mv$ the nonperturbative 
color-octet shape function must take the following asymptotic form
\begin{eqnarray}
\label{eq:shape_asymptotic}
{\cal S}_{^3S_1^{[8]}}^{\chi_{Q0}}(l_+) \Big|_{{\rm asy,}\, d=4}
=
\langle {\cal O}^{\chi_{Q0}} (^3P_0^{[1]}) \rangle \times 
\frac{4 \alpha_s C_F}{3 N_c \pi m^2} \frac{1}{l_+},
\end{eqnarray}
so that the integral 
$\int_0^{l_+^{\rm max}} dl_+ {\cal S}^{\chi_{Q0}}_{^3S_1^{[8]}} (l_+)$ 
diverges logarithmically at large $l_+^{\rm max}$
in the form $\langle {\cal O}^{\chi_{Q0}} (^3P_0^{[1]}) \rangle \times
\frac{4 \alpha_s C_F}{3 N_c \pi m^2} \log l_+^{\rm max}$. 
Because the perturbative calculation of 
${\cal S}_{^3S_1^{[8]}}^{Q\bar{Q} (^3P_0^{[1]})}(l_+)$ depends 
only on one scale $l_+$, the above expression for the color-octet 
shape function is valid at $d=4$ for all $l_+$ when the $\chi_{Q0}$ state 
is replaced by the perturbative $Q \bar{Q}(^3P_0^{[1]})$ state; hence we obtain 
eq.~(\ref{eq:octetshape_d4}). 

With the short-distance coefficients obtained from perturbative calculations, 
$P$-wave quarkonium cross sections can be computed in the shape function
formalism once the nonperturbative shape functions 
${\cal S}_{^3P_0^{[1]}}^{\chi_{Q0}}(l_+)$ and 
${\cal S}_{^3S_1^{[8]}}^{\chi_{Q0}}(l_+)$ are determined. 
In the case of the color-singlet shape function, 
the fact that effects of emissions of order $mv$ gluons 
are suppressed by $v^2$ at the amplitude level leads to the observation that 
${\cal S}_{^3P_0^{[1]}}^{\chi_{Q0}}(l_+)
= \delta(l_+) \langle {\cal O}^{\chi_{Q0}} (^3P_0^{[1]}) \rangle$
up to corrections suppressed by $v^4$ (the same conclusion can be
obtained from the vacuum-saturation approximation, as was done in
ref.~\cite{Beneke:1997qw}). 
On the other hand, it has been argued that the nonperturbative color-octet 
shape function ${\cal S}_{^3S_1^{[8]}}^{\chi_{Q0}}(l_+)$ 
cannot be computed except within models. 
Even so, the model dependence can be significantly reduced by using the 
large-$l_+$ asymptotic form~(\ref{eq:shape_asymptotic}) 
and the normalization condition 
$\int_0^\infty dl_+ {\cal S}_{^3S_1^{[8]}}^{\chi_{Q0}}(l_+) = 
\langle {\cal O}^{\chi_{Q0}} (^3S_1^{[8]}) \rangle$. 
As the normalization condition requires the integral to be infrared (IR) 
finite, 
the integrand must not grow like $1/l_+$ as $l_+ \to 0$, while it must
reproduce the asymptotic form for large $l_+$. 
Although describing the behavior of the color-octet shape function at
small $l_+$ may still require models, model parameters can be fixed 
from the color-octet NRQCD matrix element by using the normalization condition. 
As renormalization of NRQCD matrix elements is usually
done in the $\overline{\rm MS}$ scheme, we have 
\begin{equation}
\int_0^\infty dl_+
{\cal S}_{^3S_1^{[8]}}^{\chi_{Q0}}(l_+) \Big|_{d=4-2 \epsilon}
- \frac{
1}{2 \epsilon_{\rm UV}}
\frac{4 \alpha_s C_F}{3 N_c \pi m^2}
\langle {\cal O}^{\chi_{Q0}} (^3P_0^{[1]}) \rangle
= \langle {\cal O}^{\chi_{Q0}} (^3S_1^{[8]})
\rangle^{\overline{\rm MS}},
\end{equation}
where we define the $\overline{\rm MS}$ scheme at scale $\Lambda$
by subtracting the UV pole and rescaling the MS scale $\mu$ through
the relation $\mu^2 = \Lambda^2 e^{\gamma_{\rm E}}/(4 \pi)$.
This expression is not very useful as it is, because it requires use of a model
shape function in $d$ dimensions. 
Instead, we may regulate the UV divergence by cutting off the $l_+$ integral as
\begin{equation}
\int_0^\infty dl_+
{\cal S}_{^3S_1^{[8]}}^{\chi_{Q0}}(l_+) \Big|_{d=4-2 \epsilon}
=
\int_0^{l_+^{\rm max}} dl_+
{\cal S}_{^3S_1^{[8]}}^{\chi_{Q0}}(l_+) \Big|_{d=4}
+
\int_{l_+^{\rm max}}^\infty dl_+
{\cal S}_{^3S_1^{[8]}}^{\chi_{Q0}}(l_+) \Big|_{d=4-2 \epsilon},
\end{equation}
which is valid to order $\epsilon^0$. 
For large enough $l_+^{\rm max}$, the last term 
can be computed in perturbation theory, because the large-$l_+$
behavior is fixed by the renormalization of the color-octet matrix element. 
That is, 
\begin{equation}
\frac{\displaystyle
\int_{l_+^{\rm max}}^\infty dl_+
{\cal S}_{^3S_1^{[8]}}^{\chi_{Q0}}(l_+) \Big|_{d=4-2 \epsilon}
}{\langle {\cal O}^{\chi_{Q0}}(^3P_0^{[1]}) \rangle} 
= 
\frac{\displaystyle
\int_{l_+^{\rm max}}^\infty dl_+
{\cal S}_{^3S_1^{[8]}}^{Q \bar{Q} (^3P_0^{[1]})}(l_+) \Big|_{d=4-2 \epsilon}
}{\langle {\cal O}^{Q \bar{Q} (^3P_0^{[1]})}(^3P_0^{[1]}) \rangle} ,
\end{equation}
where the right-hand side can be computed form the $d$-dimensional result for 
${\cal S}_{^3S_1^{[8]}}^{Q \bar{Q} (^3P_0^{[1]})}(l_+)$. 
If we define 
\begin{equation}
\label{eq:schemeconv_def}
\Delta(\Lambda, l_+^{\rm max}) = 
\frac{\displaystyle
\int_{l_+^{\rm max}}^\infty dl_+
{\cal S}_{^3S_1^{[8]}}^{Q \bar{Q} (^3P_0^{[1]})}(l_+) \Big|_{d=4-2 \epsilon}
}{\langle {\cal O}^{Q \bar{Q} (^3P_0^{[1]})}(^3P_0^{[1]}) \rangle}
- \frac{
1}{2 \epsilon_{\rm UV}}
\frac{4 \alpha_s C_F}{3 N_c \pi m^2}, 
\end{equation}
where the last term subtracts the UV divergence in the $l_+$ integral, 
the normalization condition for the shape function can be written as 
\begin{equation}
\label{eq:normalization}
\int_0^{l_+^{\rm max}} dl_+
{\cal S}_{^3S_1^{[8]}}^{\chi_{Q0}}(l_+) \bigg|_{d=4}
+ \Delta(\Lambda, l_+^{\rm max}) 
\langle {\cal O}^{\chi_{Q0}} (^3P_0^{[1]}) \rangle
= \langle {\cal O}^{\chi_{Q0}} (^3S_1^{[8]})
\rangle^{\overline{\rm MS}},
\end{equation}
where now every term is finite at $d=4$ and the $\epsilon$ dependence can be
dropped. The term $\Delta(\Lambda, l_+^{\rm max})$, which translates cutoff 
regularization to the $\overline{\rm MS}$ scheme,
can be computed perturbatively as a series in $\alpha_s$.
At leading nonvanishing order, it satisfies 
\begin{equation}
\label{eq:schemeconv_evolution}
\frac{d}{d \log \Lambda} \Delta(\Lambda, l_+^{\rm max}) 
= 
- \frac{d}{d \log l_+^{\rm max}} \Delta(\Lambda, l_+^{\rm max}) 
= \frac{4 \alpha_s C_F}{3 N_c \pi m^2},
\end{equation}
so that $\Delta(\Lambda, l_+^{\rm max})$ is given by 
$\frac{4 \alpha_s C_F}{3 N_c \pi m^2} \log \Lambda/l_+^{\rm max}$ plus a
constant, which can be determined from eq.~(\ref{eq:schemeconv_def})
by using the $d$-dimensional result for 
${\cal S}_{^3S_1^{[8]}}^{Q \bar{Q} (^3P_0^{[1]})}(l_+)$. 
Equations~(\ref{eq:shape_asymptotic}) and (\ref{eq:normalization})
strongly constrain the model dependence of the 
nonperturbative shape function ${\cal S}_{^3S_1^{[8]}}^{\chi_{Q0}}(l_+)$ at
both large and small $l_+$. 
The predictive power of the shape function formalism can be further improved by
computing the nonperturbative shape functions in pNRQCD, 
similarly to what have been done for the NRQCD matrix elements, to express them
in terms of products of quarkonium wavefunctions and vacuum expectation values
of gluonic operators~\cite{Brambilla:2020ojz, Brambilla:2021abf,
Brambilla:2022rjd, Brambilla:2022ayc}. We will compute the shape functions in
the pNRQCD formalism in section~\ref{sec:pNRQCD}.

Before concluding this section we need to discuss the transformation of the 
shape function under Lorentz boosts. This is necessary because while the
NRQCD operator definitions of the shape functions require them to be
computed in the rest frame of the heavy quarkonium, similarly to the NRQCD
matrix elements, the short-distance coefficients often need to be
computed in a frame where the quarkonium has energetic spatial momentum. 
It is clear that from the definition of the shape functions, 
$d l_+ \, {\cal S}_N^{\chi_{Q0}} (l_+)$ is Lorentz invariant under boosts in
the $+$ direction. 
Hence, the transformation of the shape function itself can be inferred from 
the boost property of $l_+$. Under boosts from the quarkonium rest frame, 
$l_+$ transforms like 
\begin{equation}
l_+ = \frac{P_+}{P_+^*} l_+^*, 
\end{equation}
where $l^*$ is the momentum $l$ in the quarkonium rest frame, and $l_+$ on the 
left-hand side is the $+$ component of $l$ in the boosted frame; likewise, $P^*$
is the quarkonium momentum in the quarkonium rest frame, and $P_+$ is the $+$
component of $P$ in the boosted frame. Note that in the quarkonium rest frame,
$P_+^* = \sqrt{P^2}$, which coincides with the quarkonium mass. 
Throughout this paper, we will refer to the momentum
$l_+$ in the quarkonium rest frame as $l_+^*$. 
Due to the invariance of $d l_+ \, {\cal S}_N^{\chi_{Q0}} (l_+)$ under boosts, 
the first term of the left-hand side of eq.~(\ref{eq:normalization}) is
also invariant, as long as $l_+^{\rm max}$ boosts like $l_+$. 
We will also see in the calculation of the scheme conversion 
$\Delta(\Lambda, l_+^{\rm max})$ that the dependence on the boost 
cancels between the explicit dependence on $l_+^{\rm max}$ and the boost
dependence coming from the scaling violation of the perturbative shape function 
${\cal S}_{^3S_1^{[8]}}^{Q \bar Q (^3P_0^{[1]})} (l_+)$ due to nonzero 
on $\epsilon = (4-d)/2$.

\section{Shape functions in perturbative NRQCD}
\label{sec:diagram}

In this section we compute the shape functions 
${\cal S}_{^3P_0^{[1]}}^{Q\bar{Q} (^3P_0^{[1]})}(l_+)$,
${\cal S}_{^3P_0^{[1]}}^{Q\bar{Q} (^3S_1^{[8]})}(l_+)$,
${\cal S}_{^3S_1^{[8]}}^{Q\bar{Q} (^3S_1^{[8]})}(l_+)$, and
${\cal S}_{^3S_1^{[8]}}^{Q\bar{Q} (^3P_0^{[1]})}(l_+)$
in perturbation theory, which are necessary for obtaining the matching
conditions that determine the short-distance coefficients $s_N$ of the shape
function formalism. 
As we have stated in the previous section, these objects are defined in the
rest frame of the final-state $Q \bar Q$, 
and will be computed in this section in that frame as functions of $l_+^*$. 
The calculation of these shape functions are done in the same way as NRQCD
matrix elements, except that the $+$ component of the relative momentum of the 
$Q \bar Q$ produced and annihilated on the left and right of the 
projection operator is constrained to be $l_+^*$. 
At order $\alpha_s^0$, these shape functions are given by the 
squared amplitudes for $Q \bar Q \to Q \bar Q$, where the 
color and angular momentum states of the initial and final state $Q \bar Q$ are
constrained to be $^3P_0^{[1]}$ or $^3S_1^{[8]}$. 
Since they vanish unless $l^* = 0$ and the color and angular momenta of the
initial and final states are same, we obtain 
\begin{subequations}
\begin{eqnarray}
{\cal S}_{^3P_0^{[1]}}^{Q\bar{Q} (^3P_0^{[1]})}(l_+^*)
&=& \delta(l_+^*) \langle {\cal O}^{Q\bar{Q} (^3P_0^{[1]})} (^3P_0^{[1]} )\rangle
+ O(\alpha_s), 
\\
{\cal S}_{^3S_1^{[8]}}^{Q\bar{Q} (^3S_1^{[8]})}(l_+^*)
&=& \delta(l_+^*) \langle {\cal O}^{Q\bar{Q} (^3S_1^{[8]})} (^3S_1^{[8]} )\rangle
+ O(\alpha_s), 
\\
{\cal S}_{^3P_0^{[1]}}^{Q\bar{Q} (^3S_1^{[8]})}(l_+^*) &=& 0 + O(\alpha_s),
\\
{\cal S}_{^3S_1^{[8]}}^{Q\bar{Q} (^3P_0^{[1]})}(l_+^*) &=& 0 + O(\alpha_s). 
\end{eqnarray}
\end{subequations}

\begin{figure}[tbp]
\centering
\includegraphics[width=.95\textwidth]{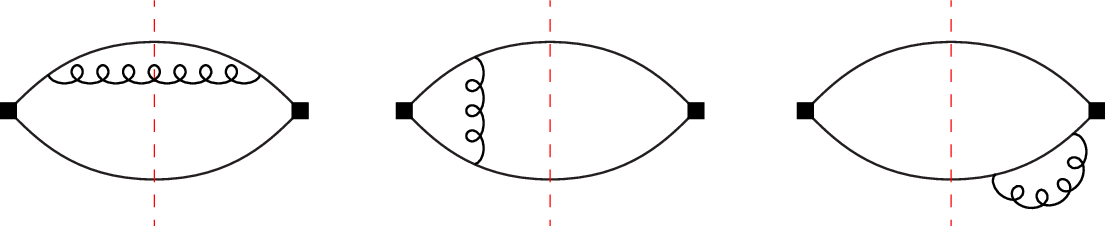}
\caption{\label{fig:ldme}
Feynman diagrams for NRQCD matrix elements and shape functions at 
order $\alpha_s$. The filled squares represent $Q \bar Q$ bilinear operators,
solid lines are heavy quark and antiquark lines, curly lines are gluons, 
and red vertical dashed lines are final-state cuts. 
There are also contributions that are obtained by exchanging 
gluon attachments on a heavy quark line to an antiquark line on the same side 
of the cut, and by exchanging the left and right sides of the cut. 
The gauge-completion Wilson line that appears in the color-octet matrix
elements and the shape functions is omitted, because diagrams that involve the
Wilson line do not appear until two-loop accuracy. 
}
\end{figure}

Now we consider the order-$\alpha_s$ contributions. 
At order $\alpha_s$, the shape functions are given by the squared amplitudes 
for $Q \bar Q \to Q \bar Q$ with a gluon exchange between any of the heavy
quark and antiquark lines. 
The Feynman diagrams for the order-$\alpha_s$ contributions are shown in
fig.~\ref{fig:ldme}.
Note that the contribution from the gluon field in 
$\overleftrightarrow{\bm{D}}$ cancels in the 
operator definition of the color-singlet shape function, 
so that there are no diagrams involving direct gluon attachments to the
$Q \bar Q$ bilinear operator. 
While same diagrams appear in the order-$\alpha_s$ calculation of 
shape functions and NRQCD matrix elements, contributions from the diagrams 
where the gluon crosses the final-state cut, like the first diagram in
fig.~\ref{fig:ldme}, differ between the NRQCD matrix elements and the shape
functions, because in the shape functions the momentum component $l_+^*$ 
is not integrated over. 
If both the initial and the final-state $Q \bar Q$ are in color-singlet states, 
such diagrams vanish due to conservation of color. In the case of 
${\cal S}_{^3S_1^{[8]}}^{Q\bar{Q} (^3S_1^{[8]})}(l_+^*)$ and 
$\langle {\cal O}^{Q\bar{Q} (^3S_1^{[8]})} (^3S_1^{[8]} )\rangle$, 
the sum of such diagrams vanish because they involve symmetric linear
combinations of the totally antisymmetric $SU(3)$ structure constants. 
The contributions from the remaining diagrams are same for both the shape
function and the NRQCD matrix element. 
Hence, the tree-level results for 
${\cal S}_{^3P_0^{[1]}}^{Q\bar{Q} (^3P_0^{[1]})}(l_+^*)$ 
and 
${\cal S}_{^3S_1^{[8]}}^{Q\bar{Q} (^3S_1^{[8]})}(l_+^*)$ 
hold to order-$\alpha_s$ accuracy:
\begin{subequations}
\label{eq:shapeDR_result1}
\begin{eqnarray}
{\cal S}_{^3P_0^{[1]}}^{Q\bar{Q} (^3P_0^{[1]})}(l_+^*)
&=& \delta(l_+^*) \langle {\cal O}^{Q\bar{Q} (^3P_0^{[1]})} (^3P_0^{[1]} )\rangle
+ O(\alpha_s^2),
\\
{\cal S}_{^3S_1^{[8]}}^{Q\bar{Q} (^3S_1^{[8]})}(l_+^*)
&=& \delta(l_+^*) \langle {\cal O}^{Q\bar{Q} (^3S_1^{[8]})} (^3S_1^{[8]} )\rangle
+ O(\alpha_s^2).
\end{eqnarray}
\end{subequations}
In the case of ${\cal S}_{^3P_0^{[1]}}^{Q\bar{Q} (^3S_1^{[8]})}(l_+^*)$, 
the contribution from the gluon attachment on the heavy quark line cancels the 
contribution from the gluon attachment on the heavy antiquark line on the same
side of the cut. 
The remaining diagrams cannot produce a color-octet $Q \bar Q$ in the
final state, and so they do not contribute to 
${\cal S}_{^3P_0^{[1]}}^{Q\bar{Q} (^3S_1^{[8]})}(l_+^*)$. Therefore, the result
\begin{equation}
\label{eq:shapeDR_result2}
{\cal S}_{^3P_0^{[1]}}^{Q\bar{Q} (^3S_1^{[8]})}(l_+^*) = 0 + O(\alpha_s^2),
\end{equation}
holds through order $\alpha_s$, 
and for the same reason, $\langle {\cal O}^{Q \bar Q (^3S_1^{[8]})}
(^3P_0^{[1]}) \rangle$ vanishes. 
While they can become nonzero from corrections of higher orders in $\alpha_s$, 
they will be suppressed by powers of $v$, because they involve dynamical gluons
of order $mv$ crossing the cut. Hence, they vanish at current order in $v$. 

\begin{figure}[tbp]
\centering
\includegraphics[width=.55\textwidth]{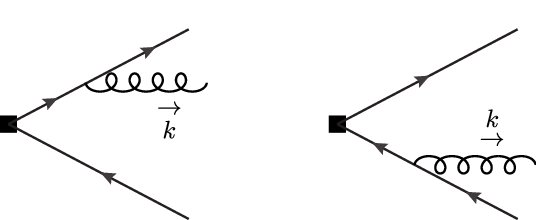}
\caption{\label{fig:gluonemission}
Feynman diagrams for emission of a soft gluon with momentum $k$ from a 
$Q \bar Q$ pair. 
}
\end{figure}

The only nontrivial contribution comes from 
${\cal S}_{^3S_1^{[8]}}^{Q\bar{Q} (^3P_0^{[1]})}(l_+^*)$. 
Similarly to ${\cal S}_{^3P_0^{[1]}}^{Q\bar{Q} (^3S_1^{[8]})}(l_+^*)$, 
we only need to consider the diagrams where a gluon crosses the cut, 
because otherwise the color-octet operator cannot produce a color-singlet 
$Q \bar Q$ in the final state. 
We set the momenta of the quark and antiquark in the final state to be 
$(E,\bm{q})$ and $(E,-\bm{q})$, respectively, with $E = \bm{q}^2/(2 m)$. 
We use the NRQCD Feynman rules in ref.~\cite{Bodwin:1998mn}. 
The contributions from gluon attachments to the $Q$ and the $\bar Q$ on one
side of the cut as shown in fig.~\ref{fig:gluonemission} involve the factor
\begin{equation}
\frac{i g(\bm{k}+2\bm{q}) T^a}{2 m} \frac{i}{k_0+i \varepsilon}
+ \frac{i g(-\bm{k}+2\bm{q}) T^a}{2 m} \frac{i}{k_0+i \varepsilon}
= 
\frac{2 i g \bm{q} T^a}{m} \frac{i}{k_0+i \varepsilon}, 
\end{equation}
where $k$ is the momentum of the gluon, and 
we neglect the terms $\bm{q}^2/(2 m)$, $(\bm{q}+\bm{k})^2/(2 m)$ and 
$(\bm{q}-\bm{k})^2/(2 m)$ compared to $k_0$ in the heavy quark and antiquark
propagators, because they are suppressed by a power of $v$. 
The product of the factor $\bm{q}$ from the gluon vertices 
and the Pauli matrix $\bm{\sigma}$ coming from the
operator definition of the shape function can be decomposed into the 
trace, antisymmetric, and traceless symmetric parts as 
\begin{equation}
\label{eq:3p012decomp}
q^i \sigma^j = \frac{\delta^{ij}}{d-1} \bm{q} \cdot \bm{\sigma}
+ \frac{q^i \sigma^j - q^j \sigma^i}{2} 
+ \left(\frac{ q^i \sigma^j + q^j \sigma^i }{2} 
- \frac{\delta^{ij}}{d-1} \bm{q} \cdot \bm{\sigma} \right), 
\end{equation}
which is valid in $d$ dimensions. In 3 spatial dimensions the three terms
correspond to the irreducible angular momentum tensors for total angular
momentum 0, 1, and 2. If the $Q \bar Q$ in the final state 
is in the $^3P_0^{[1]}$ state, only the first term survives, 
and the factor $\bm{q} \cdot \bm{\sigma}$ can be identified as the tree-level 
matrix element of the $Q \bar Q$ bilinear operator 
$\chi^\dag (-\tfrac{i}{2} \overleftrightarrow{\bm{D}} \cdot \bm{\sigma}) \psi$ 
between the vacuum and the $Q \bar{Q} (^3P_0^{[1]})$ state. 
Then, the shape function is given by 
\begin{eqnarray}
{\cal S}_{^3S_1^{[8]}}^{Q\bar{Q} (^3P_0^{[1]})}(l_+^*)
&=&
\frac{16 \pi \alpha_s}{m^2} 
\frac{d-2}{d-1}
\frac{C_F}{2 N_c}
\langle {\cal O}^{Q \bar{Q}(^3P_0^{[1]})} (^3P_0^{[1]}) \rangle
\nonumber \\ && \times
\left( \frac{\Lambda^2 e^{\gamma_{\rm E}}}{4 \pi} \right)^{\frac{4-d}{2}} 
\int \frac{d^dk}{(2 \pi)^{d}} \frac{1}{(k_0+i \varepsilon)^2} 
2 \pi \delta(k^2) \delta(l_+^* - k_+),
\end{eqnarray}
where $\Lambda$ is the $\overline{\rm MS}$ scale. 
Here, the $2 \pi \delta(k^2)$ comes from the gluon crossing the cut, and 
the $\delta(l_+^* - k_+)$ 
comes from the delta function in the definition of the shape function. 
The factor $d-2$ comes from the sum over the final-state gluon polarizations, 
and the factor $1/(d-1)$ comes from the projection onto the $^3P_0^{[1]}$
state. The color factor is computed from 
$[N_c^{-1} {\rm tr} (T^a T^b)]^2 
= T_F C_F/N_c$, with $T_F = 1/2$.
The computation of the integral over $k$ is straightforward:
by using $\int d^dk = \frac{1}{2} \int dk_+ dk_- d^{d-2} \bm{k}_\perp$, 
we obtain
\begin{equation}
\label{eq:shapeDR_result3}
{\cal S}_{^3S_1^{[8]}}^{Q\bar{Q} (^3P_0^{[1]})}(l_+^*)
= 
\frac{4 \alpha_s C_F}{3 N_c \pi m^2} 
\langle {\cal O}^{Q \bar{Q}(^3P_0^{[1]})} (^3P_0^{[1]}) \rangle
\frac{C_\epsilon}{(l_+^*)^{1+2 \epsilon}}, 
\end{equation}
where $C_\epsilon$ is given by 
\begin{equation}
\label{eq:Cepsilon}
C_\epsilon =
\frac{(1-\epsilon) (\Lambda^2 e^{\gamma_{\rm E}})^\epsilon \Gamma(1+\epsilon)}
{1-\frac{2}{3} \epsilon}. 
\end{equation}
Expanding $C_\epsilon$ in powers of $\epsilon$ to linear order yields 
$C_\epsilon = 1 + \left( \log \Lambda^2 - \frac{1}{3} \right) \epsilon +
O(\epsilon^2)$.
Equation~(\ref{eq:shapeDR_result3}) is the result for the
color-octet shape function ${\cal S}_{^3S_1^{[8]}}^{Q\bar{Q}
(^3P_0^{[1]})}(l_+^*)$ valid to order $\alpha_s$. 
Note that we recover the known result for the unrenormalized (bare)
color-octet matrix element 
$\langle {\cal O}^{Q \bar{Q}(^3P_0^{[1]})} (^3S_1^{[8]}) \rangle^{\rm bare}$
at order $\alpha_s$ by integrating over $l_+^*$:
\begin{eqnarray}
\label{eq:barematrixelement}
\langle {\cal O}^{Q \bar{Q}(^3P_0^{[1]})} (^3S_1^{[8]}) \rangle^{\rm bare}
&=&
\int_0^\infty dl_+^*
{\cal S}_{^3S_1^{[8]}}^{Q\bar{Q} (^3P_0^{[1]})}(l_+^*)
\nonumber \\ 
&=&
\frac{4 \alpha_s C_F}{3 N_c \pi m^2}
\langle {\cal O}^{Q \bar{Q}(^3P_0^{[1]})} (^3P_0^{[1]}) \rangle
\int_0^\infty dl_+^*
\frac{C_\epsilon}{(l_+^*)^{1+2 \epsilon}}
\nonumber \\
&=&
\frac{4 \alpha_s C_F}{3 N_c \pi m^2}
\langle {\cal O}^{Q \bar{Q}(^3P_0^{[1]})} (^3P_0^{[1]}) \rangle
\left( \frac{1}{2 \epsilon_{\rm UV}} - \frac{1}{2 \epsilon_{\rm IR}} \right).
\end{eqnarray}
This result also lets us compute the scheme conversion 
$\Delta(\Lambda, l^{\rm max}_+)$. 
From the definition given in eq.~(\ref{eq:schemeconv_def}), 
we have at order-$\alpha_s$ accuracy 
\begin{eqnarray}
\label{eq:schemeconvresult}
\Delta(\Lambda, l_+^{* \rm max}) &=& 
\int_{l_+^{* \rm max}}^\infty dl_+^*
\frac{4 \alpha_s C_F}{3 N_c \pi m^2}
\frac{C_\epsilon}{(l_+^*)^{1+2 \epsilon}}
- \frac{
1}{2 \epsilon_{\rm UV}}
\frac{4 \alpha_s C_F}{3 N_c \pi m^2}
\nonumber \\
&=& 
\frac{4 \alpha_s C_F}{3 N_c \pi m^2} \left( 
\log \Lambda - \log l_+^{* \rm max} -\frac{1}{6} \right)
+ O(\epsilon),
\end{eqnarray}
which is our result for the scheme conversion in the quarkonium rest frame.
Note that this result reproduces the
evolution equation in eq.~(\ref{eq:schemeconv_evolution}).
The scheme conversion vanishes at this order if we set 
$l_+^{* \rm max}= \Lambda e^{-1/6}$, that is, 
$\Delta (\Lambda, l_+^{* \rm max}= \Lambda e^{-1/6}) = 0 
+ O(\alpha_s^2)$, which implies that the cutoff result 
$\int_0^{l_+^{* \rm max}} dl_+^* {\cal S}_{^3S_1^{[8]}}^{\chi_{Q0}}(l_+^*)$
is equal to the $\overline{\rm MS}$-renormalized color-octet matrix element
$\langle {\cal O}^{\chi_{Q0}} (^3S_1^{[8]}) \rangle$ at scale 
$\Lambda = l_+^{* \rm max} e^{1/6}$. 
From $e^{-1/6} \approx 0.85$, we see that the cutoff $l_+^{* \rm max}$ is
numerically close to the $\overline{\rm MS}$ scale. In phenomenological studies
of heavy quarkonium production, the $\overline{\rm MS}$ scale $\Lambda$ for the
color-octet matrix element is usually chosen to be the heavy quark mass $m$, 
and in this case the corresponding cutoff $l_+^{* \rm max}$ equals $m \times
e^{-1/6}$. 
Consistently with the assumption that perturbative matching calculations
are valid at a suitably chosen factorization scale $\Lambda$, 
we will also assume that the behavior of the shape function is perturbative for 
$l_+^*$ at around or above $l_+^{* \rm max}$, so that the $l_+^*$ dependence 
of the scheme conversion and the shape function 
can be described by the perturbative calculation in
this section.

We can now explicitly address the boost dependence of 
$\Delta(\Lambda, l_+^{\rm max})$. 
In a boosted frame, eq.~(\ref{eq:schemeconvresult}) becomes 
\begin{eqnarray}
\label{eq:schemeconvboosted}
\Delta(\Lambda, l_+^{\rm max}) \big|_{\rm boosted} &=&
\int_{l_+^{\rm max}}^\infty dl_+
\frac{4 \alpha_s C_F}{3 N_c \pi m^2}
\frac{C_\epsilon}{(l_+)^{1+2 \epsilon}}
\left( \frac{P_+}{P_+^*} \right)^{2 \epsilon} 
- \frac{
1}{2 \epsilon_{\rm UV}}
\frac{4 \alpha_s C_F}{3 N_c \pi m^2}
\nonumber \\ &=&
\frac{4 \alpha_s C_F}{3 N_c \pi m^2} \left[
\log \Lambda - \log \left( l_+^{\rm max} \frac{P_+^*}{P_+} \right) 
-\frac{1}{6} \right]
+ O(\epsilon), 
\end{eqnarray}
which we obtain by using $l_+ = (P_+/P_+^*) l_+^*$. The factor 
$(P_+/P_+^*)^{2 \epsilon}$ in the integrand is due to the scaling violation of 
${\cal S}_{^3S_1^{[8]}}^{Q\bar{Q} (^3P_0^{[1]})}(l_+^*)$ coming from the 
$\epsilon$ dependence in $d=4-2 \epsilon$ dimensions. 
Since $l_+^{\rm max} \times (P_+^*/P_+) = l_+^{*\rm max}$, 
the result is exactly same as $\Delta(\Lambda, l_+^{* \rm max})$ computed in
the rest frame. 

Now that we have the perturbative results for the shape functions valid to
order-$\alpha_s$ accuracy, 
the short-distance coefficients $s_N$ for the shape function formalism 
can be obtained at one-loop level from the matching conditions in 
eqs.~(\ref{eq:shapematchform}) by plugging in the results for the shape
functions in eqs.~(\ref{eq:shapeDR_result1}), (\ref{eq:shapeDR_result2}), and
(\ref{eq:shapeDR_result3}). Expressions for the $s_N$ in terms of the NRQCD
short-distance coefficients $c_N$ will be given in section~\ref{sec:matching}.

\section{Nonperturbative analysis of shape functions in pNRQCD}
\label{sec:pNRQCD}

The pNRQCD formalism~\cite{Pineda:1997bj, Brambilla:1999xf,
Brambilla:2004jw} has been proven useful in analyses of inclusive
quarkonium production~\cite{Brambilla:2020ojz, Brambilla:2021abf,
Brambilla:2022rjd, Brambilla:2022ayc}. 
This formalism provides expressions for NRQCD matrix
elements as products of quarkonium wavefunctions 
and vacuum expectation values of gluonic operators, 
and reproduces the known results in terms of quarkonium wavefunctions 
in the case of color-singlet matrix elements. 
The vacuum expectation values of gluonic operators, often called gluonic
correlators, are defined in terms of Wilson lines and 
gluon field-strength tensors. 
Although first-principles calculations of gluonic correlators appearing in
NRQCD matrix elements have not yet been
done, the fact that they are universal, that is, they are 
independent of the specific quarkonium state including the heavy quark
flavor and radial excitation, 
significantly enhances the predictive power of the nonrelativistic
effective field theory formalism. 
We refer the readers to refs.~\cite{Brambilla:2020ojz, Brambilla:2021abf,
Brambilla:2022rjd, Brambilla:2022ayc} for detailed discussions of 
pNRQCD calculations of NRQCD matrix elements and
phenomenological applications in inclusive quarkonium production. 

We can expect to obtain similar results from pNRQCD calculations of shape 
functions: shape functions can be written in terms of quarkonium wavefunctions
and $l_+$-dependent gluonic correlators. 
If the $l_+$ dependences are solely contained in the gluonic correlators, then
we can expect the $l_+$ dependences of the shape functions to be universal
among shape functions of quarkonium states that differ by radial excitation or
heavy quark flavor. As the model dependence of the shape function formalism 
comes from the $l_+$ dependence of the nonperturbative shape functions, 
this would greatly enhance the
predictive power of the shape function formalism. 

In this section, we compute the shape functions 
${\cal S}_{^3P_0^{[1]}}^{\chi_{Q0}}(l_+)$
and ${\cal S}_{^3S_1^{[8]}}^{\chi_{Q0}}(l_+)$
in the pNRQCD formalism for $P$-wave quarkonium production developed in 
refs.~\cite{Brambilla:2020ojz, Brambilla:2021abf}.
Similarly to the calculations of NRQCD matrix elements, we will compute them in
the quarkonium rest frame as functions of $l_+^*$. 
We derive the results nonperturbatively following the development of the
formalism in refs.~\cite{Brambilla:2020ojz, Brambilla:2021abf}, which results
in expansions in powers of $v$ and $\Lambda_{\rm QCD}/m$, while the dynamics of
the gluon and light quarks are kept nonperturbative.  As have been done in
refs.~\cite{Brambilla:2020ojz, Brambilla:2021abf} we assume that the $\chi_Q$
states are strongly coupled, so that the scale $mv^2$ is smaller than the
typical energy gap of gluonic excitations. 
We will discuss the possibility of extending to the weak coupling case after
obtaining the results in the strongly coupled case. 

\begin{figure}[tbp]
\centering
\includegraphics[width=.4\textwidth]{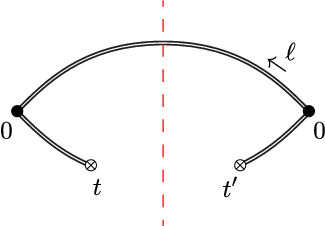}
\caption{\label{fig:E11}
Configuration of the Wilson lines appearing in the definition of the tensor 
${\cal E}_{11}^{ij}$ in eq.~(\ref{eq:E11tensor}). 
Filled circles are spacetime origins, 
double lines are adjoint Wilson lines, 
red dashed vertical line is the cut, 
and the crossed circles represent insertions of the chromoelectric
field on the temporal Wilson lines at times $t$ and $t'$. }
\end{figure}

We first briefly review the calculation of NRQCD matrix elements in the
formalism developed in refs.~\cite{Brambilla:2020ojz, Brambilla:2021abf}. 
In the pNRQCD formalism, the matrix elements are given by 
\begin{eqnarray}
\label{eq:pNRQCDformula}
\langle {\cal O}^{\cal Q}(N) \rangle 
&=& \int d^{d-1} R d^{d-1}r d^{d-1}r' \phi_{\cal Q} (\bm{x}_1-\bm{x}_2)
\nonumber \\ && \times 
\big[ -V_{{\cal Q}(N)} 
(\bm{x}_1, \bm{x}_2;\bm{\nabla}_1, \bm{\nabla}_2) 
\delta^{(d-1)} (\bm{x}_1-\bm{x}_1')
\delta^{(d-1)} (\bm{x}_2-\bm{x}_2')\big] 
\nonumber \\ && \times 
\phi_{\cal Q}^* (\bm{x}_1'-\bm{x}_2'),
\end{eqnarray}
where $\phi_{\cal Q}(\bm{x}_1-\bm{x}_2)$ is the wavefunction of the quarkonium 
${\cal Q}$ at leading order in $v$, 
$\bm{x}_1$ and $\bm{x}_2$ are the positions of the heavy quark and
antiquark, respectively, $\bm{r} = \bm{x}_1-\bm{x}_2$, and 
$\bm{R} = \frac{1}{2} (\bm{x}_1+\bm{x}_2)$. Similar relations hold for the
primed coordinates. 
The $V_{{\cal Q}(N)} (\bm{x}_1, \bm{x}_2;\bm{\nabla}_1, \bm{\nabla}_2)$ 
is a contact term that is determined by matching NRQCD and pNRQCD. 
The contact terms for the NRQCD matrix elements in eq.~(\ref{eq:nrqcdfac}) 
are given by 
\begin{subequations}
\label{eq:contactNRQCD}
\begin{eqnarray}
-V_{{\cal Q}(^3P_0^{[1]})} \Big|_{P-{\rm wave}} &=& 
- \sigma^i \otimes \sigma^j
\frac{N_c}{d-1} 
\nabla_{\bm{r}}^i 
\delta^{(d-1)} (\bm{r})
\nabla_{\bm{r}}^j , 
\\
-V_{{\cal Q}(^3S_1^{[8]})} \Big|_{P-{\rm wave}} &=& 
-\sigma^k \otimes \sigma^k 
N_c \nabla_{\bm{r}}^i \delta^{(d-1)} (\bm{r}) 
\nabla_{\bm{r}}^j 
\frac{{\cal E}_{11}^{ij}}{N_c^2 m^2}, 
\end{eqnarray}
\end{subequations}
where we discard contributions that vanish when applied to wavefunctions in
$P$-wave states. 
The tensor ${\cal E}_{11}^{ij}$ is defined by 
\begin{eqnarray}
\label{eq:E11tensor}
{\cal E}_{11}^{ij} 
&=& 
\int_0^\infty dt \, t
\int_0^\infty dt' t' \langle \bar{T} \Big[ \Phi_\ell^{\dag ab} (0)
\Phi_0^{\dag ad}(0,\bm{0};t,\bm{0})
g E^{d,i}(t,\bm{0}) \Big]
\nonumber \\
&& \hspace{20ex} \times 
T \Big[ g E^{e,j} (t',\bm{0}) \Phi_0^{ec} (0,\bm{0};t',\bm{0})
\Phi_\ell^{bc} (0) \Big] \rangle, 
\end{eqnarray}
where $T$ and $\bar T$ indicate time and anti-time orderings of the operators,
$E^{a,i}(t,\bm{x})$ is the chromoelectric field at time $t$ and spatial
position $\bm{x}$,
and $\Phi_0(t_1,\bm{x};t_2,\bm{x})$ is an adjoint Wilson line in the temporal
direction defined by 
\begin{equation}
\Phi_0(t_1,\bm{x};t_2,\bm{x}) = P \exp \left[ -i g \int_{t_1}^{t_2} dt 
A_0^{\rm adj} (t,\bm{x})
\right].
\end{equation}
The configuration of the Wilson lines in the definition of the tensor 
${\cal E}_{11}^{ij}$ is shown in fig.~\ref{fig:E11}.
Similarly to eq.~(\ref{eq:3p012decomp}), we can decompose the product 
$\sigma^k \nabla^i_{\bm{r}}$ into the trace, antisymmetric, and traceless
symmetric parts; if we apply the contact term to a wavefunction in the $^3P_0$ 
state, the antisymmetric and traceless symmetric parts vanish. Hence, we have 
\begin{equation}
-V_{{\cal Q}(^3S_1^{[8]})} \Big|_{^3P_0} =
-
\sigma^i \otimes \sigma^j
\frac{N_c}{d-1}
\nabla_{\bm{r}}^i \delta^{(d-1)} (\bm{r})
\nabla_{\bm{r}}^j
\frac{{\cal E}_{11}}{(d-1)N_c^2 m^2},
\end{equation}
where ${\cal E}_{11}$ is the trace of the tensor ${\cal E}_{11}^{ij}$, 
which can be written as 
\begin{equation}
{\cal E}_{11}
=
\int_0^\infty dt \, t
\int_0^\infty dt' t' \langle \bar{T} \left[ \Phi_\ell^{\dag ab} (0)
\Phi_0^{\dag ad}(0;t)
g E^{d,i}(t)
\right] 
T \left[
g E^{e,i} (t') \Phi_0^{ec} (0;t')
\Phi_\ell^{bc} (0) \right] \rangle.
\end{equation}
Here we suppress the spatial positions of the operators as they are all
evaluated at $\bm{x}=\bm{0}$. 
The relation between ${\cal E}_{11}$ and the quantity ${\cal E}$ defined in
refs.~\cite{Brambilla:2020ojz, Brambilla:2021abf} is given 
by ${\cal E}_{11} = \frac{(d-1) N_c}{9} {\cal E}$ in $d$ spacetime 
dimensions\footnote{
In refs.~\cite{Brambilla:2020ojz, Brambilla:2021abf} the matrix elements were
computed in 4 spacetime dimensions, because there the $d$ dependence did not
affect the phenomenological analysis.}. 
By plugging in the results for the contact terms to the 
pNRQCD formula in eq.~(\ref{eq:pNRQCDformula}), we obtain 
\begin{subequations}
\begin{eqnarray}
\langle {\cal O}^{\chi_{Q0}} (^3P_0^{[1]}) \rangle 
&=& \frac{3N_c}{2 \pi} |R'(0)|^2, 
\\
\langle {\cal O}^{\chi_{Q0}} (^3S_1^{[8]}) \rangle 
&=& \frac{3N_c}{2 \pi} |R'(0)|^2
\frac{{\cal E}_{11}}{(d-1) N_c^2 m^2}, 
\end{eqnarray}
\end{subequations}
where $R(r)$ is the radial wavefunction for the $\chi_{Q0}$ state. 
While the nonperturbative value for ${\cal E}_{11}$ has not yet been computed
from first principles, the fact that it is a universal quantity that does not
depend on the radial excitation or the heavy quark flavor enhances the
predictive power
of the nonrelativistic effective field theory formalism. For instance, we have 
\begin{equation}
\frac{m^2 \langle {\cal O}^{\chi_{Q0}} (^3S_1^{[8]}) \rangle}
{\langle {\cal O}^{\chi_{Q0}} (^3P_0^{[1]}) \rangle}
= \frac{{\cal E}_{11}}{(d-1) N_c^2}, 
\end{equation}
for any $^3P_0$ quarkonia. Note that ${\cal E}_{11}$ has a logarithmic scale
dependence that reproduces the evolution of the color-octet matrix
element~\cite{Brambilla:2020ojz, Brambilla:2021abf}.

We note that the derivatives $\bm{\nabla}_{\bm{r}}$ on the left and right of
the $\delta^{(d-1)} (\bm{r})$ in the contact terms in
eqs.~(\ref{eq:contactNRQCD}) come from the $Q \bar Q$ bilinears on the left and
right of the projection operator. 
Similarly, the two chromoelectric fields in ${\cal E}_{11}$ also come from the 
$Q \bar Q$ bilinears on the left and right of the projection operator.
By using these points, it is straightforward to obtain pNRQCD expressions 
for the shape functions. 
Rather than working directly with the definitions in eq.~(\ref{eq:shape}), 
we define the Fourier transforms
\begin{subequations}
\label{eq:shapeFT}
\begin{eqnarray}
\tilde {\cal S}^{\chi_{Q0}}_{^3P_0^{[1]}} (b_-) &=&
\int \frac{dl_+^*}{4 \pi} e^{i l^* \cdot b}
{\cal S}^{\chi_{Q0}}_{^3P_0^{[1]}} (l_+^*) 
\nonumber \\
&=& 
\frac{1}{d-1}
\langle
[\chi^\dag ( -\tfrac{i}{2} \overleftrightarrow{\bm{D}} \cdot \bm{\sigma} ) \psi
] (x=0)
{\cal P}_{\chi_{Q0}}
[ \psi^\dag ( -\tfrac{i}{2} \overleftrightarrow{\bm{D}} \cdot \bm{\sigma} )
\chi ] (x=b)
\rangle
,  \\
\tilde {\cal S}^{\chi_{Q0}}_{^3S_1^{[8]}} (b_-) &=&
\int \frac{dl_+^*}{4 \pi} e^{il^* \cdot b}
{\cal S}^{\chi_{Q0}}_{^3S_1^{[8]}} (l_+^*) 
\nonumber \\ &=&
\langle
\big( \chi^\dag \sigma^i T^a  \psi \Phi_\ell^{\dag ab} \big) (x=0)
{\cal P}_{\chi_{Q0}} 
\big(  \Phi_\ell^{bc} \psi^\dag \sigma^i T^c \chi \big) (x=b)
\rangle,
\end{eqnarray}
\end{subequations}
where now the $Q \bar Q$ bilinears on the right of the projection operator are
displaced by $b$, where $b$ lies in the $-$ direction. 
The shape functions in momentum space can be recovered from 
${\cal S}^{\chi_{Q0}}_{N} (l_+^*) = \int \frac{db_-}{2} e^{-i l^* \cdot b} 
\tilde {\cal S}^{\chi_{Q0}}_{N} (b_-)$. 
Other than that the $Q \bar Q$ bilinears on the left and right of the
projection operator are computed at different spacetime positions, the pNRQCD
calculations of the shape functions are done in the same way as NRQCD matrix
elements. We have 
\begin{eqnarray}
\label{eq:pNRQCDformulashape}
\tilde {\cal S}_N^{\cal Q} (b_-) 
&=& \int d^{d-1} R d^{d-1}r d^{d-1}r' \phi_{\cal Q} (\bm{x}_1-\bm{x}_2)
\nonumber \\ && \times
\big[ -V_{\tilde {\cal S}_N}
(\bm{x}_1, \bm{x}_2;\bm{\nabla}_1, \bm{\nabla}_2; b_-)
\delta^{(d-1)} (\bm{x}_1-\bm{x}_1'-\bm{b})
\delta^{(d-1)} (\bm{x}_2-\bm{x}_2'-\bm{b})\big]
\nonumber \\ && \times
\phi_{\cal Q}^* (\bm{x}_1'-\bm{x}_2'),
\end{eqnarray}
where the contact terms for the shape functions acquire dependence on $b_-$. 
The contact terms that survive when acting on wavefunctions in the $^3P_0$
state read 
\begin{subequations}
\label{eq:contactshape}
\begin{eqnarray}
-V_{\tilde {\cal S}_{^3P_0^{[1]}}} \Big|_{^3P_0} &=&
- \sigma^i \otimes \sigma^j
\frac{N_c}{d-1}
\nabla_{\bm{r}}^i
\delta^{(d-1)} (\bm{r})
\nabla_{\bm{r}}^j ,
\\
-V_{\tilde {\cal S}_{^3S_1^{[8]}}} \Big|_{^3P_0} &=&
- \sigma^i \otimes \sigma^j
\frac{N_c}{d-1}
\nabla_{\bm{r}}^i \delta^{(d-1)} (\bm{r})
\nabla_{\bm{r}}^j
\frac{\tilde {\cal E}\!\!\!/_{11} (b_-)}{(d-1)N_c^2 m^2},
\end{eqnarray}
\end{subequations}
where 
\begin{eqnarray}
\tilde {\cal E}\!\!\!/_{11} (b_-)
&=&
\int_0^\infty dt \, t
\int_0^\infty dt' t' \langle \bar{T} \left[ \Phi_\ell^{\dag ab} (0)
\Phi_0^{\dag ad}(0,\bm{0};t,\bm{0})
g E^{d,i}(t,\bm{0}) \right]
\nonumber \\ && \hspace{18ex} \times 
T \left[ 
g E^{e,i} (b_0+t', \bm{b}) \Phi_0^{ec} (b_0,\bm{b};b_0+t',\bm{b})
\Phi_\ell^{bc} (b) \right] \rangle.
\end{eqnarray}
Compared to the expressions for the contact terms for NRQCD matrix elements,
the only difference in the shape function formalism is that in 
$\tilde {\cal E}\!\!\!/_{11} (b_-)$, the chromoelectric field on the right 
and the Wilson lines associated with it are displaced by $b$. The displacement
does not affect the derivatives $\nabla_{\bm{r}}$, because they only depend on
the relative coordinates between the $Q$ and $\bar Q$, and they apply to
eigenstates of NRQCD. 
Since the contact term for the color-singlet shape function is independent of
$b_-$ and is the same as the contact term for the color-singlet NRQCD 
matrix element, we have 
\begin{equation}
\tilde {\cal S}^{\chi_{Q0}}_{^3P_0^{[1]}} (b_-) =
\langle {\cal O}^{\chi_{Q0}} (^3P_0^{[1]}) \rangle, 
\end{equation}
which implies 
\begin{equation}
{\cal S}^{\chi_{Q0}}_{^3P_0^{[1]}} (l_+) =
\delta(l_+) \langle {\cal O}^{\chi_{Q0}} (^3P_0^{[1]}) \rangle.
\end{equation}
On the other hand, the Fourier transformed color-octet shape function acquires 
dependence on $b_-$ through $\tilde {\cal E}\!\!\!/_{11} (b_-)$, so that 
\begin{equation}
\tilde {\cal S}_{^3S_1^{[8]}}^{\chi_{Q0}} (b_-)
= \frac{3N_c}{2 \pi} |R'(0)|^2
\frac{\tilde {\cal E}\!\!\!/_{11} (b_-)}{(d-1) N_c^2 m^2}.
\end{equation}
Then the shape function can be written as 
\begin{equation}
\label{eq:shapepNRQCDresult}
{\cal S}_{^3S_1^{[8]}}^{\chi_{Q0}} (l_+^*)
= \frac{3N_c}{2 \pi} |R'(0)|^2
\frac{{\cal E}\!\!\!/_{11} (l_+^*)}{(d-1) N_c^2 m^2},
\end{equation}
where 
\begin{equation}
{\cal E}\!\!\!/_{11} (l_+^*) = 
\int \frac{db_-}{2} e^{-il^* \cdot b} \tilde {\cal E}\!\!\!/_{11} (b_-).
\end{equation}
Similarly to the case of NRQCD matrix elements, the quantity 
${\cal E}\!\!\!/_{11} (l_+)$ is universal, and do not depend on the radial
excitation or the heavy quark flavor of the quarkonium state. 
In particular, we have 
\begin{equation}
\label{eq:shapeuniversality}
\frac{m^2 {\cal S}_{^3S_1^{[8]}}^{\chi_{Q0}} (l_+^*) }
{\langle {\cal O}^{\chi_{Q0}} (^3P_0^{[1]}) \rangle}
= \frac{{\cal E}\!\!\!/_{11} (l_+^*)}{(d-1) N_c^2},
\end{equation}
which is valid for any $^3P_0$ quarkonia. This, together with heavy quark spin
symmetry, implies that the $l_+$ dependence of the color octet shape function
is universal for all $P$-wave heavy quarkonia, including all radial excitations
of $\chi_c$ and $\chi_b$. This significantly reduces the model dependence in
the phenomenological analysis of $\chi_c$ and $\chi_b$ production 
in the shape function formalism. 

An important consistency check of the pNRQCD calculation is to compare the
result with the perturbative calculation. If we compute 
${\cal E}\!\!\!/_{11} (l_+^*)$ in perturbative QCD 
at leading nonvanishing order in $\alpha_s$, we obtain 
\begin{eqnarray}
{\cal E}\!\!\!/_{11} (l_+^*) |_{\rm pQCD}
&=&
4 \pi \alpha_s (d-2) (N_c^2-1) \mu^{2 \epsilon}
\int_0^\infty dt \,t 
\int_0^\infty dt' \,t'
\nonumber \\ && \times 
\int \frac{db_-}{2} e^{-i l^* \cdot b}
\int \frac{d^dk}{(2 \pi)^d} \frac{i \bm{k}^2}{k^2+i \varepsilon}
e^{ i k \cdot b}
e^{ i k_0 (t -t')} + O(\alpha_s^2)
\nonumber \\
&=&
4 \pi \alpha_s (d-2) (N_c^2-1) \mu^{2 \epsilon}
\int \frac{d^{d-1}k}{(2 \pi)^{d-1}} \frac{1}{2 |\bm{k}|^3}
\delta(k_+-l_+^* )+ O(\alpha_s^2)
\nonumber \\
&=&
\alpha_s (d-2) (N_c^2-1) \mu^{2 \epsilon}
\int \frac{d^{d-2}\bm{k}_\perp}{(2 \pi)^{d-2} l_+^*}
\frac{1}{(l_+^*-\bm{k}_\perp^2/l_+^*)^2/4+\bm{k}_\perp^2}+ O(\alpha_s^2)
\nonumber \\
&=&
\frac{(d-1) C_\epsilon }{3} \frac{4 \alpha_s C_F N_c}
{\pi (l_+^*)^{1+2 \epsilon}}+ O(\alpha_s^2) ,
\end{eqnarray}
where $C_\epsilon$ is defined in eq.~(\ref{eq:Cepsilon}). 
By plugging in this expression into the pNRQCD expression for the shape
function (\ref{eq:shapepNRQCDresult}) and using 
$\langle {\cal O}^{\chi_{Q0}} (^3P_0^{[1]}) \rangle
= \frac{3N_c}{2 \pi} |R'(0)|^2$ we obtain 
\begin{equation}
{\cal S}_{^3S_1^{[8]}}^{\chi_{Q0}} (l_+) \Big|_{\rm pQCD}
= 
\langle {\cal O}^{\chi_{Q0}} (^3P_0^{[1]}) \rangle 
\left(
\frac{4 \alpha_s C_F}{ 3 N_c m^2 \pi} 
\frac{C_\epsilon}{(l_+^*)^{1+2 \epsilon}} + O(\alpha_s^2) \right), 
\end{equation}
which exactly reproduces the order-$\alpha_s$ calculation of 
${\cal S}_{^3S_1^{[8]}}^{Q \bar{Q} (^3P_0^{[1]})} (l_+)$ 
in the previous section, demonstrating the validity of the pNRQCD result at
one-loop level. 
This is also consistent with the result for the scheme conversion 
$\Delta(\Lambda, l_+^{* \rm max} = \Lambda e^{-1/6}) = 0 + O(\alpha_s^2)$, 
so that we have 
\begin{equation}
\label{eq:normalization2}
\int_0^{l_+^{* \rm max}} dl_+^* {\cal S}_{^3S_1^{[8]}}^{\chi_{Q0}} (l_+^*) 
\Big|_{l_+^{* \rm max}=\Lambda e^{-1/6}} 
= \langle {\cal O}^{\chi_{Q0}} ({}^3S_1^{[8]}) \rangle^{\overline{\rm MS}}.
\end{equation}
As have been discussed in section~\ref{sec:diagram}, 
the UV divergence on the left-hand side is
regulated by the cutoff $l_+^{* \rm max}$, so that the nonperturbative shape
function ${\cal S}_{^3S_1^{[8]}}^{\chi_{Q0}} (l_+^*)$ 
can be computed in $d=4$
dimensions. This allows us to use the UV-regulated normalization condition to
constrain the model dependence of the nonperturbative shape function. 

So far the pNRQCD calculations of the NRQCD matrix elements and shape functions
presented in this section were based on the assumption that the $\chi_{Q0}$
state is strongly coupled, so that the ultrasoft scale, where the gluons have
energies and momenta of order $mv^2$, does not play a role. 
In the weakly coupled case, the chromoelectric dipole interaction will mix
color-singlet and color-octet $Q \bar Q$ states through exchange of ultrasoft
gluons, so that the contribution from
the color-octet wavefunction must be included. 
As have been discussed in refs.~\cite{Beneke:1999zr, Beneke:2013jia}, 
this interaction is suppressed by $v^{3/2}$ compared to the kinetic terms 
in the pNRQCD Lagrangian, so that for heavy quarkonium states, the contribution from 
the color-octet wavefunction is suppressed by $v^3$ compared to the
color-singlet part. The color-octet wavefunction can appear in the color-octet
NRQCD matrix element and the shape function through the contact terms at
leading order in $1/m$, which is enhanced by $1/v^2$ compared to
the contact terms that act on the color-singlet wavefunction 
[eqs.~(\ref{eq:contactNRQCD}) and (\ref{eq:contactshape})]. 
As a result, in the weak coupling case, the extra contribution coming from the
color-octet wavefunction is suppressed by at least a power of $v$ compared to
the pNRQCD expressions for NRQCD matrix elements and shape functions in the
strongly coupled case. 
Hence, we expect the pNRQCD results obtained in this section to also hold in
the weakly coupled case at leading nonvanishing order in $v$.

\section{One-loop matching conditions}
\label{sec:matching}

Based on the calculations of the perturbative shape functions, we now derive
the matching conditions that allow determination of the short-distance
coefficients $s_N$ for the shape function formalism. 

As have been discussed in section~\ref{sec:shape}, 
the matching coefficients are computed from
the cross sections for inclusive production of $Q \bar Q$ in specific color and
angular momentum states. The cross section for production of $Q \bar Q$ in the 
$^3S_1^{[8]}$ state 
[eqs.~(\ref{eq:NRQCDmatchform_octet}) and (\ref{eq:shapematchform_octet})]
gives the following matching condition
\begin{eqnarray}
c_{^3S_1^{[8]}} (P) 
\langle {\cal O}^{Q \bar{Q}(^3S_1^{[8]})} (^3S_1^{[8]}) \rangle
&=& 
\int_0^\infty dl_+ \,
s_{^3S_1^{[8]}} (P+l) {\cal S}_{^3S_1^{[8]}}^{Q\bar{Q} (^3S_1^{[8]})}(l_+)
\nonumber \\
&=& 
s_{^3S_1^{[8]}} (P) 
\langle {\cal O}^{Q \bar{Q}(^3S_1^{[8]})} (^3S_1^{[8]}) \rangle. 
\end{eqnarray}
In the last line, we used the one-loop result for the perturbative shape
function in eq.~(\ref{eq:shapeDR_result1}).
This implies that 
$s_{^3S_1^{[8]}} (P) = c_{^3S_1^{[8]}} (P)$ holds at one-loop level. 
Similarly, expressions for the $^3P_J^{[1]}$ production cross section 
[eqs.~(\ref{eq:NRQCDmatchform_singlet}) and (\ref{eq:shapematchform_singlet})]
give 
\begin{eqnarray}
\label{eq:singletmatchcond_mid}
&& \hspace{-5ex} 
c_{^3P_J^{[1]}}(P)
\langle {\cal O}^{Q \bar{Q}(^3P_0^{[1]})} (^3P_0^{[1]}) \rangle
+ c_{^3S_1^{[8]}} (P)
\langle {\cal O}^{Q \bar{Q}(^3P_0^{[1]})} (^3S_1^{[8]}) \rangle 
\nonumber \\ &=& 
\int_0^\infty dl_+
\left[ s_{^3P_J^{[1]}}(P+l) {\cal S}_{^3P_0^{[1]}}^{Q\bar{Q} (^3P_0^{[1]})}(l_+)
+ s_{^3S_1^{[8]}} (P+l) {\cal S}_{^3S_1^{[8]}}^{Q\bar{Q} (^3P_0^{[1]})}(l_+)
\right]
\nonumber \\ 
&=& 
s_{^3P_J^{[1]}}(P) 
\langle {\cal O}^{Q \bar{Q}(^3P_0^{[1]})} (^3P_0^{[1]}) \rangle
+
\int_0^\infty dl_+
c_{^3S_1^{[8]}} (P+l) {\cal S}_{^3S_1^{[8]}}^{Q\bar{Q} (^3P_0^{[1]})}(l_+), 
\end{eqnarray}
where the first line is the NRQCD expression divided by $2 J+1$, 
and the last line follows from the result for the perturbative shape
function in eq.~(\ref{eq:shapeDR_result1}) and the result 
$s_{^3S_1^{[8]}} (P) = c_{^3S_1^{[8]}} (P)$. 
In principle, we can obtain expressions for $s_{^3P_J^{[1]}}(P)$ in terms of 
$c_{^3P_J^{[1]}}(P)$ and $c_{^3S_1^{[8]}} (P)$ by plugging in the perturbative 
expressions for $\langle {\cal O}^{Q \bar{Q}(^3P_0^{[1]})} (^3S_1^{[8]}) \rangle$
and ${\cal S}_{^3S_1^{[8]}}^{Q\bar{Q} (^3P_0^{[1]})}(l_+)$. 
However, the resulting expression is not very useful, because the matrix
element $\langle {\cal O}^{Q \bar{Q}(^3P_0^{[1]})} (^3S_1^{[8]}) \rangle$ 
contains an IR pole, and hence, the NRQCD side of the expression depends on the
order-$\epsilon$ contribution to the short-distance coefficient 
$c_{^3S_1^{[8]}} (P)$ that is usually not available. 
The same IR pole that matches the one on the NRQCD side of the expression 
does appear on the shape function formalism side, which arises from the 
integral in the last line of eq.~(\ref{eq:singletmatchcond_mid}) 
in the region $l_+ \to 0$. 
Hence, the terms involving IR poles on both sides 
can be matched exactly if we rewrite this integral as
\begin{eqnarray}
\int_0^\infty dl_+
c_{^3S_1^{[8]}} (P+l) {\cal S}_{^3S_1^{[8]}}^{Q\bar{Q} (^3P_0^{[1]})}(l_+)
&=& 
\int_0^\infty dl_+
\left( c_{^3S_1^{[8]}} (P+l) - c_{^3S_1^{[8]}} (P) \right)
{\cal S}_{^3S_1^{[8]}}^{Q\bar{Q} (^3P_0^{[1]})}(l_+)
\nonumber \\ && 
+ c_{^3S_1^{[8]}} (P) \langle {\cal O}^{Q \bar{Q}(^3P_0^{[1]})} (^3S_1^{[8]})
\rangle^{\rm bare},
\end{eqnarray}
where the unrenormalized (``bare'') matrix element is given by
eq.~(\ref{eq:barematrixelement}). 
Note that the integral over $l_+$ on the right-hand side 
is IR finite because the integrand factor in
the parenthesis vanishes as $l_+ \to 0$, and the only IR pole is isolated in
the perturbative color-octet matrix element. 
This expression is still not completely satisfactory, because
NRQCD matching calculations are usually done in the $\overline{\rm MS}$
scheme, while the above expression involves an unrenormalized matrix element. 
Rather than subtracting the pole to carry out the renormalization in the 
$\overline{\rm MS}$ scheme, we can use the result for 
$\Delta(\Lambda,l_+^{\rm max})$ we obtained earlier in 
eq.~(\ref{eq:schemeconvresult}) to cut off the UV-divergent integral.
By using eqs.~(\ref{eq:normalization}) and (\ref{eq:schemeconvresult}), 
we have at one-loop level 
\begin{eqnarray}
\int_0^\infty dl_+
\theta(l_+^{\rm max}-l_+)
{\cal S}_{^3S_1^{[8]}}^{Q\bar{Q} (^3P_0^{[1]})}(l_+)
\Big|_{l_+^{*\rm max}= \Lambda e^{-1/6}}
= \langle {\cal O}^{Q \bar{Q}(^3P_0^{[1]})} (^3S_1^{[8]})
\rangle^{\overline{\rm MS}}, 
\end{eqnarray}
where $\theta(x)$ is the step function that vanishes for $x<0$ and 
equals $1$ for $x \geq 0$, and 
$\Lambda$ is the $\overline{\rm MS}$ renormalization scale 
for the color-octet matrix element 
$\langle {\cal O}^{Q \bar{Q}(^3P_0^{[1]})} (^3S_1^{[8]}) \rangle$.
By using this we can rewrite the color-octet shape-function term in the
matching condition as 
\begin{eqnarray}
&& \hspace{-5ex} 
\int_0^\infty dl_+
c_{^3S_1^{[8]}} (P+l) {\cal S}_{^3S_1^{[8]}}^{Q\bar{Q} (^3P_0^{[1]})}(l_+)
= 
c_{^3S_1^{[8]}} (P) \langle {\cal O}^{Q \bar{Q}(^3P_0^{[1]})} (^3S_1^{[8]})
\rangle^{\overline{\rm MS}}
\nonumber \\ &&
+ \int_0^\infty dl_+
\left( c_{^3S_1^{[8]}} (P+l) - c_{^3S_1^{[8]}} (P) \theta(l_+^{\rm max}-l_+) 
\right)
{\cal S}_{^3S_1^{[8]}}^{Q\bar{Q} (^3P_0^{[1]})}(l_+),
\end{eqnarray}
where now every term is UV finite, and the IR pole is isolated in the
$\overline{\rm MS}$-renormalized matrix element at the scale $\Lambda 
= l_+^{* \rm max} e^{1/6}$. 
By plugging this expression to eq.~(\ref{eq:singletmatchcond_mid}) we obtain
\begin{eqnarray}
\label{eq:singletmatchcond_mid2}
&& \hspace{-5ex}
c_{^3P_J^{[1]}}^{\overline{\rm MS}}(P)
\langle {\cal O}^{Q \bar{Q}(^3P_0^{[1]})} (^3P_0^{[1]}) \rangle
+ c_{^3S_1^{[8]}} (P)
\langle {\cal O}^{Q \bar{Q}(^3P_0^{[1]})} (^3S_1^{[8]}) \rangle^{\overline{\rm
MS}}
\nonumber \\
&=&
s_{^3P_J^{[1]}}(P)
\langle {\cal O}^{Q \bar{Q}(^3P_0^{[1]})} (^3P_0^{[1]}) \rangle
+ c_{^3S_1^{[8]}} (P) \langle {\cal O}^{Q \bar{Q}(^3P_0^{[1]})} (^3S_1^{[8]})
\rangle^{\overline{\rm MS}}
\nonumber \\ && 
+
\int_0^\infty dl_+
\left( c_{^3S_1^{[8]}} (P+l) - c_{^3S_1^{[8]}} (P) \theta(l_+^{\rm max}-l_+) 
\right)
{\cal S}_{^3S_1^{[8]}}^{Q\bar{Q} (^3P_0^{[1]})}(l_+). 
\end{eqnarray}
If we solve this for $s_{^3P_J^{[1]}}(P)$, the terms involving IR poles from
the color-octet matrix element cancel exactly, so everything can be computed at
$d=4$. 
By using the result for ${\cal S}_{^3S_1^{[8]}}^{Q\bar{Q} (^3P_0^{[1]})}(l_+)$
in eq.~(\ref{eq:shapeDR_result3}), we obtain 
\begin{equation}
\label{eq:singletmatchcond_end}
s_{^3P_J^{[1]}}(P)
= 
c_{^3P_J^{[1]}}^{\overline{\rm MS}}(P)
- 
\left(
c_{^3P_J^{[1]}}^{\rm singular}(P)
- c_{^3P_J^{[1]}}^{\rm pole}(P)
\right),
\end{equation}
where
\begin{subequations}
\begin{eqnarray}
c_{^3P_J^{[1]}}^{\rm singular}(P)
&=&
\frac{4 \alpha_s C_F}{3 N_c \pi m^2} 
\int_0^\infty dl_+
c_{^3S_1^{[8]}} (P+l) 
\frac{C_\epsilon (P_+/P_+^*)^{2 \epsilon}}{l_+^{1+2 \epsilon}}, 
\\
c_{^3P_J^{[1]}}^{\rm pole}(P)
&=&
\frac{4 \alpha_s C_F}{3 N_c \pi m^2} 
c_{^3S_1^{[8]}} (P) 
\int_0^{l_+^{\rm max}} dl_+
\frac{C_\epsilon (P_+/P_+^*)^{2 \epsilon}}{l_+^{1+2 \epsilon}}. 
\end{eqnarray}
\end{subequations}
The labels ``singular'' and ``pole'' will be explained later.
Since the poles cancel in the last line of
eq.~(\ref{eq:singletmatchcond_mid2}), 
the $\epsilon$ dependences in $c_{^3S_1^{[8]}}$ can be dropped. 
By using this result for $s_{^3P_J^{[1]}}(P)$ 
we can now write the cross section in the shape function formalism as 
\begin{equation}
\label{eq:shapefac_f1}
\sigma[{\chi_{QJ} (P)}]_{\rm shape} = 
(2 J+1) \bigg[ 
s_{^3P_J^{[1]}}(P) 
\langle {\cal O}^{\chi_{Q0}} (^3P_0^{[1]}) \rangle 
+ \int_0^\infty dl_+
c_{^3S_1^{[8]}} (P+l) {\cal S}_{^3S_1^{[8]}}^{\chi_{Q0}}(l_+) \Big|_{\rm NP}
\bigg],
\end{equation}
where $s_{^3P_J^{[1]}}(P)$ is given in eq.~(\ref{eq:singletmatchcond_end}), 
and ${\cal S}_{^3S_1^{[8]}}^{\chi_{Q0}}(l_+) \big|_{\rm NP}$ 
is the nonperturbative color-octet shape function. 

The expression for $s_{^3P_J^{[1]}}(P)$ in eq.~(\ref{eq:singletmatchcond_end})
calls for further exploration. We show that the term 
$c_{^3P_J^{[1]}}^{\rm singular}(P)$
is essentially the singular contribution to 
the $Q \bar Q(^3P_J^{[1]})$ cross section that occurs through the 
production of $Q \bar Q (^3S_1^{[8]})$, which then evolves into a $P$-wave
color-singlet $Q \bar Q$ by emitting a soft gluon. 
The soft gluon emission amplitude from a heavy quark line with momentum $p_1$ 
is given by 
\begin{equation}
\bar{u} (p_1) ig\gamma^\mu T^a \frac{i (p\!\!\!/_1 + k\!\!\!/ +m )}
{k^2+2 k \cdot p_1 +i \varepsilon}
\approx \bar{u} (p_1) ig\gamma^\mu T^a 
\frac{i (p\!\!\!/_1+m)}{2 k \cdot p_1 +i \varepsilon}
=
- g \bar{u} (p_1) T^a \frac{p_1^\mu}{k \cdot p_1 +i \varepsilon},
\end{equation}
where we retained only the leading contribution in the limit where $k$ is soft,
and in the last equality we anticommuted $p\!\!\!/_1$ to the left and used the
equation of motion. 
Similarly, the infrared divergent contribution coming from the heavy antiquark
line with momentum $p_2$ is given by 
\begin{equation}
\frac{i (- p \!\!\!/_2 - k\!\!\!/+m)}{k^2+2 k \cdot p_2 +i
\varepsilon} ig \gamma^\mu T^a v(p_2)
\approx
\frac{i (- p \!\!\!/_2+m)}{2 k \cdot p_2 +i
\varepsilon} ig \gamma^\mu T^a v(p_2)
=
+ g \frac{p_2^\mu}{k \cdot p_2 +i \varepsilon}
T^a
v(p_2).
\end{equation}
Hence, the soft-gluon emission gives rise to an infrared divergent
contribution that is given by $\sigma[Q \bar Q(^3S_1^{[8]}) (p_1+p_2+k)]$ 
times the factor 
\begin{equation}
\label{eq:irdivfac}
- g \left(\frac{p_1^\mu}{k \cdot p_1 +i \varepsilon} 
- \frac{p_2^\mu}{k \cdot p_2 +i \varepsilon} \right)T^a
\end{equation}
multiplied on both sides of the cut, and integrated over the phase space of the
soft gluon. We can see that this soft gluon factor is proportional to 
$1/|\bm{k}|$, so that the phase space integral involves the singular 
integral $\int d^{d-1} k\, |\bm{k}|^{-3}$. 
After a straightforward calculation, at leading nonvanishing order in 
expansion in powers of $q=(p_1-p_2)/2$ compared to $p_1+p_2$, 
the singular contribution to the $Q \bar Q(^3P_J^{[1]})$ cross section 
is given by 
\begin{equation}
\label{eq:singularpart}
(2 J+1) \langle {\cal O}^{Q \bar Q(^3P_0^{[1]})} (^3P_0^{[1]}) \rangle
\left(
\frac{4 \pi \alpha_s C_F}{3 N_c \pi m^2} 
\int_0^\infty \frac{d |\bm{k}|}{|\bm{k}|^{1+2 \epsilon}} c_{^3S_1^{[8]}} (P+k)
+ O(\epsilon^0) \right), 
\end{equation}
which, aside from an order-$\epsilon^0$ finite piece, matches exactly the 
contribution from the 
$c_{^3P_J^{[1]}}^{\rm singular}(P)$ 
term. 
This term does not have a UV divergence, because 
$c_{^3S_1^{[8]}} (P+l)$ vanishes when $l$ is sufficiently large, but it does
have an IR pole coming from the region $l_+ \to 0$. 
This IR pole is canceled exactly by the term 
$c_{^3P_J^{[1]}}^{\rm pole}(P)$, which can be computed as 
\begin{equation}
c_{^3P_J^{[1]}}^{\rm pole}(P)=
\frac{4 \alpha_s C_F}{3 N_c \pi m^2} c_{^3S_1^{[8]}} (P) 
\left( - \frac{1}{2 \epsilon_{\rm IR}} \right) + O(\epsilon),  
\end{equation}
where we set $l_+^{*\rm max} = \Lambda e^{-1/6}$. Hence, the term 
$c_{^3P_J^{[1]}}^{\rm pole}(P)$
subtracts exactly the IR pole in 
$c_{^3P_J^{[1]}}^{\rm singular}(P)$. 
Note that in practical calculations, $c_{^3P_J^{[1]}}^{\overline{\rm MS}}$ is
obtained by first computing the IR-divergent $Q \bar Q(^3P_J^{[1]})$ cross
section, and then subtracting the IR pole. 
As the same thing happens in the combination 
$c_{^3P_J^{[1]}}^{\rm singular}(P)-c_{^3P_J^{[1]}}^{\rm pole}(P)$, 
this is essentially the singular soft-gluon 
contribution in $c_{^3P_J^{[1]}}^{\overline{\rm MS}}(P)$. 
That is, the second term in eq.~(\ref{eq:singletmatchcond_end}) subtracts the
singular soft-gluon emission contribution from the NRQCD short-distance
coefficient $c_{^3P_J^{[1]}}^{\overline{\rm MS}}$. 
Note that this identification still holds at higher orders in $\alpha_s$ 
when radiative corrections are included in $c_{^3S_1^{[8]}} (P+l)$ and 
$c_{^3S_1^{[8]}} (P)$, because once the color-octet $Q
\bar Q$ evolves into a color-singlet state by emission of soft gluons, the
color-singlet $Q \bar Q$ no longer emits soft gluons unless corrections of
higher orders in $v$ are included, 
as we have seen in the calculation of the matrix
elements and shape functions in the previous section. 

The soft-gluon emission that is subtracted from the color-singlet 
short-distance coefficient instead appears in the color-octet shape
function contribution, where the nonperturbative effect of the soft gluon
emission is encoded in the shape function. To see this explicitly, we can
rewrite the color-octet shape function term in the cross section formula 
(\ref{eq:shapefac_f1}) as 
\begin{eqnarray}
\label{eq:sub_NP}
&& \hspace{-5ex} 
\int_0^\infty dl_+
c_{^3S_1^{[8]}} (P+l) {\cal S}_{^3S_1^{[8]}}^{\chi_{Q0}}(l_+) \Big|_{\rm NP}
=
c_{^3S_1^{[8]}} (P) \langle {\cal O}^{\chi_{Q0}} (^3S_1^{[8]})
\rangle^{\overline{\rm MS}}
\nonumber \\ && 
+
\int_0^\infty dl_+
\left( c_{^3S_1^{[8]}} (P+l) 
- c_{^3S_1^{[8]}} (P) \theta(l_+^{\rm max} - l_+) \right)
{\cal S}_{^3S_1^{[8]}}^{\chi_{Q0}}(l_+) \Big|_{\rm NP}. 
\end{eqnarray}
We see that the second term in eq.~(\ref{eq:singletmatchcond_end}) is 
equivalent to the last line of the above expression, with an opposite sign and 
the shape function replaced by its large-$l_+$ asymptotic behavior
given in eq.~(\ref{eq:shape_asymptotic}). 
By using this we can rewrite eq.~(\ref{eq:shapefac_f1}) as 
\begin{eqnarray}
\label{eq:shapefac_f2}
\sigma[{\chi_{QJ} (P)}]_{\rm shape} &=&
(2 J+1) \bigg[
c_{^3P_J^{[1]}}^{\overline{\rm MS}}(P)
\langle {\cal O}^{\chi_{Q0}} (^3P_0^{[1]}) \rangle
+
c_{^3S_1^{[8]}} (P) 
\langle {\cal O}^{\chi_{Q0}} (^3S_1^{[8]}) \rangle^{\overline{\rm MS}}
\nonumber \\ && \hspace{10ex} 
+
\int_0^\infty dl_+
\left( c_{^3S_1^{[8]}} (P+l) - c_{^3S_1^{[8]}} (P) \theta(l_+^{\rm max}-l_+)
\right)
\nonumber \\ && \hspace{22ex} \times 
\Big( {\cal S}_{^3S_1^{[8]}}^{\chi_{Q0}}(l_+) \Big|_{\rm NP}- 
{\cal S}_{^3S_1^{[8]}}^{\chi_{Q0}}(l_+) \Big|_{\rm asy} \Big)
\bigg]. 
\end{eqnarray}
Here, the first line is just the cross section in NRQCD, while the remaining
terms arise from the deviation of the nonperturbative shape function from its 
asymptotic form. That is, the integral over $l_+$ in eq.~(\ref{eq:shapefac_f2})
encode the nonperturbative corrections to the NRQCD factorization formalism
that arise from resummation of kinematical effects from the motion of 
$Q \bar Q$ relative to the quarkonium. 

Although eqs.~(\ref{eq:shapefac_f1}) and (\ref{eq:shapefac_f2}) are in
principle equivalent expressions for the $\chi_{QJ}$ production cross section
in the shape function formalism, they have different phenomenological
implications, as we will now briefly describe. 
In order to compute cross sections from 
eq.~(\ref{eq:shapefac_f1}), the short-distance coefficient 
$s_{^3P_J^{[1]}} (P)$ must be determined from
eq.~(\ref{eq:singletmatchcond_end}). 
Note that at one-loop level, 
the NRQCD short-distance coefficient $c_{^3P_J^{[1]}} (P)$
contains the soft-gluon contribution given by the second term
in eq.~(\ref{eq:singletmatchcond_end}).
This soft-gluon effect generates a logarithm of the $\overline{\rm MS}$ scale 
in $c_{^3P_J^{[1]}}$, which is canceled by the scale dependence of the
color-octet matrix element times $c_{^3S_1^{[8]}}$ at tree level. 
However, since usually in NRQCD calculations both the 
color-singlet and color-octet short-distance coefficients are computed to same
accuracy in $\alpha_s$, there is always a mismatch of the orders in $\alpha_s$
of soft-gluon effects between the color-singlet and color-octet channels,
due to the one-loop correction included in $c_{^3S_1^{[8]}}$. 
This mismatch can have a significant effect, 
as the soft gluon emission involves IR divergences which make the 
cross sections singular near threshold. If we use eq.~(\ref{eq:shapefac_f1}) to
compute $\chi_{QJ}$ cross sections, this singular contribution is subtracted
from the $^3P_J^{[1]}$ short-distance coefficient, and added back to the second
term in the square brackets of eq.~(\ref{eq:shapefac_f1}) in the form of 
eq.~(\ref{eq:sub_NP}). In this way, the singular soft-gluon emission effects
are contained entirely in the color-octet shape function contribution 
in the form given in eq.~(\ref{eq:sub_NP}), where the short-distance
coefficient $c_{^3S_1^{[8]}}$ can now be computed to same order in $\alpha_s$ 
for every term. This allows the singular soft-gluon emission effects to be
computed to consistent accuracy in $\alpha_s$. Note that, since the integral
over $l_+$ in eq.~(\ref{eq:sub_NP}) is infrared finite even when the
nonperturbative shape function is replaced by its asymptotic form, this
matching of singular soft-gluon effects can be done without knowledge of the 
nonperturbative form of ${\cal S}_{^3S_1^{[8]}}^{\chi_{Q0}}(l_+)$. 
We will investigate this point in more detail in section~\ref{sec:threshold}.

The form of the cross section given in eq.~(\ref{eq:shapefac_f2}) is suitable
for computing the nonperturbative corrections to the NRQCD factorization
formalism from the nonperturbative shape function. 
We note that if we were to neglect the deviation of the nonperturbative shape
function from its asymptotic form, the last line of eq.~(\ref{eq:shapefac_f2}) 
vanishes and we recover the cross section in NRQCD factorization. 
This is consistent with the fact that, if the color-octet shape function 
is given by the asymptotic form in eq.~(\ref{eq:shape_asymptotic}), 
all of the higher-dimensional matrix elements in eq.~(\ref{eq:highdimoctet})
are proportional to scaleless power divergences 
$\int_0^\infty dl_+ \, l_+^{n-1}$ for $n\ge 1$, 
which vanish in dimensional regularization (in the case of higher-dimensional
color-singlet matrix elements, they vanish in the form 
$\int_0^\infty dl_+ \, l_+^n \delta(l_+) = 0$).
The nonperturbative corrections to the $p_T$-differential cross sections 
coming from the last two lines in eq.~(\ref{eq:shapefac_f2}) 
will be computed in section~\ref{sec:NPmodel} using models for the
nonperturbative shape functions, which are constrained by the asymptotic form
(\ref{eq:shape_asymptotic}) and the normalization condition given by 
eq.~(\ref{eq:normalization2}).

\section{Phenomenological applications}
\label{sec:pheno}

Based on the one-loop level matching conditions and the shape functions we have
established in the previous sections, we compute $\chi_c$ and $\chi_b$ 
cross sections in the shape function formalism and compare the results with
NRQCD predictions. As we have explained in sec.~\ref{sec:intro}, understanding
the $\chi_c$ and $\chi_b$ cross sections are not only important for comparing 
predictions with cross section measurements in experiment, 
but they are also important for correctly treating
the feeddown contributions in $S$-wave production rates. 
We will show in this section that the shape function formalism can modify the
transverse-momentum dependent cross sections of $P$-wave heavy quarkonia,
which may have important implications in heavy quarkonium production
phenomenology.

\subsection{\boldmath Matching of soft-gluon emission effects at large
transverse momentum} 
\label{sec:threshold}

We first investigate the phenomenological application of the cross section
formula in eq.~(\ref{eq:shapefac_f1}) for mixing of soft-gluon emission effects 
between color-singlet and color-octet channels. 
Since the effects of soft-gluon emission near threshold is most prominent 
at large $p_T$, we consider the cross section in the fragmentation
approximation, which describes the cross section at leading power in the
expansion in powers of $m^2/p_T^2$~\cite{Collins:1981uw}. 
In this approximation, the cross section is given by 
\begin{equation}
\sigma[\chi_{QJ} (P)] = \sum_{i=g, q, \bar{q}} 
\int dz \, \hat{\sigma}_i (k) D_{i \to \chi_{QJ}+X} (z), 
\end{equation}
where the sum is over gluon, light quarks and antiquarks, 
$\hat{\sigma}_i (k)$ is the production rate of a parton $i$ with momentum 
$k$, $D_{i \to \chi_{QJ}} (z)$ is the fragmentation function,
which is a function of the momentum fraction $z = P_+/k_+$, 
with the $+$ direction defined along the quarkonium momentum. 
For notational convenience, we will write the parton cross section
$\hat{\sigma}_i$ as a function of $z$. 
The NRQCD factorization formula for the fragmentation function reads 
\begin{equation}
D_{i \to \chi_{cJ} + X} (z)
=
(2 J+1) \left[
d_{i \to {}^3S_1^{[8]}} (z)
\langle {\cal O}^{\chi_{Q0}} (^3S_1^{[8]}) \rangle
+
d_{i \to {}^3P_J^{[1]}} (z)
\langle {\cal O}^{\chi_{Q0}} (^3P_J^{[1]}) \rangle \right],
\end{equation}
where the short-distance coefficients for the fragmentation functions can be 
computed perturbatively. 
The short-distance coefficients for the fragmentation functions begin at order
$\alpha_s$, and the parton cross sections $\hat{\sigma}_i$ begin at order 
$\alpha_s^2$ for proton-proton collisions. Hence, parton cross sections to
order-$\alpha_s^3$ accuracy and fragmentation functions to order-$\alpha_s^2$ accuracy 
are needed to compute large-$p_T$ quarkonium cross sections at NLO 
in $\alpha_s$. 
The complete results for heavy quarkonium
fragmentation functions to order-$\alpha_s^2$ accuracy can be found in
refs.~\cite{Braaten:1993rw, Braaten:1993mp, Yuan:1994hn, Cho:1994gb,
Braaten:1995cj, Ma:1995vi, Braaten:2000pc, Bodwin:2012xc, Ma:2013yla,
Bodwin:2014bia, Ma:2015yka}. 
The quark fragmentation function $d_{q \to {}^3S_1^{[8]}}$ begins at 
order $\alpha_s^2$, while quark fragmentation into $Q \bar Q(^3P_J^{[1]})$
begin at order $\alpha_s^3$ and can be neglected at this accuracy. 
The antiquark fragmentation functions are same as quark fragmentation
functions, because quarkonia are charge conjugation eigenstates. 
The gluon fragmentation functions 
$d_{g \to {}^3P_J^{[1]}} (z)$ and $d_{g \to {}^3S_1^{[8]}} (z)$ to
order-$\alpha_s^2$ accuracy read 
\begin{subequations}
\begin{eqnarray}
\label{eq:FFs}
d_{g \to {}^3P_J^{[1]}} (z)
&=&
\frac{2 \alpha_s^2}{27 N_c m^5} \left[
\left( \frac{Q_J}{2J+1} \! -\! \log \frac{\Lambda}{2 m} \right) \delta(1-z)
\!+\! \frac{z}{(1-z)_+ \!\!} \! +\! \frac{P_J(z)}{2 J+1} \right] 
+ O(\alpha_s^3),
\quad\quad
\\
\label{eq:FFoctet}
d_{g \to {}^3S_1^{[8]}} (z)
&=&
\frac{\pi \alpha_s(\mu_f)}{24 m^3}
\delta (1-z) 
+ 
\frac{\alpha_s^2(\mu_f)}{24 m^3}
\bigg[
\frac{\beta_0}{2} \left( \log \frac{\mu_f}{2 m} + \frac{13}{6} \right)
\delta (1-z) 
\nonumber \\ && \hspace{2ex}
+ \left( \frac{2}{3} - \frac{\pi^2}{2} + 8 \log 2
\right)
\delta (1-z) 
+ \left( \log\frac{\mu_f}{2 m} - \frac{1}{2} \right) P_{gg} (z)
\nonumber \\ && \hspace{2ex}
+ \frac{3 (1-z)}{z}
+ 6 (2-z+z^2) \log(1-z) 
- \frac{6}{z} \left( \frac{\log(1-z)}{1-z} \right)_{\!+\!}
\bigg] + O(\alpha_s^3),
\quad \quad
\end{eqnarray}
\end{subequations}
where $\mu_f$ is the factorization scale for the fragmentation function,
$\beta_0 = \frac{11}{3} N_c -\frac{2}{3} n_f$, 
$P_{gg}(z) = 2 C_A \left[ \frac{z}{(1-z)_+} + \frac{1-z}{z} + z (1-z) +
\frac{\beta_0}{12} \delta (1-z) \right]$ is the gluon splitting
function~\cite{Gribov:1972ri, Lipatov:1974qm, Dokshitzer:1977sg,
Altarelli:1977zs}. 
The plus distribution is defined by 
\begin{equation}
\int_0^1 dz f(z) [g(z)]_+ = 
\int_0^1 dz \left[ f(z)-f(1) \right] g(z). 
\end{equation}
The $J$-dependent constants $Q_J$ and functions $P_J(z)$ read 
\begin{subequations}
\begin{eqnarray}
Q_0 &=& \frac{1}{4}, \quad 
Q_1 = \frac{3}{8}, \quad
Q_2 = \frac{7}{8}, \\
P_0 (z) &=& \frac{z (85-26 z)}{8} + \frac{9 (5-3 z)}{4} \log (1-z), \\
P_1 (z) &=& - \frac{3 z (1+4 z)}{4}, \\
P_2 (z) &=& \frac{5 z (11-4 z)}{4} + 9 (2-z) \log (1-z). 
\end{eqnarray}
\end{subequations}
Note that while $d_{g \to {}^3P_J^{[1]}} (z)$ begin at order $\alpha_s^2$,
$d_{g \to {}^3S_1^{[8]}} (z)$ begins at order $\alpha_s$, and so the
above expression contains corrections at NLO in $\alpha_s$. 
The term proportional to $\beta_0 \delta(1-z) \log \mu_f$ cancels the running of
$\alpha_s$ at lowest order, while the $\mu_f$ dependence coming from the 
$\log \mu_f P_{gg} (z)$ term cancels the $\mu_f$ dependence 
in the gluon cross
section $\hat \sigma_g$, so that the quarkonium cross section 
is independent of $\mu_f$. 
The distribution in the last line of eq.~(\ref{eq:FFoctet}) arises 
from soft-gluon emission, which is singular near the $z=1$ threshold. 
Because the gluon cross section $\hat \sigma_g (z)$ steeply rises with $z$, 
this singular term can have a significant effect on the cross section, 
especially at large $p_T$. 
Note that such effect will not appear in $d_{g \to {}^3P_J^{[1]}} (z)$ until
order $\alpha_s^3$; even if we were to compute the cross section with
fragmentation functions at order-$\alpha_s^3$ accuracy, 
$d_{g \to {}^3S_1^{[8]}} (z)$ will then have more singular corrections at 
order $\alpha_s^3$ coming from multiple soft gluon emissions near threshold, 
and the corresponding correction in $d_{g \to {}^3P_J^{[1]}} (z)$ will 
only appear at order $\alpha_s^4$. 

While corrections from soft gluon emissions near threshold can be
resummed~\cite{Kidonakis:1997gm},
it is also important that they are treated consistently throughout the
color-singlet and color-octet channels. 
This is especially the case in $\chi_{QJ}$ production at large $p_T$, where the
contribution from the color-singlet channel can turn negative at values of
$p_T$ much larger than the quarkonium mass, so that the cross section is given
by the remnant of the cancellation between the color-singlet and color-octet
channel contributions. Because of this cancellation, the cross section at large
$p_T$ can become very sensitive to soft gluon emission effects near threshold, 
so that mismatch between the treatment of soft gluon emissions between 
color-singlet and color-octet channels can spoil the reliability of the
perturbative expansion. 

As we have argued in the previous section, the formulation of the cross section
in the shape function formalism of the form given in eq.~(\ref{eq:shapefac_f1})
can be used to treat the soft-gluon emission effects consistently between the
color-singlet and color-octet channels. We will see explicitly by computing the
fragmentation function in the shape function formalism that the singular parts
of the fragmentation functions can be encoded entirely in terms of the 
color-octet short-distance coefficient. 
Let us write the fragmentation function in the shape function formalism in the
form given in eq.~(\ref{eq:shapefac_f1}) as 
\begin{eqnarray}
\label{eq:shapeFF}
D_{g \to \chi_{cJ} + X} (z) \Big|_{\rm shape}
&=&
(2 J+1) \bigg[
d^s_{g \to {}^3P_J^{[1]}} (z)
\langle {\cal O}^{\chi_{Q0}} (^3P_0^{[1]}) \rangle
\nonumber \\ && \hspace{10ex} 
+
\int_0^\infty dl_+ d_{g \to {}^3S_1^{[8]}} (z+z l_+/P_+)
{\cal S}^{\chi_{Q0}}_{^3S_1^{[8]}} (l_+) 
\bigg],
\end{eqnarray}
where $d^s_{g \to {}^3P_J^{[1]}} (z)$ is given by 
\begin{eqnarray}
d^s_{g \to {}^3P_J^{[1]}} (z)
&=& d_{^3P_J^{[1]}} (z) - \frac{4 \alpha_s C_F}{3 N_c \pi m^2}
\int_0^\infty dl_+^* \bigg(
d_{g \to {}^3S_1^{[8]}} (z + z l^*_+/P^*_+)
\nonumber \\ && \hspace{30ex}
-d_{g \to {}^3S_1^{[8]}} (z) \theta(l_+^{*\rm max}- l^*_+) \bigg)
\frac{C_\epsilon}{(l_+^*)^{1+2 \epsilon}}
,
\end{eqnarray}
which we obtain from eq.~(\ref{eq:singletmatchcond_end}). 
We used the boost invariance of $l_+/P_+$ to write the integral in terms of 
$l_+^*$ in the quarkonium rest frame. 
We need this expression at order-$\alpha_s^2$ accuracy, which comes from the
order-$\alpha_s$ contribution in $d_{^3S_1^{[8]}} (z)$. 
Explicit calculation of the subtraction term is carried out as follows:
\begin{eqnarray}
&& \hspace{-10ex} \frac{4 \alpha_s C_F}{3 N_c \pi m^2} 
\int_0^\infty dl_+^* \bigg( 
d_{g \to {}^3S_1^{[8]}} (z + z l^*_+/P^*_+)  
-d_{g \to {}^3S_1^{[8]}} (z) \theta(l_+^{* \rm max}- l^*_+) \bigg) 
\frac{C_\epsilon}{(l^*_+)^{1+2 \epsilon}} 
\nonumber \\
&=& 
\frac{4 \alpha_s C_F}{3 N_c \pi m^2} 
\times \frac{\pi \alpha_s}{24 m^3} 
\left( \frac{C_\epsilon z^{2 \epsilon}}{(2 m)^{2 \epsilon} (1-z)^{1+2 \epsilon}}
+ \frac{1}{2 \epsilon_{\rm IR}} \delta (1-z) \right)
+ O(\alpha_s^3)
\nonumber \\
&=& 
\frac{2 \alpha_s^2}{27 N_c m^5} 
\left[ \frac{1}{(1-z)_+} 
- \delta(1-z) \left( \log \frac{\Lambda}{2 m} -\frac{1}{6} \right) 
+ O(\epsilon)
\right]
+ O(\alpha_s^3), 
\end{eqnarray}
where in the last line we used $C_F = 4/3$ for $N_c = 3$ and 
\begin{equation}
\int_0^1 dz \frac{f(z)}{(1-z)^{1+ 2\epsilon}} 
=
\int_0^1 dz f(z) \left[ -\frac{\delta(1-z)}{2 \epsilon_{\rm IR}} 
+ \frac{1}{(1-z)_+} 
+ O(\epsilon)\right] , 
\end{equation}
for any function $f(z)$. From this we obtain 
\begin{equation}
\label{eq:shapesingletFF}
d^s_{^3P_J^{[1]}} (z)
=
\frac{2 \alpha_s^2}{27 N_c m_c^5} \left\{
\left( \frac{Q_J}{2J+1} - \frac{1}{6} \right) \delta(1-z)
- 1 + \frac{P_J(z)}{2 J+1} \right\} + O(\alpha_s^3).
\end{equation}
We see that now the singular plus function contribution in $d_{g \to
^3P_J^{[1]}} (z)$ has disappeared, along with the dependence on the NRQCD
factorization scale $\Lambda$. As we have shown in the previous section in
eq.~(\ref{eq:sub_NP}), this contribution reappears in the color-octet 
shape function term in eq.~(\ref{eq:shapeFF}). 
If we rewrite eq.~(\ref{eq:shapeFF}) using eq.~(\ref{eq:sub_NP}) as 
\begin{eqnarray}
\label{eq:shapeFF2}
D_{g \to \chi_{cJ} + X} (z) \Big|_{\rm shape}
&=&
(2 J+1) \bigg[
d^s_{g \to {}^3P_J^{[1]}} (z)
\langle {\cal O}^{\chi_{Q0}} (^3P_0^{[1]}) \rangle
+d_{g \to {}^3S_1^{[8]}} (z)
\langle {\cal O}^{\chi_{Q0}} (^3S_1^{[8]}) \rangle
\nonumber \\ && \hspace{8ex}
+ \int_0^\infty dl_+ \bigg( d_{g \to {}^3S_1^{[8]}} (z+z l_+/P_+)
\nonumber \\ && \hspace{20ex}
-d_{g \to {}^3S_1^{[8]}} (z) \theta(l_+^{\rm max}- l_+) 
\bigg)
{\cal S}^{\chi_{Q0}}_{^3S_1^{[8]}} (l_+)
\bigg],
\end{eqnarray}
we can now use the order-$\alpha_s^2$ expression for $d_{g \to {}^3S_1^{[8]}}$
everywhere, so that the soft-gluon emission effects are consistently taken
into account. Note that if we are only interested in the reorganization of the 
perturbative corrections arising from soft-gluon emission effects, 
we can replace the color-octet shape function in eq.~(\ref{eq:shapeFF2}) by its
asymptotic form ${\cal S}^{\chi_{Q0}}_{^3S_1^{[8]}} (l_+) \big|_{\rm asy}
= \frac{4 \alpha_s C_F}{3 N_c \pi m^2 l_+}
\langle {\cal O}^{\chi_{Q0}} (^3P_0^{[1]}) \rangle$, 
because the integrand factor in the parenthesis vanishes as $l_+ \to 0$,
rendering the integral over $l_+$ IR finite. 
In this way, the contribution proportional to 
$\langle {\cal O}^{\chi_{Q0}} (^3P_0^{[1]}) \rangle$ given by 
\begin{eqnarray}
\label{eq:shapeFF_singlet}
\hspace{-10ex} 
D_{g \to \chi_{cJ} + X} (z) \Big|_{\textrm{singlet, shape}}
&=&
(2 J+1) 
\langle {\cal O}^{\chi_{Q0}} (^3P_0^{[1]}) \rangle
\bigg[
d^s_{g \to {}^3P_J^{[1]}} (z)
\nonumber \\ && \hspace{-20ex} 
+ \frac{4 \alpha_s C_F}{3 N_c \pi m^2} 
 \int_0^\infty \frac{dl_+}{l_+} \bigg( d_{g \to {}^3S_1^{[8]}} (z+z l_+/P_+)
-d_{g \to {}^3S_1^{[8]}} (z) \theta(l_+^{\rm max}- l_+)
\bigg) \bigg],
\end{eqnarray}
where $d^s_{g \to {}^3P_J^{[1]}}$ is given by eq.~(\ref{eq:shapesingletFF}) 
and $d_{g \to {}^3S_1^{[8]}}$ is computed to order-$\alpha_s^2$ accuracy, 
can be interpreted as the color-singlet contribution to the fragmentation
function augmented with order-$\alpha_s^3$ soft-gluon effects. 
Here, we have now set $\epsilon = 0$, because the integral over $l_+$ is
IR and UV finite. 

To quantify the effect of the inclusion of the order-$\alpha_s^3$ soft-gluon
effects to the color-singlet contribution, we compute the color-singlet
contributions to the cross sections in NRQCD and in the shape function
formalisms defined by 
\begin{subequations}
\begin{eqnarray}
\sigma[\chi_{QJ}]_{\textrm{singlet, NRQCD}}
&=& \int_0^1 dz \, \hat{\sigma}_g (z) 
D_{g \to \chi_{cJ} + X} (z) \Big|_{\textrm{singlet, NRQCD}}, 
\\
\sigma[\chi_{QJ}]_{\textrm{singlet, shape}}
&=& \int_0^1 dz \, \hat{\sigma}_g (z) 
D_{g \to \chi_{cJ} + X} (z) \Big|_{\textrm{singlet, shape}}, 
\end{eqnarray}
\end{subequations}
where 
$D_{g \to \chi_{cJ} + X} (z)|_{\textrm{singlet, NRQCD}} = (2 J+1) 
\langle {\cal O}^{\chi_{Q0}} (^3P_0^{[1]}) \rangle d_{g \to ^3P_J^{[1]}} (z)$
for NRQCD 
and $D_{g \to \chi_{cJ} + X} (z) |_{\textrm{singlet, shape}}$ is given
in eq.~(\ref{eq:shapeFF2}) for the shape function formalism, 
and compare them with $\sigma[\chi_{QJ}]_{\textrm{octet}}$ defined by 
\begin{eqnarray}
\sigma[\chi_{QJ}]_{\textrm{octet}} 
&=& 
(2 J+1) \langle {\cal O}^{\chi_{Q0}} (^3S_1^{[8]}) \rangle
\nonumber \\ && \times
\int_0^1 dz \, \bigg[ \hat{\sigma}_g (z) d_{g \to ^3S_1^{[8]}} (z)
+ 
\big( \hat{\sigma}_q (z)+\hat{\sigma}_{\bar{q}} (z) \big)
d_{q \to ^3S_1^{[8]}} (z) \bigg], 
\end{eqnarray}
as functions of $p_T$. 
Note that at this accuracy there is no quark fragmentation contribution to the
color-singlet channel, because as we previously mentioned 
$d_{q \to {}^3P_J^{[1]}}(z)$ begin at order $\alpha_s^3$. 
In order to facilitate comparisons independently of the 
determinations of NRQCD matrix elements, we define the dimensionless ratios
\begin{equation}
R^J
=
\frac{m^2 \sigma[\chi_{QJ}]_{\textrm{singlet}}/
\langle {\cal O}^{\chi_{Q0}} (^3P_0^{[1]}) \rangle}
{\sigma[\chi_{QJ}]_{\textrm{octet}}/
\langle {\cal O}^{\chi_{Q0}} (^3S_1^{[8]}) \rangle}. 
\end{equation}
We compute the ratios $R^J$ for 
$\chi_{c1}$ and $\chi_{c2}$ production as functions of $p_T$ 
at central rapidity 
from proton-proton collisions at $\sqrt{s}=13$~TeV. 
We set the charm quark masses to be $m = 1.5$~GeV and choose the
renormalization scale for the color-octet matrix element to be 
$\Lambda = 1.5$~GeV, as is usually done in phenomenological studies of
charmonium production. 
We compute the gluon cross section $\hat{\sigma}_g$ at NLO accuracy 
by using the code in ref.~\cite{Aversa:1988vb}. 
In order to make contact with existing NLO calculations of quarkonium 
cross sections in NRQCD, we neglect the order-$\alpha_s^5$ cross term that
comes from the order-$\alpha_s^3$ gluon cross section $\hat{\sigma}_g$ and 
the order-$\alpha_s^2$ color-octet fragmentation function. 
That is, we compute the products $\hat{\sigma}_g (z) 
d_{g \to {}^3S_1^{[8]}}(z)$ 
in the integrands of $\sigma[\chi_{QJ}]_{\textrm{singlet, shape}}$ and 
$\sigma[\chi_{QJ}]_{\textrm{octet}}$ as 
\begin{eqnarray}
\hat{\sigma}_g (z) d_{g \to ^3S_1^{[8]}} (z)
&=& 
\alpha_s^3 
\hat{\sigma}_g^{(2)} (z) d^{(1)}_{g \to ^3S_1^{[8]}} (z)
\nonumber \\ && 
+ \alpha_s^4 
\left( 
\hat{\sigma}_g^{(2)} (z) d^{(2)}_{g \to ^3S_1^{[8]}} (z)
+ 
\hat{\sigma}_g^{(3)} (z) d^{(1)}_{g \to ^3S_1^{[8]}} (z) \right) 
+ O(\alpha_s^5), 
\end{eqnarray}
where $\hat{\sigma}_g = \alpha_s^2 \hat{\sigma}^{(2)}_g + \alpha_s^3 
\hat{\sigma}^{(3)}_g + O(\alpha_s^4)$ 
and $d_{g \to {}^3S_1^{[8]}} (z)
= \alpha_s d^{(1)}_{g \to {}^3S_1^{[8]}} (z) +
\alpha_s^2 d^{(2)}_{g \to {}^3S_1^{[8]}} (z) + O(\alpha_s^3)$. 
Similarly, in the products 
$\hat{\sigma}_g (z) d_{g \to {}^3P_J^{[1]}}(z)$ and 
$\hat{\sigma}_g (z) d^s_{g \to {}^3P_J^{[1]}}(z)$, 
we can neglect the order-$\alpha_s^3$ piece in $\hat{\sigma}_g (z)$ because
$d_{g \to {}^3P_J^{[1]}}(z)$ and $d^s_{g \to {}^3P_J^{[1]}}(z)$ 
begin at order $\alpha_s^2$. In the same way, $\hat{\sigma}_q$ can be computed
to order-$\alpha_s^2$ accuracy because $d_{q \to {}^3S_1^{[8]}}$ begins at
order $\alpha_s^2$. 
We compute $\hat{\sigma}_g$ at NLO in $\alpha_s$ with 
CTEQ6M parton distribution functions~\cite{Pumplin:2002vw} 
at scale $\mu_f = m_T = \sqrt{p_T^2 + 4 m^2}$, and compute $\alpha_s$ at the
same scale using the two-loop formula provided in ref.~\cite{Pumplin:2002vw}
with $n_f = 5$ light quark flavors and $\Lambda_{\rm QCD}^{(5)} = 226$~MeV. 
This calculation of the short-distance coefficients at leading
power has been shown to agree well with the fixed-order calculation at large
$p_T$, and the fragmentation contributions completely dominate the cross
section for $p_T$ above 100~GeV~\cite{Bodwin:2014gia, Bodwin:2015iua}. 

\begin{figure}[tbp]
\centering
\includegraphics[width=.99\textwidth]{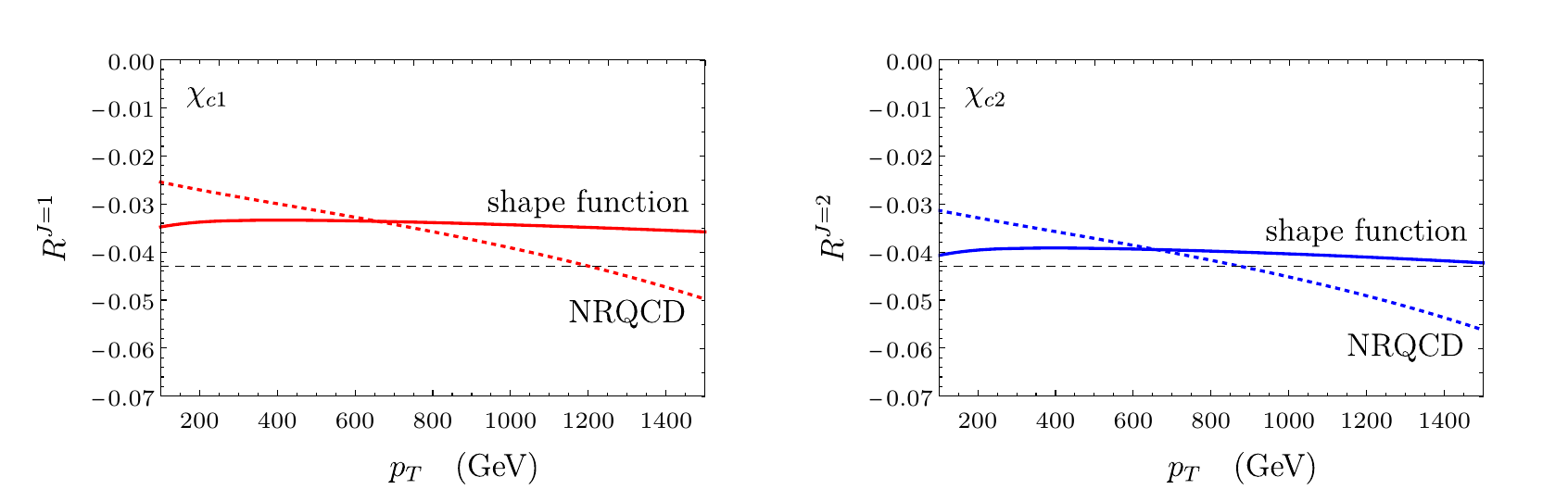}
\caption{\label{fig:figratio}
Dimensionless singlet to octet ratios $R^J$ for $J=1$ and $2$ 
from the shape function (solid
lines) and NRQCD factorization (dotted lines) formalisms for production of
$\chi_{c1}$ (left) and $\chi_{c2}$ (right) at central rapidity 
from $pp$ collisions at $\sqrt{s}=13$~TeV. 
The $\chi_{cJ}$ cross sections turn negative if
the $R^J$ go below the matrix element ratio 
$- m^2 \langle {\cal O}^{\chi_{c0}} (^3S_1^{[8]}) \rangle/ 
\langle {\cal O}^{\chi_{c0}} (^3P_0^{[1]}) \rangle$, whose best-fit value 
$-0.043$ is shown as dashed black lines. 
}
\end{figure}

The results for $R^{J=1}$ and $R^{J=2}$ from NRQCD and the shape function
formalism are shown in fig.~\ref{fig:figratio}. 
We concentrate on the $J=1$ and $J=2$ cases, because $\chi_{c0}$ is usually not
measured in hadroproduction due to the tiny branching fraction for 
$\chi_{c0} \to J/\psi+\gamma$. 
Note that these ratios take negative
values at large $p_T$, because the color-singlet cross section involves a 
large negative contribution coming from the singular plus function term 
$z/(1-z)_+$ in $d_{g \to {}^3P_J^{[1]}}(z)$ 
associated with singlet-octet mixing induced by soft-gluon emission. 
To ensure the positivity of the $\chi_{cJ}$ cross sections, 
the ratios must be greater than the ratio of matrix elements 
given by $- m^2 \langle {\cal O}^{\chi_{c0}} (^3S_1^{[8]}) \rangle/ 
\langle {\cal O}^{\chi_{c0}} (^3P_0^{[1]}) \rangle$. 
The best-fit value for 
$m^2 \langle {\cal O}^{\chi_{c0}} (^3S_1^{[8]}) \rangle/ 
\langle {\cal O}^{\chi_{c0}} (^3P_0^{[1]}) \rangle$ at $\Lambda = 1.5$~GeV
determined from cross section measurements at hadron colliders 
is $0.043$ at NLO accuracy~\cite{Ma:2010vd, Brambilla:2021abf}. 
We see that in the case of NRQCD, the ratios 
$R^J$ become more negative as $p_T$ increases and eventually 
make the cross sections turn negative, as they go below $-0.043$ when $p_T$
exceeds about 1.2~TeV for $J=1$ and 0.9~TeV for $J=2$. 
This problem does not occur until much larger $p_T$ 
in the shape function calculation, 
as the ratios $R^J$ are almost constant over a wide range of $p_T$, 
and the positivity of the cross sections is ensured as the values of 
$R^J$ stay safely above $-0.043$ for much larger values of $p_T$. 
As we have argued, this happens because the shape function formalism allows us
to match the accuracy in $\alpha_s$ of the soft-gluon emission effects between
the color-singlet and color-octet channels, so that the cancellation between
the two channels can occur consistently without mismatch. 

We note that although the shape function formalism eliminates the mismatch of
soft-gluon emission effects between channels that mix under renormalization, 
this does not necessarily alleviate the need for resummation of threshold
logarithms associated with soft-gluon emission, as current results are 
still based on fixed order-calculations. Even when carrying out the threshold
resummation, the shape function formalism in the form of 
eq.~(\ref{eq:shapefac_f1}) employed in this section will allow us to consider 
the resummation effects in the color-octet and color-singlet channels in a
consistent manner, and ensure that there is no mismatch of soft-emission
effects.

\subsection{\boldmath Nonperturbative corrections from shape functions}
\label{sec:NPmodel}

We now focus on the effect of nonperturbative corrections arising from the
nonperturbative shape function that are neglected in usual applications of 
NRQCD factorization.
That is, we make use of eq.~(\ref{eq:shapefac_f2}) to compute the
additional nonperturbative contribution coming from the 
deviation of the nonperturbative shape function from its asymptotic form. 
Because first-principles determination of the color-octet shape functions 
have not yet been done, this requires the use of model functions 
for the nonperturbative shape function. 
However, as we have argued previously, 
the normalization condition (\ref{eq:normalization2}) and 
the requirement that the nonperturbative
shape function must coincide with the asymptotic form at large $l_+$ 
severely constrain the model dependence. 
Additionally, the pNRQCD results for the shape functions imply that the 
$l_+$ dependence of the shape functions must be same for any
$P$-wave heavy quarkonia, independently of radial excitation or heavy quark
flavor; this lets us use a single model function for all shape functions for 
production of any $P$-wave heavy quarkonium state. 

We first attempt a crude estimate of the size of the nonperturbative
corrections at large transverse momentum $p_T$. We note that the factor 
${\cal S}_{^3S_1^{[8]}}^{\chi_{Q0}}(l_+) \big|_{\rm NP}-
{\cal S}_{^3S_1^{[8]}}^{\chi_{Q0}}(l_+) \big|_{\rm asy}$ 
in eq.~(\ref{eq:shapefac_f2}) 
diminishes as $l_+$ increases, as the nonperturbative shape function must
coincide with its asymptotic form when $l_+$ is in the perturbative regime. 
In the opposite case, when $l_+$ approaches zero, 
${\cal S}_{^3S_1^{[8]}}^{\chi_{Q0}}(l_+) \big|_{\rm asy}$ diverges like 
$1/l_+$, while the nonperturbative shape function must either be finite or at
least diverge slower than $1/l_+$ in order to ensure the IR finiteness
of its normalization. This implies that at very small values of $l_+$, 
we can approximate 
${\cal S}_{^3S_1^{[8]}}^{\chi_{Q0}}(l_+) \big|_{\rm NP}- 
{\cal S}_{^3S_1^{[8]}}^{\chi_{Q0}}(l_+) \big|_{\rm asy}$ by 
$-{\cal S}_{^3S_1^{[8]}}^{\chi_{Q0}}(l_+) \big|_{\rm asy}$.
Hence, the additional nonperturbative contribution in
eq.~(\ref{eq:shapefac_f2}) can be estimated as 
\begin{eqnarray}
\sigma[\chi_{QJ}(P)]_{\rm NP} 
&=& (2 J+1)\int_0^\infty dl_+
\left( c_{^3S_1^{[8]}} (P+l) - c_{^3S_1^{[8]}} (P) \theta(l_+^{\rm max}-l_+)
\right)
\nonumber \\ && \hspace{20ex} \times
\Big( {\cal S}_{^3S_1^{[8]}}^{\chi_{Q0}}(l_+) \Big|_{\rm NP}-
{\cal S}_{^3S_1^{[8]}}^{\chi_{Q0}}(l_+) \Big|_{\rm asy} \Big)
\nonumber \\
\hspace{5ex} 
&\approx& - \int_0^{l_+^{\rm NP}} dl_+
\left( c_{^3S_1^{[8]}} (P+l) - c_{^3S_1^{[8]}} (P) \right)
{\cal S}_{^3S_1^{[8]}}^{\chi_{Q0}}(l_+) \Big|_{\rm asy}, 
\end{eqnarray}
where $l_+^{\rm NP}$ is chosen so that 
${\cal S}_{^3S_1^{[8]}}^{\chi_{Q0}}(l_+) \big|_{\rm NP}-
{\cal S}_{^3S_1^{[8]}}^{\chi_{Q0}}(l_+) \big|_{\rm asy}
\approx -{\cal S}_{^3S_1^{[8]}}^{\chi_{Q0}}(l_+) \big|_{\rm asy}$
for $l_+ \lesssim l_+^{\rm NP}$. 
Note that $\sigma[\chi_{QJ}(P)]_{\rm NP}$ is positive in this case,
because usually $c_{^3S_1^{[8]}} (P+l) < c_{^3S_1^{[8]}} (P)$ as the extra
momentum $l$ reduces the available phase space. 
In the case of inclusive production at large $p_T$, the $p_T$-differential 
color-octet cross sections fall off like $1/p_T^n$, with $n \approx 4$. 
This, and the fact that at large $p_T$ the color-octet $Q \bar Q$ is produced
predominantly near threshold through gluon fragmentation leads to the following
approximation for the $l_+$ dependence of the short-distance coefficient:
\begin{equation}
c_{^3S_1^{[8]}} (P+l) \approx 
c_{^3S_1^{[8]}} (P) 
\left( \frac{P_+}{P_+ + l_+} \right)^n
\approx 
c_{^3S_1^{[8]}} (P) 
\left( 1- \frac{n l_+}{P_+} \right), 
\end{equation}
where in the last equality we expanded in powers of $l_+$ to linear order
compared to $P_+$, because $l_+^{\rm NP}$ is much smaller than $P_+$. 
By using this result, we can make the following estimate 
\begin{eqnarray}
\sigma[\chi_{QJ}(P)]_{\rm NP} 
&\approx& 
(2 J+1)c_{^3S_1^{[8]}} (P)
\int_0^{l_+^{\rm NP}} dl_+
\frac{n l_+}{P_+}
\times 
\frac{4 \alpha_s C_F}{3 N_c \pi m^2 l_+} 
\langle {\cal O}^{\chi_{Q0}} (^3P_0^{[1]}) \rangle 
\nonumber \\
&=& 
(2 J+1) c_{^3S_1^{[8]}} (P)
\frac{n l_+^{\rm NP}}{P_+}
\times 
\frac{4 \alpha_s C_F}{3 N_c \pi m^2} 
\langle {\cal O}^{\chi_{Q0}} (^3P_0^{[1]}) \rangle. 
\end{eqnarray}
Note that the boost-invariant ratio $l_+^{\rm NP}/P_+$ must be much smaller 
than $1/2$, because $P_+^* = 2 m$ gives $l_+^{\rm max}/P_+ = e^{-1/6}/2$ when 
$\Lambda = m$. If we set $l_+^{\rm NP}/P_+ = \lambda/2$, with $\lambda \ll 1$,
the additional nonperturbative contribution to the 
cross section from the $P$-wave quarkonium cross section is estimated to be 
\begin{equation}
\label{eq:NPestimate}
\sigma[\chi_{QJ}(P)]_{\rm NP} 
\approx 
(2 J+1) c_{^3S_1^{[8]}} (P)
\frac{4 \alpha_s C_F}{3 N_c \pi m^2}
\langle {\cal O}^{\chi_{Q0}} (^3P_0^{[1]}) \rangle \times \frac{n \lambda}{2}.
\end{equation}
Compare this with the color-octet contribution to the cross section in the
NRQCD factorization formalism given by 
\begin{equation}
\sigma[\chi_{QJ}(P)]_{\rm octet} = 
(2 J+1) c_{^3S_1^{[8]}} (P)
\frac{{\cal E}_{11}}{3 N_c^2 m^2}
\langle {\cal O}^{\chi_{Q0}} (^3P_0^{[1]}) \rangle, 
\end{equation}
where we used the pNRQCD result for the color-octet matrix element to write the
cross section in terms of the color-singlet matrix element and the universal
quantity ${\cal E}_{11}$. In the case of $\chi_{cJ}$, best-fit values of 
$P$-wave quarkonium matrix elements give ${\cal E}_{11} = 1.17$ at 
scale $\Lambda = 1.5$~GeV~\cite{Ma:2010vd, Brambilla:2021abf}. 
By setting $\alpha_s = 0.25$ at the scale of the charmonium mass, 
$\lambda = 0.5$, and $n=4$, we find that $\sigma[\chi_{QJ}(P)]_{\rm NP}$ 
is of roughly the same order of magnitude as $\sigma[\chi_{QJ}(P)]_{\rm octet}$
in the case of charmonium production. In the case of bottomonium, 
$\sigma[\chi_{QJ}(P)]_{\rm NP}$ is smaller than 
$\sigma[\chi_{QJ}(P)]_{\rm octet}$, because ${\cal E}_{11}$ becomes larger at
the scale of the bottom quark mass. 

Even though the additional nonperturbative effects from nonperturbative 
shape functions estimated in eq.~(\ref{eq:NPestimate}) can be as large as order
one compared to NRQCD calculations of the cross section, this by itself is
inconsequential in large-$p_T$ quarkonium production phenomenology, 
because the additional contributions do not alter the $p_T$ dependence 
of the cross section and only brings in changes in the overall normalization. 
That is, the additional nonperturbative contributions can be absorbed into a
redefinition of the color-octet matrix element. 
Because in phenomenological analyses the color-octet 
matrix elements are obtained from measured cross sections,
and not computed from first principles, this redefinition does not affect the
large-$p_T$ phenomenology in any way. 
It should be noted, however, that this estimate is valid only for the
large-$p_T$ asymptotic behavior of the cross section. 
At smaller values of $p_T$, 
the shapes of the color-octet cross sections change, 
which would invalidate the assumptions that
lead us to the estimate in eq.~(\ref{eq:NPestimate}). Hence, at smaller values
of $p_T$, the additional nonperturbative contributions may change the $p_T$
shapes of the cross section significantly. In order to investigate these
effects quantitatively, we need to compute $\sigma[\chi_{QJ}(P)]_{\rm NP}$
explicitly with model shape functions. 

Before we go on with the computation of the cross section, 
we discuss another nonperturbative effect associated with the heavy quarkonium
mass that can be incorporated in the shape function formalism. 
As have been discussed in ref.~\cite{Beneke:1997qw}, 
the shape function formalism allows matching
the $Q \bar Q$ phase space with the physical one involving the heavy quarkonium
mass. In NRQCD factorization, the invariant mass of the $Q \bar Q$ produced in
the short-distance process can be different from the heavy quarkonium mass by
order $mv$, due to the fact that the $Q \bar Q$ can emit order-$mv$ gluons
before evolving into a quarkonium. 
In the shape function formalism, the kinematics of the
emission of order-$mv$ gluons is explicitly taken into account by the $l_+$
dependences in the shape functions and the corresponding short-distance
coefficients. Hence, as have been suggested in ref.~\cite{Beneke:1997qw}, 
in the calculation of the short-distance coefficients 
$s_N (P+l)$ and $s_N(P)$, the $Q \bar Q$ mass $m_{Q \bar Q} = \sqrt{P^2}$ 
can be chosen to be the heavy quarkonium mass, instead of $m_{Q \bar Q} = 2 m$. 
In principle, in the nonrelativistic power counting the difference between 
$2 m$ and the heavy quarkonium mass is of order $mv^2$. 
Because the shape functions describe the emission of order-$mv$ gluons, 
and not necessarily order-$mv^2$ effects, the order-$mv^2$ effects can be
consistently neglected from the short-distance coefficients by setting 
$m_{Q \bar Q} = 2 m$ in the same way as done in NRQCD. However, because of the 
renormalon ambiguity in the heavy quark pole mass, the difference between the
heavy quarkonium mass and $2 m$ can exceed $mv^2$ and 
become numerically significant depending on the 
choice of the pole mass $m$. Setting $m_{Q \bar Q}$ to be the quarkonium mass 
avoids this issue and ensures that the $Q \bar Q$ phase space matches the
physical one within order $mv^2$. 

We now compute the $\chi_{c}$ and $\chi_{b}$ cross sections in NRQCD and
shape function formalisms. Since we are interested in the nonperturbative
effects from the shape function formalism at values of $p_T$ lower than the
fragmentation regime, we compute the cross sections for the LHCb kinematics 
from 7~TeV $pp$ collisions at the LHC, where data for $P$-wave quarkonium 
production rates relative to $1S$ quarkonium production rates are available 
for $p_T \geq 2$~GeV for $\chi_c$~\cite{LHCb:2012af} and 
$p_T \geq 6$~GeV for $\chi_b$~\cite{LHCb:2014ngh}, 
which go below the quarkonium masses. 
The available $p_T$ ranges are, however, above the heavy quark pole masses 
which we set to be $m=1.5$~GeV for charm and $m=4.75$~GeV for bottom.
The cross sections are computed in NRQCD and shape function formalisms as 
\begin{subequations}
\begin{eqnarray}
\frac{d \sigma[\chi_{QJ}]}{dp_T} \Big|_{\rm NRQCD}
&=& (2 J+1) \bigg(
\frac{d \sigma[Q \bar Q (^3P_J^{[1]})]}{dp_T} 
\langle {\cal O}^{\chi_{Q0}} (^3P_0^{[1]}) \rangle
+ 
\frac{d \sigma[Q \bar Q (^3S_1^{[8]})]}{dp_T} 
\langle {\cal O}^{\chi_{Q0}} (^3S_1^{[8]}) \rangle
\bigg), 
\nonumber \\
\\ 
\frac{d \sigma[\chi_{QJ}]}{dp_T} \Big|_{\rm shape}
&=& (2 J+1) \bigg(
\frac{d \sigma[Q \bar Q (^3P_J^{[1]})]}{dp_T}
\langle {\cal O}^{\chi_{Q0}} (^3P_0^{[1]}) \rangle
+
\frac{d \sigma[Q \bar Q (^3S_1^{[8]})]}{dp_T}
\langle {\cal O}^{\chi_{Q0}} (^3S_1^{[8]}) \rangle
\bigg)
\nonumber \\  && 
+
\frac{d \sigma[\chi_{QJ}]}{d p_T} \bigg|_{\rm NP}, 
\end{eqnarray}
\end{subequations}
where 
\begin{eqnarray}
\label{eq:NPcontribution_def}
\frac{d \sigma[\chi_{QJ}]}{d p_T}\bigg|_{\rm NP} &=& 
(2 J+1) \int_0^\infty dl_+
\left( \frac{d \sigma[Q \bar Q (^3S_1^{[8]})]}{dp_T} \Big|_{P \to P+l}-
\frac{d \sigma[Q \bar Q (^3S_1^{[8]})]}{dp_T} 
\theta (l_+^{\rm max}-l_+) \right)
\nonumber \\ && \hspace{20ex} \times
\Big( {\cal S}_{^3S_1^{[8]}}^{\chi_{Q0}}(l_+) \Big|_{\rm NP}-
{\cal S}_{^3S_1^{[8]}}^{\chi_{Q0}}(l_+) \Big|_{\rm asy} \Big). 
\end{eqnarray}
We compute the NRQCD short-distance coefficients 
at NLO in $\alpha_s$ 
numerically by using the FDCHQHP package~\cite{Wan:2014vka} 
with CTEQ6M parton distribution functions~\cite{Pumplin:2002vw}. 
Note that $d \sigma[\chi_{QJ}]/dp_T |_{\rm shape}$ involve the same
short-distance coefficients as NRQCD, but in this case we compute them with the 
$Q \bar Q$ mass set to be the physical $\chi_{QJ}$ mass, 
while in the NRQCD cross section the $Q \bar Q$ mass is $2 m$, as is usually
done in phenomenological studies. 
We compute the short-distance coefficients that enter the nonperturbative 
correction term $d \sigma[\chi_{QJ}]/d p_T |_{\rm NP}$ 
at leading order in $\alpha_s$ from the known parton cross sections 
$d \hat{\sigma} [ ij \to Q \bar Q (^3S_1^{[8]})+X]/d \hat{t}$ 
where $i,j = g$, $q$, and $\bar q$, and $\hat{t}$ is a partonic Mandelstam
variable. 
For the term involving $\theta (l_+^{\rm max}-l_+)$, 
we can compute the $d \sigma[Q \bar Q (^3S_1^{[8]})]/dp_T$ at tree level as 
\begin{equation}
\label{eq:LOcrosssectionformula}
\frac{d \sigma[Q \bar Q (^3S_1^{[8]})]}{dp_T} 
= \sum_{i,j = g,q,\bar{q}} \int dy dy' \left( 2 p_T x_1 x_2 \right)
f_{i/p} (x_1) f_{j/p} (x_2) 
\frac{d \hat{\sigma} [ ij \to Q \bar Q (^3S_1^{[8]}) +X]}{d \hat{t}} , 
\end{equation}
where $f_{i/p}(x)$ is the parton distribution function for finding a parton 
$i$ with momentum fraction $x$ in a proton, 
$y$ is the rapidity of the $Q \bar Q$,
and $y'$ is the rapidity of the recoil $X$. 
The factor $2 p_T x_1 x_2$ is the Jacobian arising from the change of
integration variables from $x_1$, $x_2$, and $\hat{t}$ to 
$p_T$, $y$, and $y'$. The parton cross sections 
$d \hat{\sigma} [ ij \to Q \bar Q (^3S_1^{[8]})+X]/d \hat{t}$ 
are available as functions of the partonic Mandelstam variables 
$\hat{s}$, $\hat{u}$, and $\hat{t}$, 
which can be written in terms of $s$, $p_T$, $y$, $y'$, 
and $P^2 = m_{Q \bar Q}^2$ as
\begin{subequations}
\begin{eqnarray}
\hat{s} &=& x_1 x_2 s,
\\
\hat{t} &=& m_{Q \bar Q}^2 - x_1 \sqrt{s} m_T e^{-y}, 
\\
\hat{u} &=& m_{Q \bar Q}^2 -\hat{s}-\hat{t}, 
\end{eqnarray}
\end{subequations}
where $m_T = \sqrt{m_{Q \bar Q}^2 + p_T^2}$, 
$x_1 = ( m_T e^y + p_T e^{y'} )/\sqrt{s}$, and
$x_2 = ( m_T e^{-y} + p_T e^{-y'} )/\sqrt{s}$. 
Analytical expressions for the 
tree-level $2 \to 2$ parton cross sections for $gg$, $gq$, and $q \bar q$
initial states can be found in ref.~\cite{Cho:1995vh}.
The calculation of $d \sigma[Q \bar Q (^3S_1^{[8]})]/dp_T \big|_{P \to P+l}$
is carried out in a similar way, except now the partonic Mandelstam variables 
are given by 
\begin{subequations}
\label{eq:Mandelstamshape}
\begin{eqnarray}
\hat{s} &=& x_1 x_2 s,
\\
\hat{t} &=& m_{Q \bar Q}^2 + (m_T-p_T) l_T - x_1 \sqrt{s} (m_T + l_T)e^{-y},
\\
\hat{u} &=& m_{Q \bar Q}^2 + (m_T-p_T) l_T -\hat{s}-\hat{t},
\end{eqnarray}
\end{subequations}
where $x_1 = [ (m_T+l_T) e^{y} + (p_T+l_T) e^{y'}]/\sqrt{s}$, 
$x_2 = [ (m_T+l_T) e^{-y} + (p_T+l_T) e^{-y'}]/\sqrt{s}$, 
and $l_T$ is the transverse component of $l$. 
Note that there is some ambiguity in choosing the direction of the
momentum $l$, because boosts along the beam direction will change the
directions of $\bm{P}$ and $\bm{l}$ differently. 
We choose the directions of $\bm{l}$ and $\bm{P}$ to coincide in the frame 
where $|\bm{P}| = p_T$ and $|\bm{l}| = l_T$. In this frame, 
$l_+ = 2 |\bm{l}| = 2 l_T$, and the boost to this frame from the rest frame of 
the $\chi_{Q}$ is given by $l_+ = l_+^* \times (m_T+p_T)/m_{Q \bar Q}$. 
The short-distance coefficient 
$d \sigma[Q \bar Q (^3S_1^{[8]})]/dp_T \big|_{P \to P+l}$ 
can be computed in a similar form as eq.~(\ref{eq:LOcrosssectionformula}),
where now the $l$-dependent partonic Mandelstam variables are given in 
eqs.~(\ref{eq:Mandelstamshape}), and the Jacobian now takes a more complicated
$l$-dependent form. 

In order to compute $d \sigma[\chi_{QJ}]/d p_T \big |_{\rm NP}$, 
the parton cross sections $d \hat{\sigma}/d \hat{t}$ for the color-octet
channel must be known as functions of the partonic Mandelstam variables. The
analytical expressions for $gg$, $gq$, and $q \bar q$ initial states are
available at tree level (order $\alpha_s^3$) in ref.~\cite{Cho:1995vh}. 
At NLO (order
$\alpha_s^4$), publicly available results are given only as $p_T$-differential
cross sections convolved with parton distribution functions and integrated over
rapidity ranges, so that they cannot be used directly in
eq.~(\ref{eq:NPcontribution_def}) to compute $d \sigma[\chi_{QJ}]/d p_T \big
|_{\rm NP}$. 
Fortunately, in the kinematical ranges that
we are interested in this section, the effect of the NLO correction is small
for the color-octet channel: as can be seen in ref.~\cite{Ma:2010yw}, 
the NLO $K$ factor for the $^3S_1^{[8]}$ channel 
is close to $1$ and is almost constant in $p_T$. 
Hence, we neglect the NLO corrections to 
$d \sigma[Q \bar Q (^3S_1^{[8]})]/dp_T$ when we compute the nonperturbative 
corrections $d \sigma[\chi_{QJ}]/d p_T \big |_{\rm NP}$, while we compute the
short-distance coefficients at NLO accuracy everywhere else. 

\begin{figure}[tbp]
\centering
\includegraphics[width=.55\textwidth]{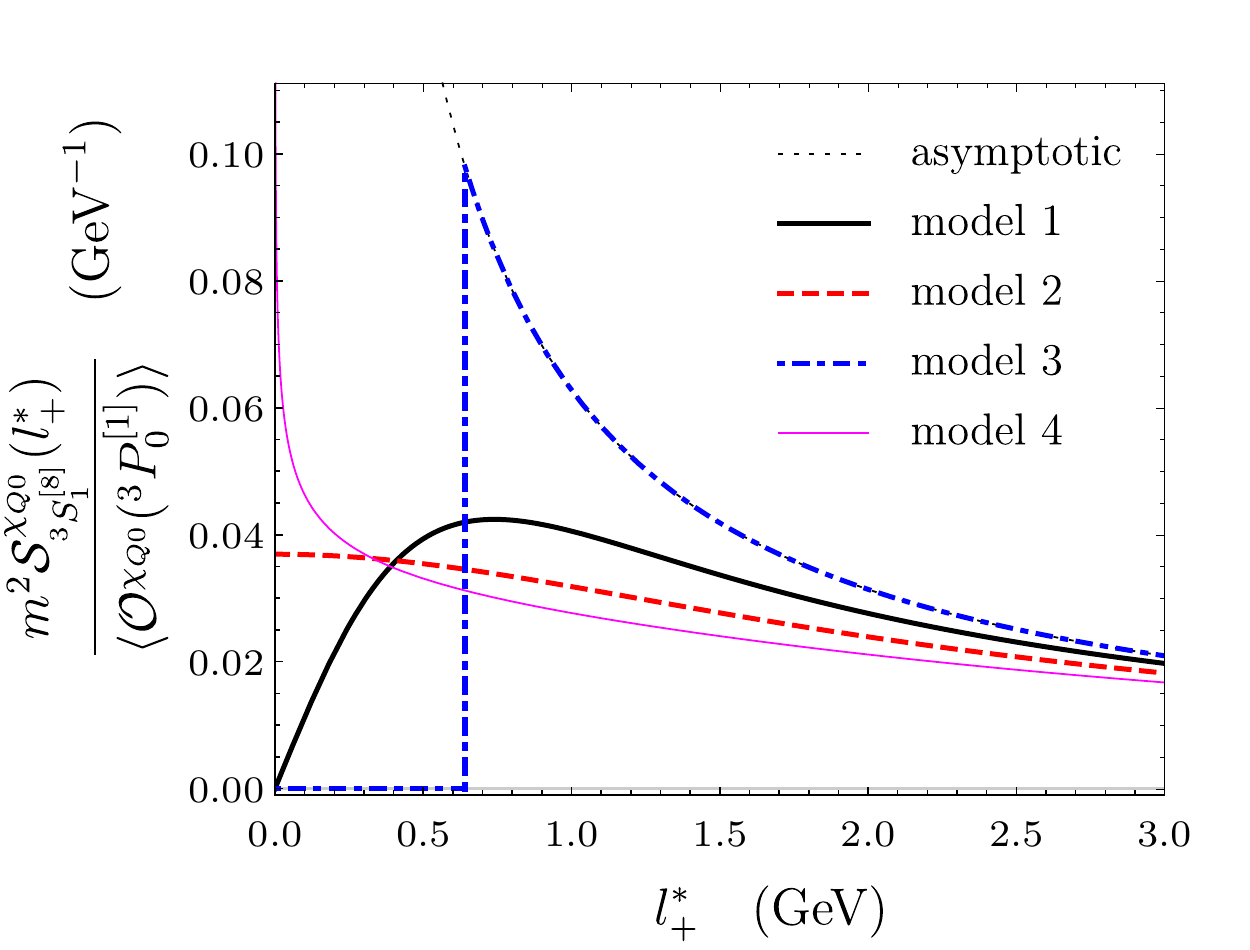}
\caption{\label{fig:models}
Shapes in $l_+^*$ of the model functions for the color-octet shape function 
${\cal S}_{^3S_1^{[8]}}^{\chi_{Q0}}(l^*_+)$ divided by 
$\langle {\cal O}^{\chi_{Q0}} (^3P_0^{[1]})\rangle/m^2$ for model 1 (black
solid line), model 2 (red dashed line), model 3 (blue dot-dashed line),
and model 4 (thin magenta line), compared to the asymptotic form (thin dotted
line). The model parameter for each model is determined from the normalization 
eq.~(\ref{eq:normalization2}) and the NLO best-fit value 
$m^2 \langle {\cal O}^{\chi_{c0}} (^3S_1^{[8]}) \rangle/
\langle {\cal O}^{\chi_{c0}} (^3P_0^{[1]}) \rangle = 0.043$ at 
$\Lambda = 1.5$~GeV. 
}
\end{figure}

Now we need a model for the nonperturbative shape function. 
As we have stated earlier, the nonperturbative shape function 
${\cal S}_{^3S_1^{[8]}}^{\chi_{Q0}}(l_+) \big|_{\rm NP}$
must take the asymptotic form 
${\cal S}_{^3S_1^{[8]}}^{\chi_{Q0}}(l_+) \big|_{\rm asy}$
for $l_+ \gtrsim l_+^{\rm max}$, 
and normalized by 
$\int_0^{l_+^{\rm max}} dl_+ 
{\cal S}_{^3S_1^{[8]}}^{\chi_{Q0}}(l_+) \big|_{\rm NP}
= \langle {\cal O}^{\chi_{Q0}} (^3S_1^{[8]}) \rangle^{\overline{\rm MS}}$. 
For the normalization integral to be IR finite, the nonperturbative shape
function must either be finite, or at least diverge slower than $1/l_+$
as $l_+ \to 0$. 
We consider the following model functions 
\begin{subequations}
\begin{eqnarray}
{\cal S}_{^3S_1^{[8]}}^{\chi_{Q0}}(l_+) \big|_{\textrm{model 1}}
&=& 
\frac{4 \alpha_s C_F}{3 N_c \pi m^2} 
\langle {\cal O}^{\chi_{Q0}} (^3P_0^{[1]}) \rangle
\frac{l_+}{l_+^2 +a_+^2}, 
\\
{\cal S}_{^3S_1^{[8]}}^{\chi_{Q0}}(l_+) \big|_{\textrm{model 2}}
&=& 
\frac{4 \alpha_s C_F}{3 N_c \pi m^2} 
\langle {\cal O}^{\chi_{Q0}} (^3P_0^{[1]}) \rangle
\frac{1}{(l_+^2 +a_+^2)^{1/2}}, 
\\
{\cal S}_{^3S_1^{[8]}}^{\chi_{Q0}}(l_+) \big|_{\textrm{model 3}}
&=& 
\frac{4 \alpha_s C_F}{3 N_c \pi m^2} 
\langle {\cal O}^{\chi_{Q0}} (^3P_0^{[1]}) \rangle
\frac{1}{l_+} \theta(a_+-l_+), 
\\
{\cal S}_{^3S_1^{[8]}}^{\chi_{Q0}}(l_+) \big|_{\textrm{model 4}}
&=& 
\frac{4 \alpha_s C_F}{3 N_c \pi m^2} 
\langle {\cal O}^{\chi_{Q0}} (^3P_0^{[1]}) \rangle
\frac{1}{l_+ [(a_+/l_+)^2+1]^{\alpha/2}}, 
\quad 0<\alpha<1. 
\end{eqnarray}
\end{subequations}
In model 1, the model shape function vanishes linearly as $l_+ \to 0$, 
while in model 2 the model function becomes a finite constant. 
The model 3 shape function coincides the asymptotic form for $l_+ > a_+$, 
while it vanishes for $l_+ < a_+$. 
In model 4, the model function diverges as $l_+ \to 0$ like 
$1/l_+^{1-\alpha}$, more slowly than the asymptotic form, to ensure the IR
finiteness of the normalization integral. We use the value $\alpha=0.8$ so that 
the model function slowly diverges at $l_+=0$, and recovers the asymptotic form
at large $l_+$. 
The model parameters $a_+$, which transforms under boosts like $l_+$, can be
determined from the normalization condition. 
By using the value $\alpha_s = 0.33$ at scale $1.5$~GeV, 
and the best-fit value for $\chi_{cJ}$ matrix elements giving 
$m^2 \langle {\cal O}^{\chi_{c0}} (^3S_1^{[8]}) \rangle/ 
\langle {\cal O}^{\chi_{c0}} (^3P_0^{[1]}) \rangle = 0.043$ at 
$\Lambda = 1.5$~GeV~\cite{Ma:2010vd, Brambilla:2021abf}, 
we obtain the model parameters at the rest frame of the
$\chi_{Q0}$ given by $a_+^* = 0.74$, 1.7, 0.64, and 2.6 in units of GeV for 
models 1, 2, 3, and 4, respectively. 
The shapes in $l_+^*$ of theses model functions for
fixed values of the model parameters are shown in fig.~\ref{fig:models} 
compared to the asymptotic form. 

In order to investigate the model dependence of the cross section in the shape
function formalism, we compute $d \sigma[\chi_{QJ}]/d p_T \big |_{\rm NP}$
from the four model shape functions. In order to assess the impact of the
nonperturbative corrections, we compute the dimensionless ratios of the 
color-octet channel contributions in the NRQCD and shape function formalisms 
\begin{equation}
r^Q_{\rm octet} = 
\frac{\displaystyle 
\frac{d \sigma[\chi_{QJ}]}{dp_T} \Big|_{\textrm{octet, shape}}
}{\displaystyle 
\frac{d \sigma[\chi_{QJ}]}{dp_T} \Big|_{\textrm{octet, NRQCD}} },
\end{equation}
where the numerator and denominator are defined by 
\begin{subequations}
\begin{eqnarray}
\frac{d \sigma[\chi_{QJ}]}{dp_T} \Big|_{\textrm{octet, shape}}
&=& (2 J+1) 
\frac{d \sigma[Q \bar Q (^3S_1^{[8]})]}{dp_T}
\langle {\cal O}^{\chi_{Q0}} (^3S_1^{[8]}) \rangle
+
\frac{d \sigma[\chi_{QJ}]}{d p_T} \bigg|_{\rm NP}.
\\
\frac{d \sigma[\chi_{QJ}]}{dp_T} \Big|_{\textrm{octet, NRQCD}}
&=& (2 J+1) 
\frac{d \sigma[Q \bar Q (^3S_1^{[8]})]}{dp_T}
\langle {\cal O}^{\chi_{Q0}} (^3S_1^{[8]}) \rangle,
\end{eqnarray}
\end{subequations}
Since we are interested in the behavior of $d \sigma[\chi_{QJ}]/dp_T$ compared
to the color-octet channel contribution in the usual NRQCD factorization
formalism, we compute the short-distance coefficients 
$d \sigma[Q \bar Q (^3S_1^{[8]})/dp_T$ at tree level (order $\alpha_s^3$) for
the calculation of $r^Q_{\rm octet}$ (we will use the full NLO results in the
calculation of cross sections for comparison with measurements). 
As we have stated earlier, we set the $Q \bar Q$ mass to be $2 m$ for the cross
section in NRQCD, while we set it to be the quarkonium mass in the shape
function formalism. The ratios depend on the ratio of NRQCD matrix elements 
$m^2 \langle {\cal O}^{\chi_{Q0}} (^3S_1^{[8]}) \rangle/
\langle {\cal O}^{\chi_{Q0}} (^3P_0^{[1]}) \rangle$. For charmonium we set it
to be the best-fit value 0.043 at $\Lambda = 1.5$~GeV. 
For bottomonium, we compute the ratio for $\chi_{b0}$ at scale $4.75$~GeV 
by using the universality of the gluonic quantity ${\cal E}_{11}$ and 
its evolution equation~\cite{Brambilla:2021abf}. That is, 
\begin{eqnarray}
\label{eq:octetevolve}
\frac{m^2 \langle {\cal O}^{\chi_{b0}} (^3S_1^{[8]}) \rangle
(\Lambda = 4.75\textrm{~GeV}) 
}{
\langle {\cal O}^{\chi_{b0}} (^3P_0^{[1]}) \rangle}
&=& 
\frac{m^2 \langle {\cal O}^{\chi_{c0}} (^3S_1^{[8]}) \rangle
(\Lambda = 1.5\textrm{~GeV}) 
}{
\langle {\cal O}^{\chi_{c0}} (^3P_0^{[1]}) \rangle}
\nonumber \\ && 
+
\frac{1}{(d-1) N_c^2} \frac{24 C_F}{\beta_0} \log \frac{\alpha_s
(1.5\textrm{~GeV})}{\alpha_s (4.75\textrm{~GeV})}. 
\end{eqnarray}
where the left-hand side equals ${\cal E}_{11}/[(d-1) N_c^2]$ at scale 
4.75~GeV, and the last line comes from the evolution of this quantity from
scale 1.5~GeV to 4.75~GeV. Numerically, we obtain similar numbers if we
use the model shape functions and integrate over $l_+^*$ up to 
$l_+^{* \rm max} = 4.75\times e^{-1/6}$~GeV. 
This result is compatible with the NLO fit results 
for $\chi_b$ matrix elements~\cite{Han:2014kxa}, and has been used
successfully in phenomenological analyses of $\chi_b$ cross sections based on 
the pNRQCD formalism~\cite{Brambilla:2021abf}. 

\begin{figure}[tbp]
\centering
\includegraphics[width=.49\textwidth]{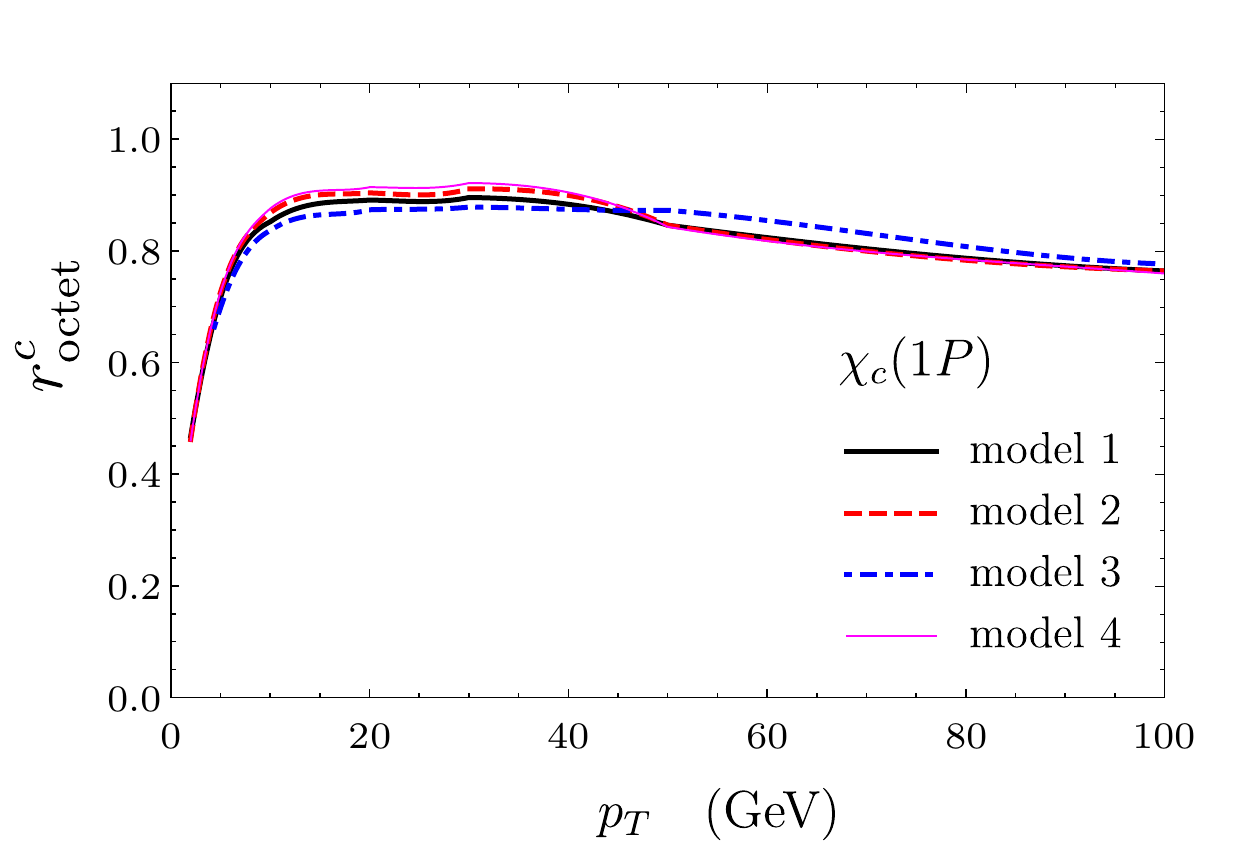}
\caption{\label{fig:rcoctet}
The ratio of the color-octet channel contributions in the shape function and
NRQCD formalisms $r^c_{\rm octet}$ for production of $\chi_c$ 
computed from model shape functions with 
model 1 (black solid line), model 2 (red dashed line), model 3 (blue dot-dashed
line), and model 4 (thin magenta line). 
}
\end{figure}

\begin{figure}[tbp]
\centering
\includegraphics[width=.49\textwidth]{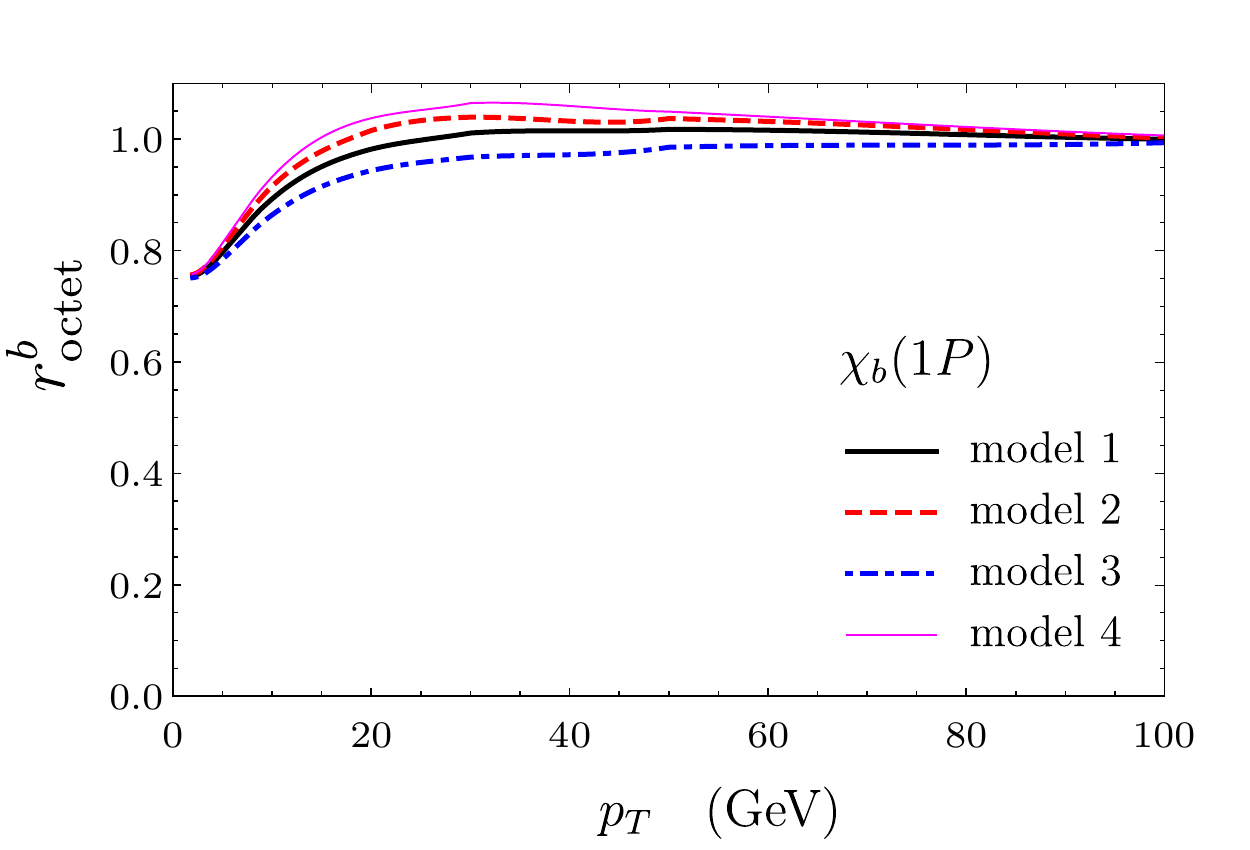}
\includegraphics[width=.49\textwidth]{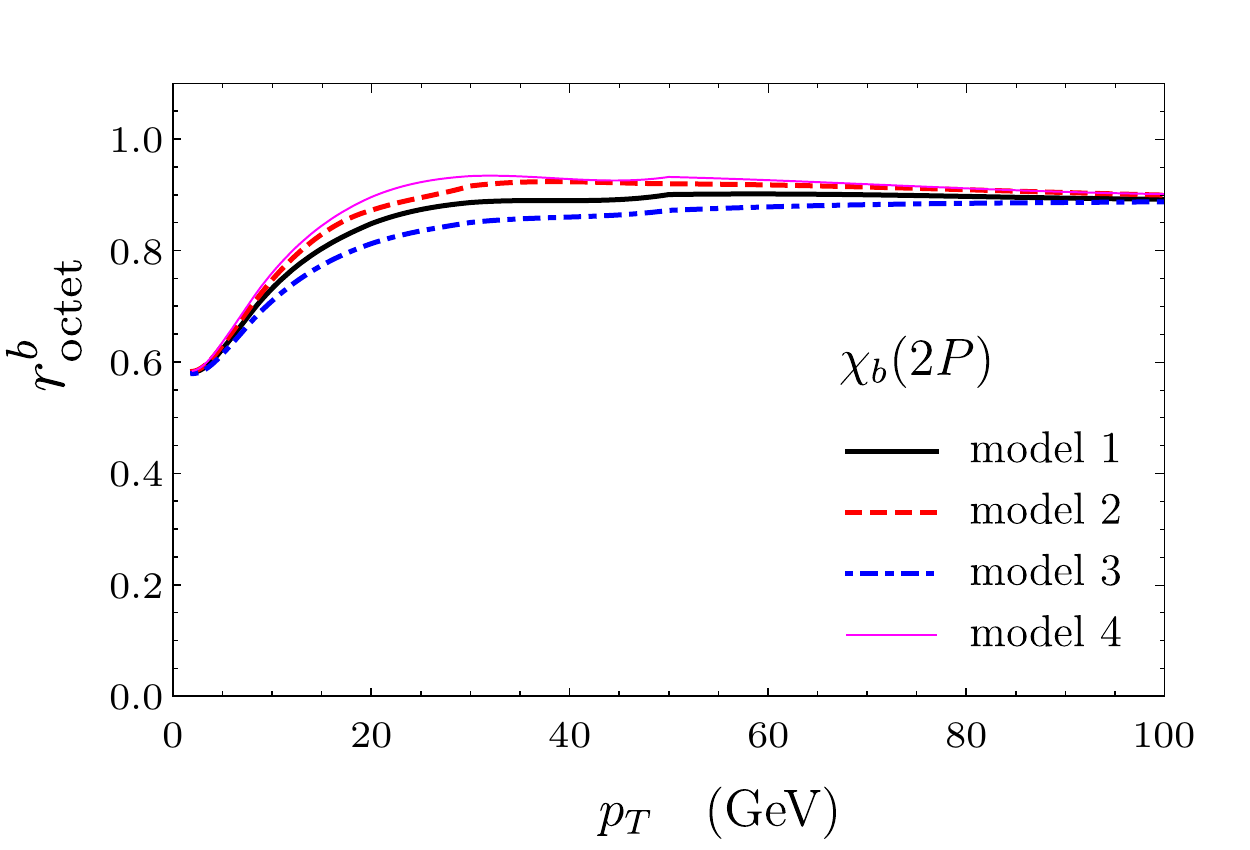}
\caption{\label{fig:rboctet}
The ratio of the color-octet channel contributions in the shape function and
NRQCD formalisms $r^b_{\rm octet}$ for production of $\chi_b(1P)$ (left) 
and $\chi_b(2P)$ (right) computed from model shape functions with
model 1 (black solid line), model 2 (red dashed line), model 3 (blue dot-dashed
line), and model 4 (thin magenta line).
}
\end{figure}
The results for $r^c_{\rm octet}$ for the model shape functions are shown in 
fig~\ref{fig:rcoctet}. 
The ratio is below 1, even though 
$d \sigma[\chi_{QJ}]/d p_T|_{\rm NP}$ is positive; 
this happens because the short-distance coefficient 
$d \sigma[Q \bar Q (^3S_1^{[8]})]/dp_T$
is smaller for the shape function formalism due to the use of a larger value 
of the 
$Q \bar Q$ mass, which reduces the overall normalization of the short-distance
coefficient through the nonrelativistic normalization of the $Q \bar Q$ state. 
The model dependence is small, amounting to less than $\pm 0.022$. 
Regardless of the model chosen to compute 
$d \sigma[\chi_{QJ}]/d p_T|_{\rm NP}$, the ratio 
$r^c_{\rm octet}$ is nearly flat for values of $p_T$ above about 20~GeV,
consistently with the estimate we made in the beginning of this section.
At lower values of $p_T$, the ratio changes slope and steadily decreases as
$p_T$ drops below from about 10~GeV. This is a combined effect of the reduction
of the relative size of $d \sigma[\chi_{QJ}]/d p_T|_{\rm NP}$ compared to 
$d \sigma[\chi_{QJ}]/d p_T|_{\textrm{octet, NRQCD}}$, 
and the reduction of $d \sigma[Q \bar Q (^3S_1^{[8]})]/dp_T$ at 
$m_{Q \bar Q} = m_{\chi_c}$ compared to the $m_{Q \bar Q} = 2 m$ case.
This behavior is similar for $r^b_{\rm octet}$ for $1P$ and $2P$ states, 
as shown in fig.~\ref{fig:rboctet}. 
The deviation of the ratio $r^b_{\rm octet}$ from one is less severe 
than the charmonium case. In the case of $r^b_{\rm octet}$, the model
dependence is less than about $\pm 0.049$ for both $\chi_b(1P)$ and
$\chi_b(2P)$. The values for $r^b_{\rm octet}$ are smaller for $\chi_b(2P)$
compared to $\chi_b(1P)$ at similar values of $p_T$, 
because the $2P$ state is heavier than the $1P$ state. 
Because in all cases model dependences are smaller
than the usual estimates of theoretical uncertainties, 
we neglect the model dependence and compute 
$d \sigma[\chi_{QJ}]/d p_T|_{\rm NP}$ using shape function model 1, 
which gives results that are close to the average of all models considered
here. Model 1 seems to be a reasonable choice, since if we were to interpret 
the nonlocal gluonic operator vacuum expectation value ${\cal E}\!\!\!/_{11}
(l_+)$ as the probability for a gluon with momentum $l$ to propagate to
spacetime infinity, we would expect the probability to vanish as $l \to 0$ due
to confinement. 

\begin{figure}[tbp]
\centering
\includegraphics[width=.59\textwidth]{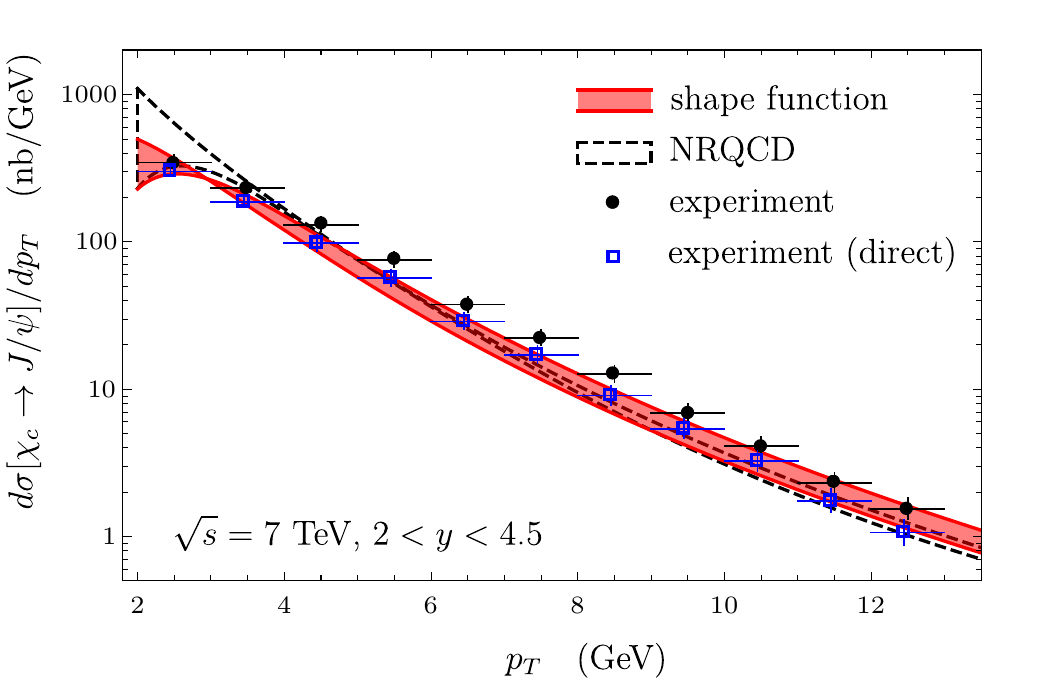}
\caption{\label{fig:chic}
Production rate of $J/\psi$ from $\chi_c$ decays in $p p$ collisions 
at $\sqrt{s} =7$~TeV in the rapidity range $2 < y < 4.5$ computed from the
shape function (red band) and NRQCD (dashed outlined band) formalisms as
functions of the $J/\psi$ transverse momentum. 
Experimental values (black filled circles) are obtained from the
prompt $J/\psi$ cross section and the feeddown fraction measurements from
LHCb~\cite{LHCb:2011zfl, LHCb:2012af}.
The ``direct'' experimental values (blue open squares) are obtained by 
subtracting the contribution from the cascade decay 
$\psi(2S) \to \chi_c + X$ followed by $\chi_c \to J/\psi+\gamma$, which is 
computed from the prompt $\psi(2S)$ cross section measurements from
LHCb~\cite{LHCb:2012geo} and the measured branching fractions in
ref.~\cite{ParticleDataGroup:2022pth}. 
}
\end{figure}
We now compute the cross sections for $\chi_c$ at $pp$ collisions at
center-of-mass energy $\sqrt{s}=7$~TeV with rapidity cut $2<y<4.5$. 
For this we need to determine the values of NRQCD matrix elements. 
We compute the color-singlet matrix element from the well-known result 
$\langle {\cal O}^{\chi_{c0}} (^3P_0^{[1]}) \rangle 
=\frac{3 N_c}{2 \pi} |R'(0)|^2$, which is also consistent with pNRQCD
calculations of the color-singlet matrix element, and use the determination 
$|R'(0)|^2=0.057$~GeV$^{5}$ obtained in ref.~\cite{Brambilla:2021abf} 
from two-photon decay rates of $\chi_{c0}$ and $\chi_{c2}$. 
For the color-octet matrix element, we use the NLO best-fit value 
$m^2 \langle {\cal O}^{\chi_{Q0}} (^3S_1^{[8]}) \rangle/
\langle {\cal O}^{\chi_{Q0}} (^3P_0^{[1]}) \rangle = 0.043$ at 
$\Lambda = 1.5$~GeV~\cite{Brambilla:2021abf}, 
as we have done for the determination of the parameters 
of our model shape functions. 
As we have stated earlier, we compute the NLO short-distance coefficients for
the color-singlet and color-octet channels using the FDCHQHP Fortran
package~\cite{Wan:2014vka} 
with CTEQ6M parton distribution functions~\cite{Pumplin:2002vw}. 
We set the scales for $\alpha_s$ and the parton distributions functions 
to be $m_T = \sqrt{p_T^2 + m_{Q \bar Q}^2}$, and vary them simultaneously 
between $\tfrac{1}{2} m_T$ and $2 m_T$, as have been done in many
phenomenological studies of quarkonium production. 
To match what have been measured in experiments, we compute the combined cross 
sections 
\begin{equation}
\frac{d \sigma[\chi_{c} \to J/\psi]}{dp_T}
= 
\sum_{J=1,2} 
B_{\chi_{cJ} \to J/\psi + \gamma} \frac{d \sigma[\chi_{cJ}]}{dp_T}, 
\end{equation}
as functions of $J/\psi$ transverse momentum $p_T^{J/\psi}$. 
We estimate $p_T^{J/\psi}$ by using the formula 
\begin{equation}
p_T^{J/\psi} = \frac{m_{J/\psi}}{m_{\chi_c}} p_T^{\chi_c}, 
\end{equation}
which follows from the fact that the processes $\chi_{cJ} \to J/\psi + \gamma$
for $J=1$ and 2 are mainly E1 transitions. 
We take the $\chi_{cJ} \to J/\psi + \gamma$ branching fractions 
from ref.~\cite{ParticleDataGroup:2022pth}.
We neglect the contribution from $\chi_{c0}$, because 
the branching fraction $B_{\chi_{c0} \to J/\psi + \gamma}$ is very small.
The results for $d \sigma[\chi_{c} \to J/\psi]/dp_T$ from NRQCD and the shape
function formalisms are shown in fig.~\ref{fig:chic}, 
compared to experimental values obtained from LHCb measurements. 
The experimental values for absolute cross sections have been obtained from the
measurements of the ratio of $\chi_c$ to $J/\psi$ production rates in
ref.~\cite{LHCb:2012af} and
the measured prompt $J/\psi$ production rates in ref.~\cite{LHCb:2011zfl}.
Note that the prompt cross section measurements in ref.~\cite{LHCb:2011zfl} 
include feeddowns from decays of $\psi(2S)$, while our calculation does not; 
to facilitate a better comparison, we also show experimental values
for feeddown-subtracted $\chi_c$ cross sections, which we obtain from the 
measured $\psi(2S)$ cross sections in ref.~\cite{LHCb:2012geo} and the 
branching fractions $B_{\psi(2S) \to \chi_{cJ} + X}$ for $J=1$ and 2 from
ref.~\cite{ParticleDataGroup:2022pth}. 
The difference between results from the NRQCD and the shape function formalisms
are rather small and are within the theoretical uncertainties, except when
$p_T \approx m_{\chi_c}$: for $p_T$ around 3--5~GeV, the cross section in the
shape function formalism is smaller than the result from the NRQCD formalism,
and agrees better with the feeddown-subtracted experimental values. 
While this difference slightly improves agreement with experiment, 
it only amounts to about 30\%, which is the nominal size of
corrections suppressed by $v^2$ that are neglected in both NRQCD and shape
function formalisms. 

\begin{figure}[tbp]
\centering
\includegraphics[width=.49\textwidth]{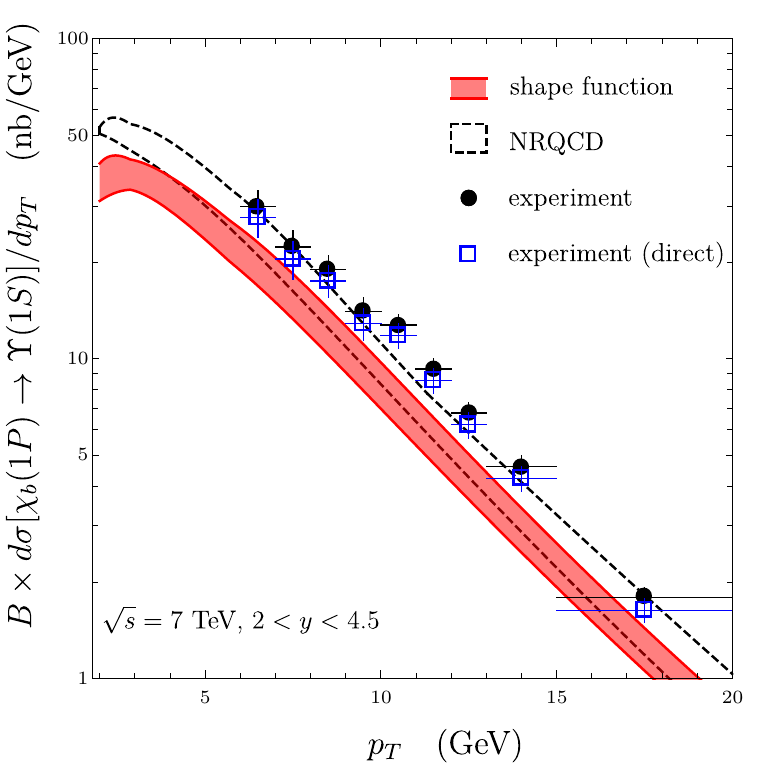}
\includegraphics[width=.49\textwidth]{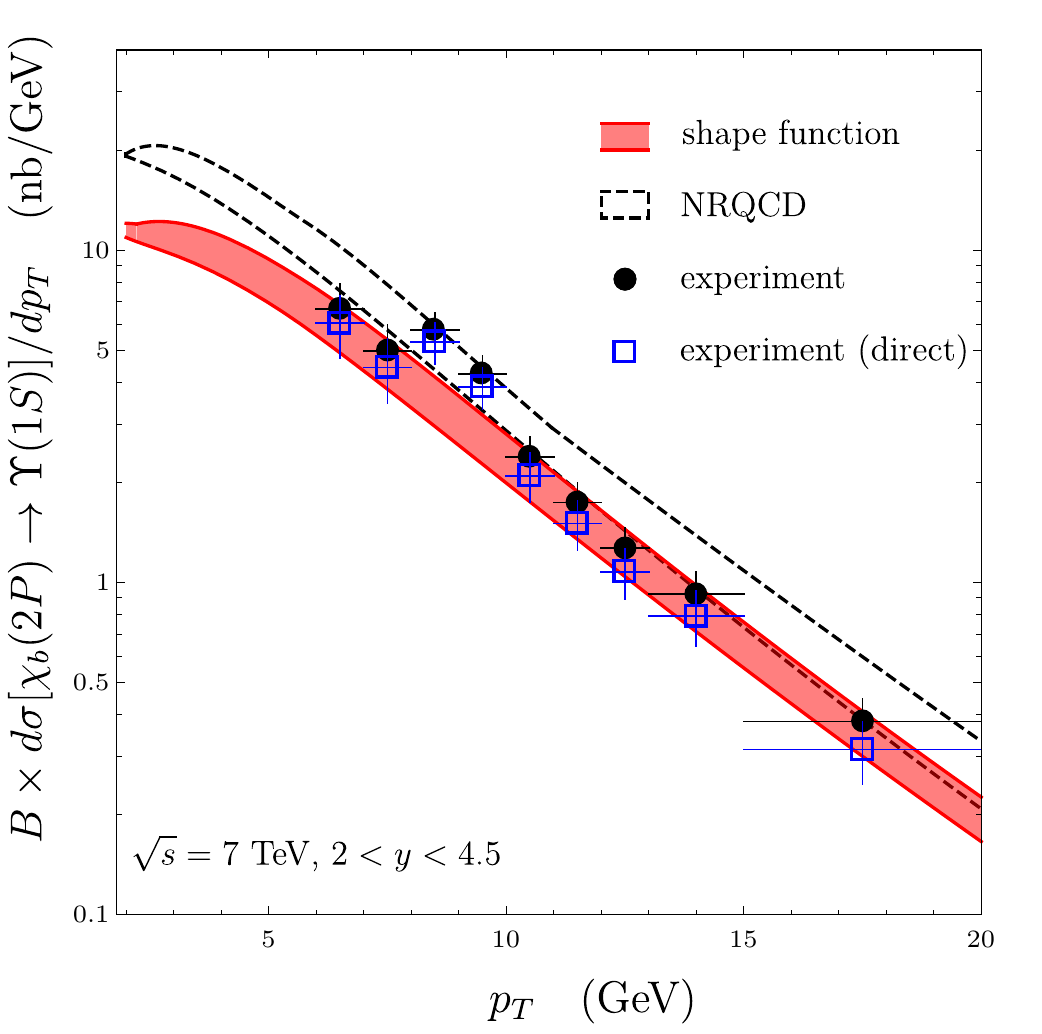}
\caption{\label{fig:chib}
Production rate of $\Upsilon(1S)$ from $\chi_b (1P)$ (left) and 
$\chi_b(2P)$ (right) decays 
times the branching fraction $B \equiv B_{\Upsilon(1S) \to \mu^+ \mu^-}$ 
in $p p$ collisions
at $\sqrt{s} =7$~TeV in the rapidity range $2 < y < 4.5$ computed from the
shape function (red band) and NRQCD (dashed outlined band) formalisms as
functions of the $\Upsilon$ transverse momentum.
Experimental values (black filled circles) are obtained from the
inclusive $\Upsilon$ cross section and the feeddown fraction measurements from 
LHCb~\cite{LHCb:2014ngh, LHCb:2015log}.
The ``direct'' experimental values (blue open squares) are obtained by
subtracting the cascade decay contributions from 
$\Upsilon(2S) \to \chi_b (1P) + X$, 
$\Upsilon(3S) \to \chi_b (1P) + X$, 
and 
$\Upsilon(3S) \to \chi_b (2P) + X$, 
which are computed from the inclusive 
$\Upsilon(2S)$ and $\Upsilon(3S)$ cross section measurements from
LHCb~\cite{LHCb:2015log} and measured branching fractions in
ref.~\cite{ParticleDataGroup:2022pth}.
}
\end{figure}
\begin{figure}[tbp]
\centering
\includegraphics[width=.49\textwidth]{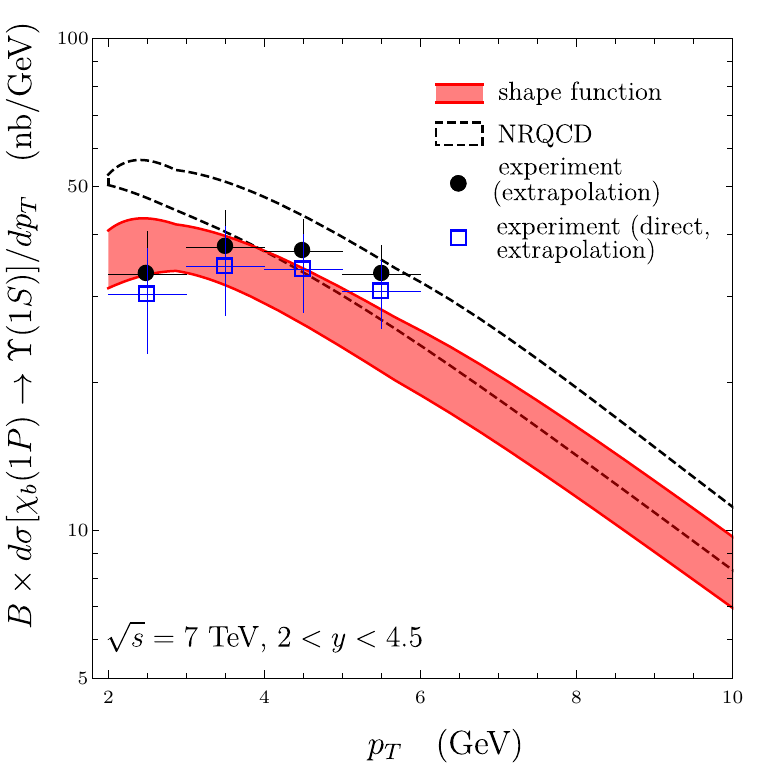}
\includegraphics[width=.49\textwidth]{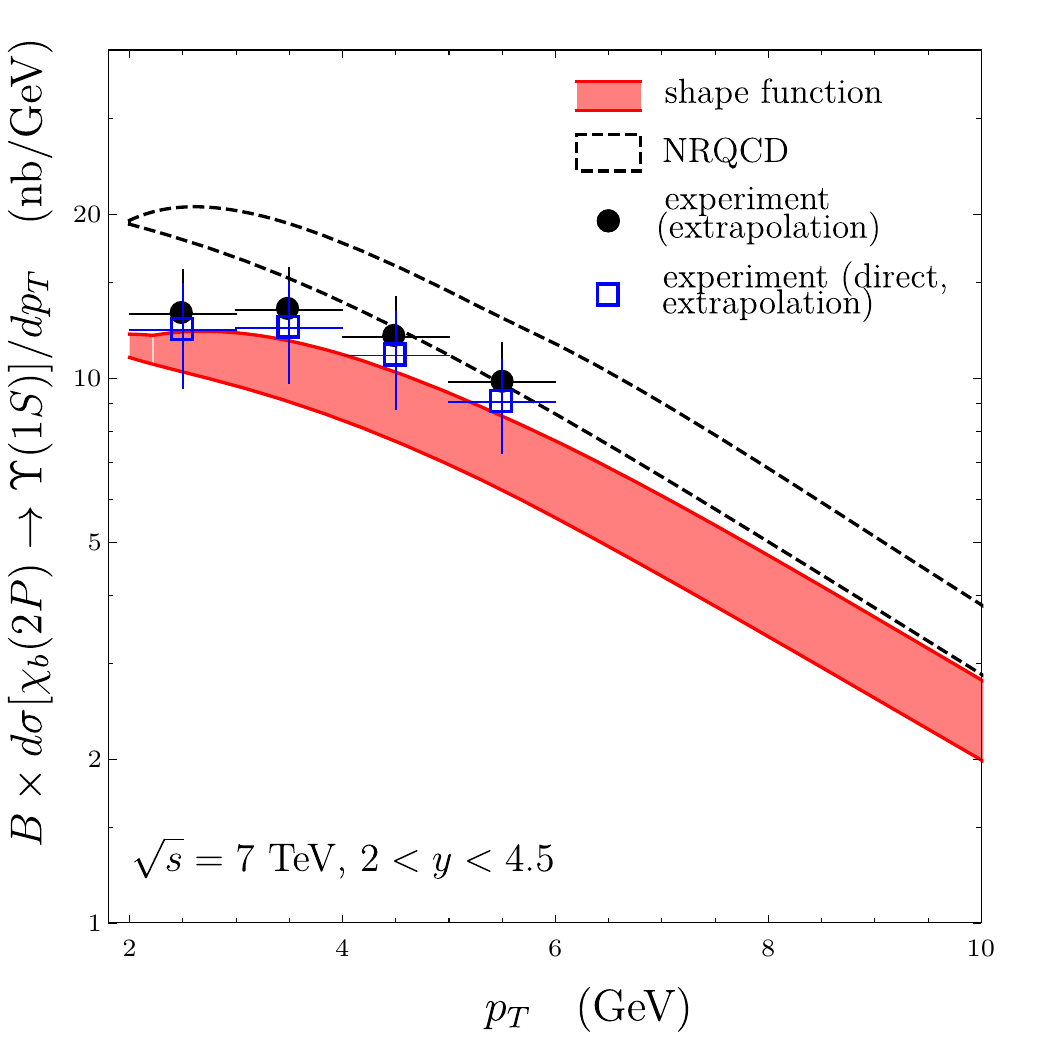}
\caption{\label{fig:chibextra}
Production rate of $\Upsilon(1S)$ from $\chi_b (1P)$ (left) and
$\chi_b(2P)$ (right) decays
times the branching fraction $B \equiv B_{\Upsilon(1S) \to \mu^+ \mu^-}$
in $p p$ collisions
at $\sqrt{s} =7$~TeV in the rapidity range $2 < y < 4.5$ computed from the
shape function (red band) and NRQCD (dashed outlined band) formalisms as
functions of the $\Upsilon$ transverse momentum. 
Experimental values (black filled circles) for $p_T \leq 6$~GeV 
are obtained by using the measured inclusive $\Upsilon$ cross section in 
ref.~\cite{LHCb:2015log} and 
extrapolating the feeddown fraction measurements from LHCb~\cite{LHCb:2014ngh} 
to smaller values of $p_T$. 
The ``direct'' experimental values (blue open squares) are obtained by
subtracting the cascade decay contributions from
$\Upsilon(2S) \to \chi_b (1P) + X$, $\Upsilon(3S) \to \chi_b (1P) + X$, 
and $\Upsilon(3S) \to \chi_b (2P) + X$, 
which are computed from the inclusive
$\Upsilon(2S)$ and $\Upsilon(3S)$ cross section measurements from LHCb~\cite{LHCb:2015log} and measured branching fractions in
ref.~\cite{ParticleDataGroup:2022pth}.
}
\end{figure}
We can repeat the same calculation for bottomonium: we compute cross sections
for $\chi_b(1P)$ and $\chi_b(2P)$ for LHCb kinematics at $\sqrt{s}=7$~TeV
and $2<y<4.5$.
Similarly to the case of $\chi_c$ we need values for the NRQCD matrix elements.
For the color-singlet matrix elements we again use the relation
$\langle {\cal O}^{\chi_{b0}} (^3P_0^{[1]}) \rangle 
=\frac{3 N_c}{2 \pi} |R'(0)|^2$, with 
$|R'(0)|^2= 1.47$~GeV$^5$ for $1P$ and 
$|R'(0)|^2= 1.74$~GeV$^5$ for $2P$ states, which are averages of
potential-model calculations considered in ref.~\cite{Brambilla:2020xod}. 
For the color-octet matrix element, we use the evolution equation in 
eq.~(\ref{eq:octetevolve}) and the best-fit value for the matrix element ratio
for $\chi_c$ to determine $\langle {\cal O}^{\chi_{b0}} (^3S_1^{[8]}) \rangle$
at scale $\Lambda =4.75$~GeV. 
On the experimental side, LHCb measurements are available for the feeddown
fractions in ref.~\cite{LHCb:2014ngh}
for $\Upsilon(nS)$ from $\chi_b(n'P)$ decays for $n$ and $n'$ 
between 1 and 3 with $n'\geq n$, and the absolute cross sections for 
$\Upsilon(nS)$ as functions of $p_T$ of $\Upsilon$ in ref.~\cite{LHCb:2015log}. 
We use the measured values for $\Upsilon(1S)$ cross sections and the feeddown
fractions to obtain the experimental values for $\chi_{bJ}(n'P)$ cross sections
times branching fraction into $\Upsilon(1S)$, summed over $J$. 
To compare with the experimental values, we compute 
\begin{equation}
\frac{B \times d \sigma[\chi_{b}(n'P) \to \Upsilon(1P)]}{dp_T}
= B \times \sum_{J=1,2}
B_{\chi_{bJ}(n'P) \to \Upsilon(1P)+\gamma} 
\frac{d \sigma[\chi_{bJ}(n'P)]}{dp_T},
\end{equation}
where $B \equiv B_{\Upsilon(1S) \to \mu^+ \mu^-}$, 
as functions of $\Upsilon$ transverse momentum $p_T^\Upsilon$. 
We compute $p_T^\Upsilon$ by using 
\begin{equation}
p_T^{\Upsilon} = \frac{m_{\Upsilon}}{m_{\chi_b(n'P)}} p_T^{\chi_b(n'P)}.
\end{equation}
We take the measured branching fractions from
ref.~\cite{ParticleDataGroup:2022pth}.
Similarly to the charmonium case, we neglect the contribution from $\chi_{b0}$
on the account that its branching fraction into $\Upsilon$ is small. 
The results for $d \sigma[\chi_{b}(n'P) \to \Upsilon(1P)]/dp_T$ from NRQCD and
shape function formalisms are shown in fig.~\ref{fig:chib}, 
compared to experimental values. 
As we have done for $\chi_c$, we also show the experimental results with
feeddowns from $\Upsilon(2S)$ and $\Upsilon(3S)$ subtracted by using cross
section measurements in ref.~\cite{LHCb:2015log} and branching fractions from 
ref.~\cite{ParticleDataGroup:2022pth}.
For both $1P$ and $2P$, both formalisms yield similar results for $p_T \gtrsim
10$~GeV, while the shape function formalism predicts smaller cross sections as 
$p_T$ decreases. For $p_T$ around 2 -- 4~GeV, the difference can exceed the
theoretical uncertainties estimated from scale variations. The difference
between the central values of the results from the two formalisms can be almost
20\% for $1P$ and more than 30\% for $2P$ states, which also exceed the nominal
size of the order-$v^2$ corrections that are neglected in both formalisms. 
Unfortunately, the measurements for the feeddown fractions in
ref.~\cite{LHCb:2014ngh} are only available for $p_T \geq 6$~GeV, 
so it is currently not possible to tell which formalism 
would give better descriptions of small-$p_T$ behaviors of the cross sections. 
Since the measured feeddown fractions drop slowly as $p_T$ decreases, 
we could attempt to estimate the experimental values of the 
small-$p_T$ cross sections 
$d \sigma[\chi_{b}(n'P) \to \Upsilon(1P)]/dp_T$ at $p_T < 6$~GeV by 
linearly extrapolating the feeddown fractions and multiplying by the measured
$\Upsilon(1S)$ cross sections in ref.~\cite{LHCb:2015log}. 
The result is shown in fig.~\ref{fig:chibextra} compared to predictions from
NRQCD and shape function formalisms. 
It can be seen that for both $1P$ and $2P$ states, 
the shape function formalism leads to results that agree better with the
extrapolation of the experimental results. 
Nevertheless, actual measurements at small $p_T$ will be needed 
to tell whether the shape function formalism provides better descriptions of 
low-$p_T$ cross sections, although this can be challenging
due to diminishing experimental efficiency as $p_T$ decreases. 

Before concluding this section let us discuss the possibility of further
extending the formalism to even lower values of $p_T$. In this case, there are
contributions from processes where the $Q \bar Q$ is produced with vanishing
transverse momentum, and evolves into a quarkonium by emitting soft gluons. 
The color-singlet channel involves the contribution where 
$Q \bar Q(^3S_1^{[8]})$ is first produced at rest, which evolves into 
$Q \bar Q(^3P_J^{[1]})$ by emitting a gluon with momentum $k$. 
This produces a singular plus distribution in $p_T$ where the $p_T$
differential cross section diverges as $p_T \to 0$, although the integrated
cross section is finite. Because the effect of gluon emission can become
nonperturbative, it is tempting to replace the perturbative calculation of the
gluon emission with a nonperturbative shape function, in a way similar to 
eq.~(\ref{eq:shapefac_f1}). This way, the effect of the
soft-gluon emission at small gluon momentum $k$ would be described by the
product of the $Q \bar Q(^3S_1^{[8]})$ cross section at vanishing transverse
momentum, and the color-octet shape function with $l_+ = 2 |\bm{k}|$. 
Because the shape functions are isotropic, the shape function times the 
$Q \bar Q(^3S_1^{[8]})$ cross section predicts that the singular part of the 
$Q \bar Q(^3P_J^{[1]})$ cross section is isotropic in the quarkonium momentum. 
It is easy to see that this is not the case: the simplest way is to 
look into the parton cross sections for $q \bar q \to Q \bar Q(^3P_J^{[1]})+g$,
which are available as functions of the partonic Mandelstam variables in
ref.~\cite{Baier:1983va}. 
From the analytical expressions for $d \hat{\sigma}/d\hat{t}$, 
we obtain the singular parts of the 
differential cross sections at the parton CM frame given by 
\begin{equation}
\label{eq:qqQQsinglet}
\hat\sigma[q \bar q \to Q \bar Q(^3P_J^{[1]})]\big|_{\rm singular}
= (2 J+1) \langle {\cal O}^{Q \bar Q (^3P_0^{[1]})} (^3P_0^{[1]})\rangle
\frac{8 \alpha_s^3 \pi^2}{729 m^6} \frac{1}{|\bm{P}|}
C_J (\cos \theta) d \cos \theta, 
\end{equation} 
where we neglect the terms that are finite when the size of the quarkonium
three-momentum $|\bm{P}|$ vanishes, $\theta$ is the scattering angle, 
and $C_J(\cos \theta)$ are the angular distributions given by 
\begin{subequations}
\begin{eqnarray}
C_{J=0} (x) &=&  
\frac{3 (1+x^2)}{8}, 
\\
C_{J=1} (x) &=&  
\frac{3(3-x^2) }{16}, 
\\
C_{J=2} (x) &=&  
\frac{3(13+x^2) }{80}, 
\end{eqnarray}
\end{subequations}
which are normalized as 
$\int_{-1}^{+1} C_J(\cos \theta) d \cos \theta = 1$.  
From the expression for the $d$-dimensional two-body phase space available in 
ref.~\cite{Petrelli:1997ge}, we can see that the factor $|\bm{P}|^{-1}$ 
in eq.~(\ref{eq:qqQQsinglet}) becomes 
$|\bm{P}|^{-1-2 \epsilon}$ in $d=4-2 \epsilon$ dimensions, 
so that the cross sections contain an IR pole at $|\bm{P}| = 0$ given by 
\begin{equation}
\label{eq:qqQQoctet}
\hat\sigma[q \bar q \to Q \bar Q(^3P_J^{[1]})] \big|_{\rm pole} 
= (2 J+1) \langle {\cal O}^{Q \bar Q (^3P_0^{[1]})} (^3P_0^{[1]})\rangle
\frac{8 \alpha_s^3 \pi^2}{729 m^6}
\frac{\delta(|\bm{P}|) }{- 2 \epsilon_{\rm IR}} 
C_J(\cos \theta) d \cos \theta. 
\end{equation}
Upon integrating these expressions over $\cos \theta$ we reproduce the IR pole 
at $|\bm{P}| = 0$ and the singular contribution at $|\bm{P}| \to 0$ given 
in eq.~(186) of ref.~\cite{Petrelli:1997ge}. 
Compare this with the $\sigma[q \bar q \to Q \bar Q(^3S_1^{[8]})]$ cross
section at order $\alpha_s^2$ given by~\cite{Petrelli:1997ge}
\begin{equation}
\hat\sigma[q \bar q \to Q \bar Q(^3S_1^{[8]})] 
= \frac{\alpha_s^2 \pi^3}{54 m^4} \delta(|\bm{P}| )
\langle {\cal O}^{Q \bar Q(^3S_1^{[8]})} (^3S_1^{[8]}) \rangle, 
\end{equation}
which implies that the process $q \bar q \to Q \bar Q(^3S_1^{[8]})$, 
followed by $Q \bar Q(^3S_1^{[8]}) \to Q \bar Q(^3P_J^{[1]})$ by soft gluon
emission, involves the IR pole 
\begin{equation}
(2 J+1) \frac{\alpha_s^2 \pi^3}{54 m^4} \delta(|\bm{P}| )
\times \frac{4 \alpha_s C_F}{3 N_c \pi m^2} 
\langle {\cal O}^{Q \bar Q (^3P_0^{[1]})} (^3P_0^{[1]})\rangle 
\left( - \frac{1}{2 \epsilon_{\rm IR}} \right), 
\end{equation}
which is isotropic, and reproduces the IR pole in the $P$-wave color-singlet 
cross section {\it only after completely integrating over the scattering angle}. 
In actual experiments, cuts are usually made so that a
complete integration over the scattering angle is not possible; 
especially, for $p_T$-differential cross sections $p_T = |\bm{P}| \sin \theta$
is fixed. Hence, NRQCD factorization fails at vanishingly small values of 
quarkonium momentum. 
In contrast, at a nonzero quarkonium momentum $\bm{P}$, 
there is always a finite neighborhood of $\bm{P}$ in phase space where the
angles of the soft gluon momentum $\bm{k}$ can be completely integrated over, 
so that the cancellation of IR poles always occur. Even in this case, the size 
of the neighborhood of $\bm{P}$, given by $\bm{P}+\bm{k}$, is limited by the 
maximum available energy $|\bm{k}|$ of the soft gluon emission. 
If this maximum energy is too small, such as the case
where $p_T$ is smaller than the NRQCD factorization scale, radiative
corrections due to soft gluon emission will involve large logarithms of the
ratio of the NRQCD factorization scale and the maximum available soft gluon
energy. 
Therefore, NRQCD factorization will become unreliable for
values of the transverse momentum below the NRQCD factorization scale, and
completely break down at zero transverse momentum. 
The shape function formalism fails for the same reason: convolving the
order-$\alpha_s^2$ color-octet cross section 
(\ref{eq:qqQQoctet}) with the perturbative shape function coincides with
the singular part of the color-singlet cross section (\ref{eq:qqQQsinglet})
only after completely integrating over the scattering angle, 
which in practice cannot be done when comparing with experiment. 

\begin{figure}[tbp]
\centering
\includegraphics[width=.55\textwidth]{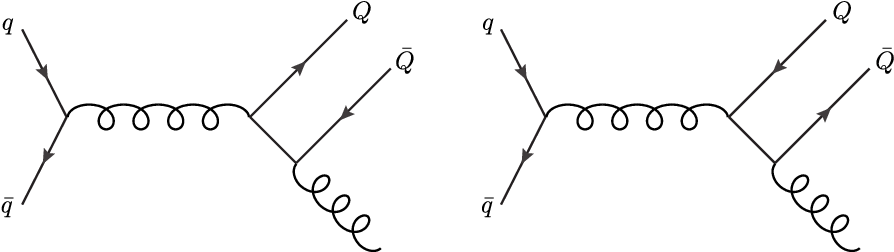}
\caption{\label{fig:QQg}
Tree-level Feynman diagrams for the process 
$q \bar q \to Q \bar{Q} (^3P_J^{[1]}) + g$. }
\end{figure}
The root cause of the breakdown of factorization is that 
the infrared divergent factor in eq.~(\ref{eq:irdivfac}) contains a dependence
on the relative momentum $q = (p_1 - p_2)/2$ between
the heavy quark and antiquark coming from the heavy quark propagator
denominator. This dependence cannot be dropped, as doing so will lead to an
incorrect result for the singular contribution to the cross section 
in eq.~(\ref{eq:singularpart}). 
As can be seen from the Feynman diagram for the process 
$q \bar q \to Q \bar Q(^3P_J^{[1]})+g$ shown in fig.~\ref{fig:QQg}, 
the denominators of eq.~(\ref{eq:irdivfac}) come from the 
virtual heavy quark lines, which have virtualities much smaller than $m$ 
when the gluon momentum $k$ is soft. Hence, this 
propagator must be included in the long-distance matrix element. 
If the $Q \bar Q$ is produced with transverse momentum much larger than the 
soft gluon momentum $k$, we can take the average over the angles of $\bm{k}$ 
because the effect of variation in $k$ on the kinematics of the quarkonium is
at most order $m v$ and can be neglected, 
and we obtain the usual result in eq.~(\ref{eq:singularpart}). 
However, in the case where the $Q \bar Q(^3S_1^{[8]})$ is produced at rest and 
evolves into $Q \bar Q(^3P_J^{[1]})$ plus a soft gluon, the gluon momentum is 
strongly entangled with the kinematics of the quarkonium as 
$\bm{k} = - \bm{P}$. 
In this case, the long-distance matrix element is sensitive to the 
details of the propagator between the heavy quark and antiquark, 
which does not allow a description in terms of a local operator product of 
the heavy quark and antiquark fields. That is, a non-local operator product of 
the heavy quark and antiquark fields on the same side of the projection
operator is needed to describe such processes. To make matters worse, in this
case it is not even clear whether it would be possible to isolate the nonlocal
operator product of the heavy quark and antiquark fields into a vacuum
expectation value, because soft interactions between initial and final state 
partons will no longer become suppressed, unlike the case of large-$p_T$ 
production. 
A similar conclusion was obtained in the analysis of 
transverse-momentum dependent factorization of $P$-wave quarkonium production
in ref.~\cite{Ma:2014oha}, where the authors considered the process 
$gg \to Q \bar Q(^3P_J^{[1]}) + X$, which is more complicated because 
the process $gg \to Q \bar Q(^3S_1^{[8]})$ vanishes at tree level. 

As we have shown, this breakdown of NRQCD factorization at $p_T \gtrsim 0$ 
signals a failure of the factorization formalism in terms of 
local operator matrix elements. This reveals an important distinction
between the description of $B$ hadron production in heavy quark effective
theory~\cite{Isgur:1989vq} 
and quarkonium production in NRQCD. 
In heavy quark effective theory, where long-distance matrix elements involve 
the $b$ quark field in place of the $Q \bar Q$ bilinears, the $b$ quark field 
is always a local field. In contrast, in quarkonium production nothing 
prevents the heavy quark and antiquark from being produced 
nonlocally; this nonlocality can be neglected if it is disentangled from the
short-distance process, and in this case we recover the standard form of NRQCD
factorization. However this is not always guaranteed to happen, 
and can fail for production with small transverse momentum 
as we have seen from the example above.

\section{Summary and conclusion}
\label{sec:summary}

In this work we investigated the application of the shape function formalism 
for resumming kinematical effects in inclusive production rates of $P$-wave
heavy quarkonia. In this formalism, the products of short-distance coefficients
and NRQCD matrix elements are replaced by convolutions of 
short-distance coefficients and shape functions. 
The shape functions encode the kinematical effects arising from the motion of
the heavy quark and antiquark pair in the heavy quarkonium rest frame. 
They are nonperturbative functions of a lightlike 
momentum $l$, which corresponds
to the difference between the momentum of the quarkonium and the momentum of
the heavy quark-antiquark pair. While the formalism was originally developed
in refs.~\cite{Beneke:1997qw, Fleming:2003gt} 
based on tree-level calculations of quarkonium production, 
we have extended the application of the formalism to next-to-leading order in
the strong coupling, which requires treatment of the ultraviolet divergences
that arise from renormalization of NRQCD matrix elements. 
Due to the mixing between color-singlet and color-octet
channels induced by the infrared divergence in the color-singlet $P$-wave
production cross section, the short-distance coefficients differ between the
NRQCD and shape function formalisms. 
Based on the perturbative calculations of the shape functions at
next-to-leading order accuracy in sec.~\ref{sec:diagram}, 
we have derived in sec.~\ref{sec:matching} the 
matching conditions that are valid at one-loop level which allow determination
of the short-distance coefficients in the shape function formalism from the 
standard NRQCD ones. This allows calculation of the cross
sections in the shape function formalism at one-loop level using the known
NRQCD short-distance coefficients in the $\overline{\rm MS}$ scheme, 
once the nonperturbative shape functions are known. 

In sec.~\ref{sec:pNRQCD}
we presented an improved analysis of the nonperturbative shape functions
for $P$-wave heavy quarkonia. By using the potential NRQCD formalism
developed in refs.~\cite{Brambilla:2020ojz, Brambilla:2021abf,
Brambilla:2022rjd, Brambilla:2022ayc}, we reproduced the proposed form of the
color-singlet shape function in ref.~\cite{Beneke:1997qw} 
that is proportional to $\delta(l_+)$, where $l_+$ is the $+$
component of the lightlike momentum $l$ associated with the shape function
formalism. Together with the requirement that the normalization of the shape
function must be equal to the corresponding NRQCD matrix element, this
completely fixes the color-singlet shape function up to corrections suppressed
by powers of $v$. 
In the case of the color-octet shape function, in existing studies it was
assumed that it would not be possible to calculate them except within models. 
While first-principles determinations of the nonperturbative 
color-octet shape functions may be currently out of reach, we find that the 
large-$l_+$ asymptotic form fixed by the perturbative calculation, 
and the normalization condition 
that its integral over $l_+$ must reproduce the 
NRQCD matrix element, strongly constrain the $l_+$ dependence of the shape
function. Furthermore, we showed through a
nonperturbative potential NRQCD calculation of the shape functions
that the $l_+$ dependence of the color-octet shape 
functions must be same for all $P$-wave heavy quarkonia, independently of the
radial excitation or the heavy quark flavor, up to corrections suppressed by
powers of $v$. This strongly constrains the behavior of the shape function, 
and the model dependence is constrained in the small $l_+$ region.
As the contribution from the small $l_+$ region to the cross section 
is suppressed, 
this model dependence has only a small effect on the $P$-wave quarkonium 
cross section. Hence, the analysis in this work diminishes the model dependence
of the shape function formalism for production of $P$-wave heavy quarkonia. 

Based on the results for the one-loop perturbative matching conditions and
nonperturbative shape functions obtained in this work, 
we presented two phenomenological applications. 
First, in section~\ref{sec:threshold} we used the fact that the color-octet
shape function describes soft-gluon emissions 
to match the accuracies in $\alpha_s$ of the singular radiative 
corrections near threshold between the color-singlet and color-octet
states. This alleviates the problem in fixed-order calculations that, when the 
short-distance coefficients are computed to same accuracy in the strong
coupling, the singular threshold effects are truncated to different orders in
$\alpha_s$ in the color-octet and color-singlet channels. 
This can be problematic
especially at very large transverse momentum, where the threshold effects 
can be enhanced. As we have shown in section~\ref{sec:threshold}, 
standard NRQCD calculations can yield negative production rates of $\chi_c$; 
the situation is much more improved when we match the singular threshold
effects between the two channels 
using the results from the shape function formalism, and the positivity
of the cross sections are ensured to very large values of transverse
momentum. While this does not eliminate the need for threshold resummation, 
the matching of the threshold effects considered in this work can be useful for
consistent treatment of threshold resummation throughout the singlet and octet
channels. 

In section~\ref{sec:NPmodel}, we investigated the impact of the nonperturbative
shape function on the transverse-momentum dependent cross sections of $\chi_c$
and $\chi_b$. We employed a variety of model functions for the color-octet shape
function to compute the nonperturbative corrections, where the model parameters
are constrained by the phenomenologically obtained values of the color-octet
matrix elements. As we have argued above, the model dependence only affects the
$l_+ \to 0$ behaviors of the shape functions, 
and have little effect on the cross sections, 
so that the model dependence can even be neglected compared to other
theoretical uncertainties. 
We find that at large $p_T$, the effect of nonperturbative corrections is
mainly to change the overall normalization of the color-octet contribution,
which is inconsequential in phenomenology because a change in normalization
can be reabsorbed into the color-octet matrix element. 
This situation gradually changes as we go to lower transverse
momentum, and at values of the quarkonium transverse momentum similar or
smaller than the heavy quarkonium mass, the cross section in the shape function
formalism undershoots the NRQCD prediction by more than the usual size of 
estimated theoretical uncertainties obtained from scale variations. 
Although this result is interesting, it is currently not possible to tell 
whether the shape
function formalism gives a better description of the $P$-wave heavy quarkonium 
cross sections at small $p_T$; 
in the case of $\chi_c$, the difference between the two
formalisms is not too large compared to the nominal size of relativistic 
corrections suppressed by $v^2$, which is usually taken to be about $0.3$ for
charmonium. 
In the case of $\chi_b$, the results from the shape function formalism does
tend to agree better with experimental values obtained by extrapolating the
feeddown fractions of $\Upsilon$ from $\chi_b$ decays to $p_T \leq 6$~GeV; 
nevertheless, actual measurements will be necessary to tell if this is indeed
the case.

As we have mentioned, 
the $P$-wave quarkonium production cross sections are also important for
understanding the feeddown contributions in $S$-wave quarkonium production 
rates. It would also be interesting to apply the shape function formalism for
production of $S$-wave quarkonia, such as $J/\psi$, $\psi(2S)$, and $\Upsilon$ 
states. Production of $S$-wave quarkonia is phenomenologically more
significant, and measurements are available for a wider range of transverse
momentum. 
As have been discussed in ref.~\cite{Brambilla:2022ayc}, the
color-octet NRQCD matrix elements that appear in $S$-wave quarkonium production
rates not only involve logarithmic UV divergences that induce mixing between
different color-octet channels, but also quadratic power divergences as well,
unlike the $P$-wave case. 
We can expect that these quadratic power divergences will appear in color-octet 
shape functions as contributions proportional to $l_+$. Since in perturbative
QCD these contributions will lead to scaleless power UV divergences, they
will need to be subtracted in order to make contact with NRQCD short-distance
coefficients that are computed in dimensional regularization. 
Due to this the nonperturbative shape functions may affect the $S$-wave cross
sections differently from $P$-wave cross sections. 
It would be interesting to see how the application of the shape function 
formalism will affect $S$-wave quarkonium cross sections.

\acknowledgments

I thank Geoffrey Bodwin for helpful discussions on the perturbative analysis of
NRQCD matrix elements and resummation of threshold logarithms. 
This work is supported by the National Research Foundation of Korea
(NRF) Grant funded by the Korea government (MSIT) under Contract No.
NRF-2020R1A2C3009918 and by a Korea University grant.

\bibliography{shapechipaper.bib}
\bibliographystyle{JHEP}

\end{document}